\documentclass[aps,rmp,reprint,groupedaddress]{revtex4-1}

\usepackage{graphicx}
\usepackage{dcolumn}
\usepackage{bm}

\begin{document}


\title{Looking at cosmic near-infrared background radiation anisotropies}


\author{A. Kashlinsky}
\email[]{Alexander.Kashlinsky@nasa.gov}
\altaffiliation{Code 665, Observational Cosmology Lab, Goddard Space Flight Center, 
Greenbelt, MD 20771 and
SSAI, Lanham, MD 20770, USA}
\author{R. G. Arendt}
\altaffiliation{Code 665, Observational Cosmology Lab, Goddard Space Flight Center, 
Greenbelt, MD 20771 and CRESST/University of Maryland, Baltimore County, USA}
\author{F. Atrio-Barandela}
\altaffiliation{Dept of Theoretical Physics, University of Salamanca, Spain}
\author{N. Cappelluti}
\altaffiliation{Yale Center for Astronomy and Astrophysics, P.O. Box 208120, New Haven, CT 06520, USA and Department of Physics, University of Miami, Coral Gables, Florida 33124, USA}
\author{A. Ferrara}
\altaffiliation{Scuola Normale Superiore, Piazza dei Cavalieri 7, I-56126 Pisa, Italy}
\author{G. Hasinger}
\altaffiliation{Institute for Astronomy, 2680 Woodlawn Drive, University of Hawaii, Honolulu, HI 96822, USA}



\begin{abstract}
The cosmic infrared background (CIB) contains emissions accumulated over the entire history of the Universe, including from objects inaccessible to individual telescopic studies. The near-IR ($\sim1-10$ $\mu$m) part of the CIB, and its fluctuations, reflects emissions from nucleosynthetic sources and gravitationally accreting black holes (BHs). If known galaxies are removed to sufficient depths the source-subtracted CIB fluctuations at  near-IR can reveal sources present in the first-stars-era and possibly new stellar populations at more recent times. This review discusses the recent progress in this newly emerging field which identified, with new data and methodology, significant source-subtracted CIB fluctuations substantially in excess of what can be produced by remaining known galaxies. The CIB fluctuations further appear coherent with unresolved cosmic X-ray background (CXB) indicating a very high fraction of BHs among the new sources producing the CIB fluctuations. These observations have led to intensive theoretical efforts to explain the measurements and their properties. While current experimental configurations have limitations in decisively probing these theories, their potentially remarkable implications will be tested in the upcoming CIB measurements with the ESA's {\it Euclid} dark energy mission. We describe the goals and methodologies of LIBRAE (Looking at Infrared Background Radiation with {\it Euclid}), a NASA-selected project for CIB science with {\it Euclid}, which has the potential for transforming the field into a new area of precision cosmology. 
\end{abstract}



\def\plotone#1{\centering \leavevmode
\epsfxsize=\columnwidth \epsfbox{#1}}

\def\wisk#1{\ifmmode{#1}\else{$#1$}\fi}

\def\wm2sr {Wm$^{-2}$sr$^{-1}$ }		
\def\nw2m4sr2 {nW$^2$m$^{-4}$sr$^{-2}$\ }		
\def\nwm2sr {nWm$^{-2}$sr$^{-1}$\ }		
\def\nw2m4sr {nW$^2$m$^{-4}$sr$^{-1}$\ }
\def\Ncut {$N_{\rm cut}$\ }
\def\lt     {\wisk{<}}
\def\gt     {\wisk{>}}
\def\le     {\wisk{_<\atop^=}}
\def\ge     {\wisk{_>\atop^=}}
\def\lsim   {\wisk{_<\atop^{\sim}}}
\def\gsim   {\wisk{_>\atop^{\sim}}}
\def\kms    {\wisk{{\rm ~km~s^{-1}}}}
\def\Lsun   {\wisk{{\rm L_\odot}}}
\def\Msun   {\wisk{{\rm M_\odot}}}
\def\um     { $\mu$m\ }
\def\sig    {\wisk{\sigma}}
\def\etal   {{\sl et~al.\ }}
\def\eg	    {{\it e.g.\ }}
\def\ie     {{\it i.e.\ }}
\def\bsl    {\wisk{\backslash}}
\def\by     {\wisk{\times}}
\def\cosec {\wisk{\rm cosec}}
\def\mic {\wisk{ \mu{\rm m }}}

\def\amin   {\wisk{^\prime\ }}
\def\asec   {\wisk{^{\prime\prime}\ }}
\def\cc     {\wisk{{\rm cm^{-3}\ }}}
\def\deg     {\wisk{^\circ}}
\def\ddeg   {\wisk{{\rlap.}^\circ}}
\def\damin  {\wisk{{\rlap.}^\prime}}
\def\dasec  {\wisk{{\rlap.}^{\prime\prime}}}
\def\approxeq{$\sim \over =$}
\def\abouteq{$\sim \over -$}
\def\percm{cm$^{-1}$}
\def\percmsq{cm$^{-2}$}
\def\percmcub{cm$^{-3}$}
\def\perhz{Hz$^{-1}$}
\def\perpc{$\rm pc^{-1}$}
\def\persec{s$^{-1}$}
\def\peryr{yr$^{-1}$}
\def\te{$\rm T_e$}
\def\tenup#1{10$^{#1}$}
\def\to{\wisk{\rightarrow}}
\def\thin{\thinspace}
\def\uk{$\rm \mu K$}
\def\p{\vskip 13pt}

\def\HI {H$\;${\small I}}
\def\HII {H$\;${\small II}}
\def\HeI {He$\;${\small I}}
\def\HeII {He$\;${\small II}}

\def\aj       {Astron. J.}
\def\apj      {Astrophys. J.}
\def\apjs     {Astrophys. J. Suppl. Ser.}
\def\apjl     {Astrophys. J. Lett.}
\def\aap      {Astron. Astrophys.}
\def\aaps     {Astron. Astrophys. Suppl. Ser.}
\def\araa     {Annu. Rev. Astron. Astrophys.}
\def\physrep  {Phys. Rep.}
\def\mnras    {Mon. Not. R. Astron Soc.}
\def\procspie {Proc. SPIE}
\def\pasj     {PASJ}
\def\pasa     {PASA}
\def\aapr     {AApR}
\def\ssr      {SSR}

\maketitle

\tableofcontents

\clearpage
\setcounter{page}{1}

\section{Introduction}
\label{sec:intro}

Development of modern physical cosmology began in earnest in the early 20th century, when Hubble's insight into the radial velocities of
galaxies yielded the first observational evidence for expanding Universe models
proposed as an immediate consequence of Einstein's gravitational field equations. The now accepted theory of the Universe's origin is known as the Big Bang model, named so in jest in 1950 by one of the proponents of the alternative ``steady-state" cosmology, Sir Fred Hoyle.  The
discovery of the Cosmic Microwave Background  (CMB) radiation in 1964 \citep{Penzias:1965}, anticipated by \citet{Alpher:1948a,Alpher:1948}, added a firm
observational pillar in support of the Big Bang theory \citep{Dicke:1965}.  The {\it COBE}/FIRAS measurements \citep{Mather:1990} revealed a highly accurate black-body energy spectrum for the CMB confirming  its origin in the hot dense early phase of the Big Bang. The CMB angular structure, uncovered first with the {\it COBE}/DMR measurements \citep{Smoot:1992},
provided an unprecedented insight into the density field of the Universe a mere $\sim$400,000 years after the Big Bang.

This century has so far marked the emergence of precision cosmology, when the fundamental cosmological parameters and the contributions of the Universe's basic 
constituents to its matter and energy budget have been accurately determined. A standard cosmological model describing the evolution of structure has been 
established in agreement with the observations of CMB angular fluctuations on sub-degree scales as probed by balloons \citep{de-Bernardis:2000,Lange:2001} and post-{\it COBE} finer resolution {\it WMAP} \citep{Bennett:2013} and Planck \citep{Planck-Collaboration:2015} satellites. In addition, our understanding of high-energy physics is now sufficiently advanced for connecting cosmological phenomena to the quantum 
physics of the primordial Universe. 

The widely accepted cosmological concordance model requires large amounts of dark matter (DM), as well as dark energy (DE)
of unknown nature and origin, and -- in broad terms -- explains the Universe's structure as follows. The large-scale isotropy of the Universe, as well as the 
small-scale inhomogeneities that evolved into galaxies and galaxy clusters, are thought to be the result of a period of early accelerated expansion, termed inflation \citep{Kazanas:1980,Guth:1981,Linde:1982}. 
While the precise mechanism driving inflation and the underlying pre-inflationary structure of space-time are still unknown, the matter
density field predicted by inflation is now established observationally on scales $\gsim 10$Mpc, which encompass masses $\gsim 10^{14}M_\odot$. Smaller scales subtend structures presently in nonlinear regime where the original density field is not probed directly.

Much of the progress has been made through observational and theoretical studies of the CMB. It was only recently, however, that a lesser known relative of the CMB, the cosmic infrared background (CIB) started getting attention. CIB contains emissions over the entire history of the Universe, including from sources inaccessible to direct telescopic studies. The latter category includes the epoch when first stars were born as well as possible new populations at later times. The near-IR (1-10\mic) CIB, the subject of this review, probes emissions from early stars and black holes (BHs) \cite{Partridge:1967,Bond:1986,McDowell:1986,Santos:2002,Salvaterra:2003,Cooray:2004,Kashlinsky:2004}. To isolate the part of the CIB from new, potentially interesting cosmological sources, resolved galaxies must be excised from the maps to sufficiently faint levels. The remaining, source-subtracted CIB can then be compared to that expected after ``reasonable" extrapolations, based on other data, from remaining known galaxies. An excess, if significant, would potentially reveal important cosmological information on the nature of the new sources, their epochs, abundances, and the density field in which they reside.

Measurements over the past decade from analyses of {\it Spitzer} by \citet{Kashlinsky:2005a,Kashlinsky:2007a,Kashlinsky:2012}, \citet{Cooray:2012} and {\it AKARI} \cite{Matsumoto:2011} satellite data identified near-IR CIB fluctuations remaining in deep integrations on sub-degree and degree scales.   It appears that these fluctuations cannot originate from remaining known galaxy populations \citep{Helgason:2012a,Kashlinsky:2005a}. It was further found that the CIB fluctuations are coherent with unresolved soft cosmic X-ray background (CXB) at levels much higher than expected from remaining known populations suggesting significantly greater BH proportions among the CIB sources than in known populations \cite{Cappelluti:2013,Cappelluti:2017,Helgason:2014,Mitchell-Wynne:2016}. The extensive list of empirical properties for these CIB fluctuations, discussed below, provides a further important set of clues to the origin of these sources. While some of the CIB properties, such as its amplitude, can be modeled with {\it new} populations at intermediate redshifts, other empirical evidence points toward the fluctuations originating at early epochs, possibly the ``first stars era".  New programs, specifically on the upcoming dark energy {\it Euclid} mission, will probe with unprecedented accuracy and scope the CIB from high redshifts, and its properties, enabling unique insight into the era of the first luminous sources, identifying their nature and the properties of the underlying density field at those epochs.

This review summarizes the current state of the near-IR CIB fluctuation measurements, their potentially remarkable theoretical implications, and discusses the future prospects of this rapidly developing field. Wherever the context permits, we will plot results in terms of the original quantities displayed in the corresponding measurement papers.

Common acronyms and abbreviations used throughout the review are listed in Table \ref{tab:abbreviations} in Sec.\ \ref{sec:appendix} (Appendix).

\section{Background cosmology and definitions}
\label{sec:cosmology}

Brightnesses of resolved sources are given in the AB magnitude system 
\citep{Oke:1983}, where the flux density, $S_\nu$, of a source is related 
to the AB magnitude by $S_\nu = S_0\ 10^{-0.4 m_{AB}}$ with $S_0 = 3631$ Jy 
$ = 3631\times10^{-26}$ W~m$^{-2}$~Hz$^{-1}$. The surface brightnesses, $I_\nu$, 
of extended sources are often given in units of MJy sr$^{-1}$. It is common 
practice to express surface brightness per $\log{\nu}$ instead of $\nu$ by 
defining a flux as 
$F = I_\nu (d\nu/d\log{\nu}) = \nu I_\nu = (c/\lambda)I_\nu = \lambda I_\lambda$. 
Thus, in commonly used units, 
$F\ {\rm [nW\ m^{-2}\ sr^{-1}]} = (3000/\lambda\ [\mu{\rm m}])\ I_\nu\ \rm{[MJy\ sr^{-1}]}$.
It is instructive to convert CIB surface brightness levels into their comoving photon number density:  $n_{\rm CIB}(\nu)=\frac{4\pi}{c}I_\nu/h_{\rm Planck} = 0.63 \left(\frac{I_\nu}{\rm MJy/sr}\right)$ cm$^{-3}$. For comparison the CMB photons are orders-of-magnitude more abundant with $n_{\rm CMB}=413$ cm$^{-3}$.

CMB observations 
established the flat geometry of the Universe. We will thus adopt the Friedman-Robertson-Walker flat metric for the Universe with the interval given by $ds^2=c^2dt^2-(1+z)^{-2}(dx^2+x^2d\omega)$ where $z, x,t,\omega$ are the redshift, comoving coordinate distance, cosmic time, and solid angle. Photons move along null geodesics, $ds^2=0$.  The Friedman equations with the matter, dark energy ($\Lambda$),  radiation/relativistic component and curvature density parameters $\Omega_m, \Omega_\Lambda, \Omega_\gamma, \Omega_K$ lead to
 $c(1+z)dt/dz = R_H/E(z)$, where $E(z)\equiv [\Omega_\gamma(1+z)^4+ \Omega_m(1+z)^3+\Omega_K(1+z)^2+\Omega_\Lambda f(z)]^{1/2}$ with $f(z)$ describing the evolution of DE and $R_H\equiv cH_0^{-1}$.
The Hubble constant is $H(z)=H_0E(z)$ and the distance measures become: the coordinate distance $x(z)=c\int(1+z) dt = R_H \int_0^z dz/E(z)$, the comoving angular diameter distance $d_A(z)=x(z)$ and the luminosity distance $d_L(z)=(1+z)x(z)$.  Proper distances/scales are $(1+z)^{-1}\times$(comoving distances/scales) and proper time intervals are $(1+z)\times$(cosmic time).
 
We adopt the cosmological parameters $\Omega_K=0$, $\Omega_\Lambda=0.72$, $\Omega_{\rm matter}=0.28$, $\Omega_{\rm baryon}=0.045$, $h=0.71$, and $\sigma_8=0.9$ where the present-day Hubble constant is $H_0=100h$ km/sec/Mpc. $\sigma_8$ is the present-day linear matter density contrast over a sphere of comoving radius $r_8=8h^{-1}$Mpc \citep{Davis:1983}. The total mass contained on average within the comoving radius $r$ is $M(r)=4.9\times 10^{11}(r/1h^{-1}{\rm Mpc})^3$ and the baryonic mass is $M_{\rm baryon}(r)=7.8\times 10^{10}(r/1h^{-1}{\rm Mpc})^3$. We also adopt $\Omega_\gamma=0$ and, for the bulk of the review, $f(z)=1$, equivalent to the DE reflecting the vacuum energy density resulting in a cosmological constant $\Lambda$ or equation of state with pressure ${\cal P}=-\rho c^2$. 

The origin of structures in the Universe is tied to matter density fluctuations, $\delta_{\rm m}(\vec{x})$, produced during inflationary expansion. During matter-dominated era, these fluctuations feel stronger gravitational field and expand at a slower rate than the average Universe: fluctuations grow until they become non-linear, separate from the comoving frame and collapse. The fluctuations represent a stochastic random field, which can then be decomposed into independent Fourier modes. The variance of each mode is defined as the power spectrum. The correlation function, $C(|\vec{x}_1-\vec{x}_2|)\equiv\langle\delta_{\rm m}(\vec{x}_1)\delta_{\rm m}(\vec{x_2})\rangle$,  is then the Fourier transform of the power spectrum. The isotropy of the Universe requires that the correlation function depends only on the absolute value of the separation distance and the power spectrum only on the absolute wavenumber. The power spectrum defines all properties of Gaussian random fields, such as inflation-produced density fluctuations.
 
\begin{figure}[t]
\includegraphics[width=3.in]{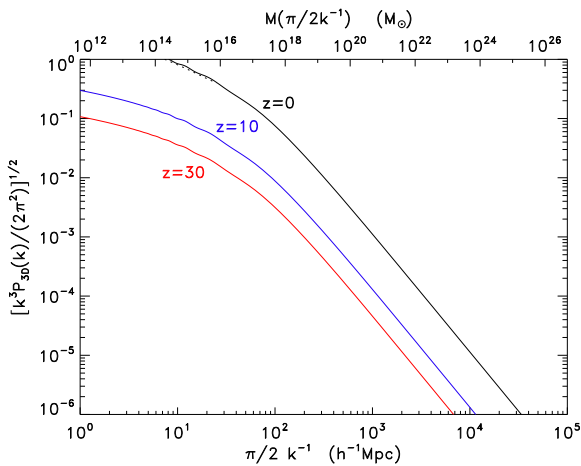}
\caption[]{\small{ The rms spectrum of primordial adiabatic density field in $\Lambda$CDM concordance model at $z=0,10,30$.} At still higher $z$, the Universe is in the Einstein-de Sitter regime and the density field can be scaled $\propto (1+z)^{-1}$. At smaller scales the power spectrum approaches the regime $P_{\rm 3D}\propto k^{-3}$. In spherical collapse model, fluctuations containing mass $M$ with $\sqrt{k^3P_{\rm 3D}/(2\pi^2)}\gsim \delta_{\rm col}=1.68$ would collapse by $z$. Dotted line shows $\sigma_M(r)$ vs $r$ at $z$=0.}
\label{fig:p_lcdm}
\end{figure}

The standard cosmological model gives the power spectrum of the primordial adiabatic component of matter density fluctuations produced during the inflationary roll-over, which is in agreement with CMB observations. This component of the matter density fluctuations starts with an approximately Harrison-Zeldovich (HZ) slope of the 3-D power spectrum $P_{3D,{\rm initial}}(k)\propto k$ \citep{Guth:1982}, which is preserved on scales above the horizon at matter-radiation equality, but gets modified by differential growth of smaller wavelength harmonics \citep{Bond:1984}. The resultant 3-dimensional power spectrum. $P_{3D}(k)$, normalized to $\sigma_8$ at $z$=0,  is shown in Fig.\ \ref{fig:p_lcdm}. The inflationary density field is highly Gaussian and so is fully specified by its power spectrum.
  An important scale, to serve as a standard ruler, imprinted in the spatial spectrum is that of the baryonic acoustic oscillations (BAOs) at the acoustic horizon at decoupling, $r_{\rm BAO}\simeq$150Mpc \citep{Eisenstein:1998,Eisenstein:1999}.   
  The mean squared amplitude of density fluctuations over a given radius $r$ is given by $\sigma_M^2(r)=\frac{1}{2\pi^2}\int k^2P_{\rm 3D} (k) W_{\rm TH}(kr) dk$ with $W_{\rm TH}(x)=[3j_1(x)/x]^2$, $j_n$ being the spherical Bessel function of order $n$; it is shown in Fig.\ \ref{fig:p_lcdm} with dotted line at $z$=0. Such measurements on a scale $r = \pi/2\:k^{-1}$ correspond to an effective 
sampling interval of $\Delta = 2r$, and thus a (minimum) spatial wavelength 
of $2\Delta = 2\pi k^{-1}$. Later, measurements of angular fluctuations are 
characterized in terms of the equivalent angular wavelength $\theta = 2\pi q^{-1}$.

If the dark matter (DM) is made up of primordial black holes (PBHs), as motivated recently \citep{Bird:2016,Clesse:2016,Kashlinsky:2016} by the LIGO gravitational wave (GW) discovery \citep{Abbott:2016,Abbott:2016a}, the Poissonian fluctuations due to PBHs would provide an extra isocurvature density fluctuation component discussed first by \citet{Meszaros:1974,Meszaros:1975} before the inflationary paradigm was introduced. The addition to power in density fluctuations  at the time of the PBH formation from that component would be a constant $P_{\rm PBH, initial}=n_{PBH}^{-1}$  in comoving units. This component would add to small scale power of the density field, increasing the efficiency of early collapse of first halos \citep{Kashlinsky:2016}.

Starting at matter-radiation equality, matter fluctuations grow $\delta\rho/\rho \propto (1+z)^{-1}$ until the epoch when cosmological constant dominates, $(1+z)\gsim 3$, and the growth slows down. As fluctuations turn non-linear they separate from the comoving frame of expansion and collapse to form virialized halos which can host luminous sources forming out of the collapsing baryonic gas, provided it can efficiently cool to below the halo virial temperature. 

Once the luminous sources form at high redshift, $z \gsim 10$, their UV emission
near 0.1 $\mu$m would contribute to the present CIB near $\gsim 1 \mu$m, and their cosmological power spectrum of clustering would be reflected in the CIB angular anisotropies. Hard X-ray emission from high-$z$ BHs would contribute to the 
soft X-ray CXB, and related UV emission would make this coherent with the CIB. 
 
The primordial density field in the standard $\Lambda$CDM cosmological model, as shown in Fig.\ \ref{fig:p_lcdm}, is such that after the Universe recombines at $z_{\rm rec}\sim 1,000$, there is an extended period, nicknamed the ``Dark Ages'', when no luminous sources existed and everything was made up of neutral hydrogen (\HI) until the dark halos collapsing at $z\lsim 40-50$ produced the first luminous sources. The near-IR CIB provides a new powerful tool to study the emergence of the Universe from the Dark Ages and the nature of the early luminous sources; recent CIB fluctuation results may have already produced tantalizing insight into these questions.
 
 The nature of the first luminous sources, and proportions of BHs among them, are currently unknown together with the luminosity density they produced as the Universe started emerging from the Dark Ages. If dominated by massive stars and/or accreting BHs, the first luminous sources would radiate at the Eddington limit, $L_{\rm Edd}\propto M$, so that the net bolometric luminosity density produced by them is insensitive to the details of their mass function, $n(M)$, since $\int n(M)L dM \propto \rho$(in sources) \citep{Rees:1978}. This leads to the net bolometric flux roughly equal to the maximal luminosity of any gravitating object, $L_{\rm max}=c^5/G$, distributed over the Hubble radius ($R_H=cH_0^{-1}$) sphere, or $F\sim L_{\rm max}/(4\pi R_H^2)$ times model-dependent parameters, such as the fraction of baryons in these sources, the redshift of emission and their radiation efficiency \citep{Kashlinsky:2004}. For sources at $z\gsim 10$, these emissions would go primarily into the near-IR CIB and with a net flux which is significant for realistic parameter values. While this part of the mean CIB is challenging to isolate from other components, it would have substantial fluctuations with a distinct spatial distribution reflecting the underlying matter power spectrum at those epochs.
 
 There are intuitive reasons why there would be potentially significant, measurable CIB fluctuations from the first stars era \cite{Cooray:2004,Kashlinsky:2004}: 1)  first stars are predicted to have been massive, with luminosity per unit mass larger than present-day stellar populations by a factor $\sim 10^4$; a similar factor applies to accreting Eddington-limited BHs;
2) their relative CIB fluctuations would be larger as they span a
relatively short time-span in the evolution of the universe; and
3) these sources formed at the peaks of the underlying
density field, amplifying their clustering properties.

\section{Mean levels of Background Light}
\label{sec:meancib}

We define the Extragalactic Background Light (EBL) at wavelengths from UV to 10 \mic\ to be the sum of all emissions from extragalactic sources. The CIB 
is the EBL at IR wavelengths, and in 
this review we focus on the near-IR CIB at 1 to 10 \mic. 
The cosmic optical background (COB, 0.1-1 \mic) has similar origins as the CIB, 
but is restricted to sources at $z\lesssim7$ \citep{Bernstein:2007,Kawara:2017,Mattila:2017,Mattila:2017a}.
The CXB is the net diffuse emissions from 0.5 to 100 keV, with the soft X-ray CXB referring to the range of [0.5--2] keV.

In this section we discuss the status of the mean levels of the backgrounds. Previous reviews by \citet{Hauser:2001,Kashlinsky:2005} covered the status of the measurements prior to 2004 and the reader is referred to these papers for overviews. Here we will mainly discuss the progress and the new results obtained since that time, referring to the earlier results only briefly when required for clarity and completion.

\subsection{Galaxy counts and resolved EBL/CIB}
\label{sec:gcounts}

The total flux from counted galaxies in deep surveys gives a direct lower bound on the CIB, identifying the contribution to it from the known resolved populations. A possible CIB excess over that component would contain contributions from new extragalactic populations.  Since populations' energy emissions are cut off below the Lyman-cutoff wavelength, $\sim 0.1(1+z)\mic$, the wavelength dependence of the CIB would indicate the epochs when it arises. This situation from optical to near-IR bands has been discussed in review by \citet{Kashlinsky:2005} using the counts data available at the time. The updated discussion is presented below.
 
\begin{figure*}
\includegraphics[width=7in, trim= 1.5cm 0 0 0]{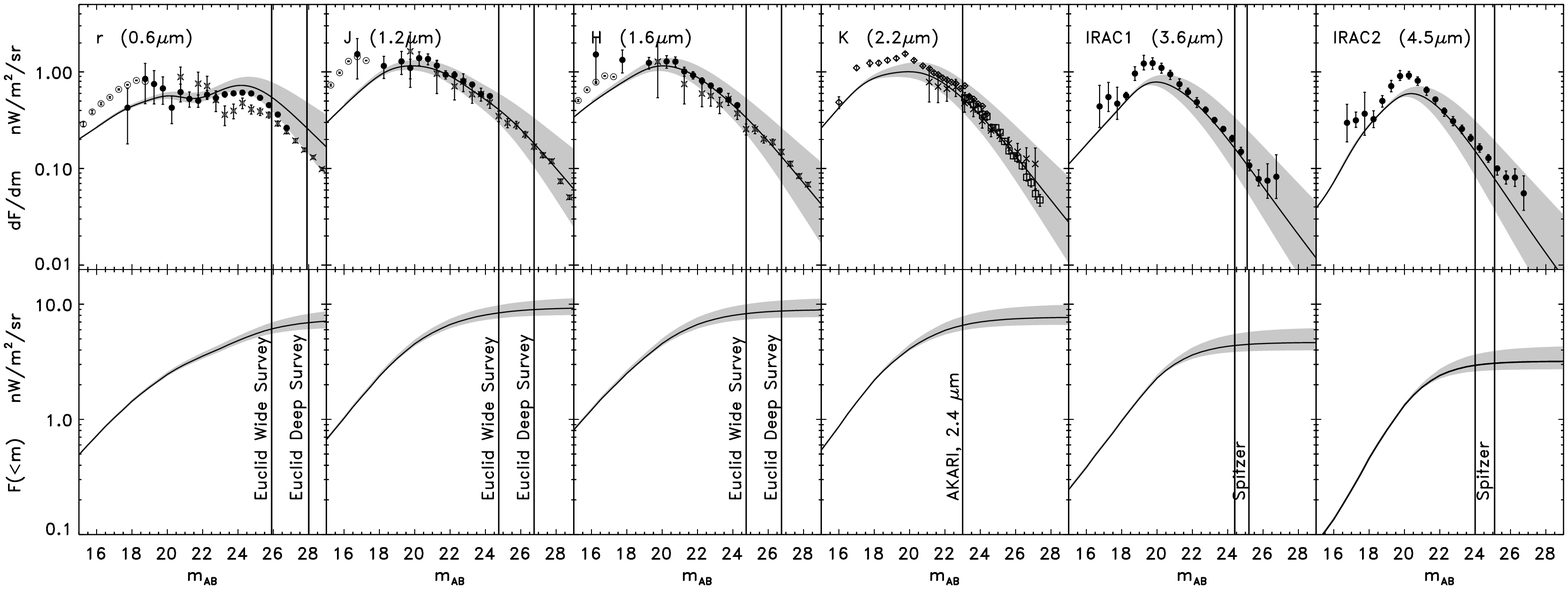}
\caption{\small  Differential and cumulative flux from data vs HRK12 reconstruction (shaded bands span the range from HFE to LFE with solid line showing the default reconstruction). Diffuse fluxes at r, J, H bands are obtained using counts from Fig. 12 of \citet{Windhorst:2011} in the same notation, K-counts from \citet{Keenan:2010} (open squares) and \citet{Maihara:2001} (crosses), and at 3.6 and 4.5 \mic\ the IRAC counts from \citet{Ashby:2013,Ashby:2015} are used at $m_{\rm AB}>16$ where they are less polluted by Galaxy star counts. In K-band, \citet{Fontana:2014} present more updated counts, which are consistent with what is shown, but extend to shallower magnitudes ($m_{\rm AB}<26$) and are not displayed here for clarity.  
}
\label{fig:cib_counts}
\end{figure*}
Following earlier determinations of deep counts in the near-IR  
\citep[e.g.][and many other authors]{Madau:2000,Fazio:2004,Totani:2001,Totani:2001a,Gardner:1993,Glazebrook:1994} further deeper counts covering wider range of wavelengths have been obtained with new ground and spaceborne instruments  \citep[][etc]{Thompson:2005,Keenan:2010,Ashby:2013,Ashby:2015,Windhorst:2011,Driver:2016}. Fig.\ \ref{fig:cib_counts} shows the build-up of the EBL and CIB from the deepest optical (represented by r-band) and near-IR counts available as of this review. The CIB contributions from known galaxy populations peak at AB mag $\sim 20-21$ with little additional contribution out to AB$\gsim 28$ \citep[e.g.\ see detailed discussion in][]{Kashlinsky:2005,Driver:2016}.
The figure also shows the reconstructed CIB from known populations with the methodology of \citet[][hereafter HRK12]{Helgason:2012a} discussed later in Sec.\ \ref{sec:hrk12}. This heuristic reconstruction follows the counts data very accurately, especially at the faint end relevant here; the small deviations at the bright end may be due to pollution from star counts and other systematics. 
\begin{table}
\caption{HRK reconstruction of diffuse flux (nW/m$^2$/sr).}
\begin{tabular} {| p{0.475in} | p{0.375in} | p{0.375in} |  p{0.375in} | p{0.45in} | p{0.375in} | p{0.45in} |}
 \hline 
{ \footnotesize } & { \footnotesize r} & {\footnotesize J} & { \footnotesize H} & {\footnotesize K/2.4\mic} & { \footnotesize 3.6\mic} & { \footnotesize 4.5\mic}\\
 \hline
{ \footnotesize Net $F$} & { \footnotesize $7.5_{-1.2}^{+2.5}$} & { \footnotesize $9.4_{-1.3}^{+2.4}$} & {\footnotesize $9.0_{-1.2}^{+2.5}$} &  {\footnotesize $7.7_{-1.1}^{+2.4}$} & {\footnotesize $4.7_{-0.7}^{+1.6}$}  & { \footnotesize $3.2_{-0.5}^{+1.2}$} \\
\hline
{\footnotesize $m_0$} & { \footnotesize 26} & { \footnotesize 24.5} & {\footnotesize 24.5} &  {\footnotesize 23} & {\footnotesize 25}  & { \footnotesize 25} \\
{ \footnotesize $F(>\!m_0)$} & { \footnotesize $1.4_{-0.6}^{+1.6}$} & { \footnotesize $1.0_{-0.4}^{+1.0}$} & {\footnotesize $0.8_{-0.3}^{+0.9}$} &  {\footnotesize $1.1_{-0.4}^{+1.1}$} & {\footnotesize $0.2_{-0.1}^{+0.4}$}  & { \footnotesize $0.14_{-0.08}^{+0.31}$} \\
 \hline
\end{tabular}
\label{tab:cib_counts}
\end{table}
Table \ref{tab:cib_counts} gives the CIB estimates in the HRK12 reconstruction for all galaxies and for those remaining below current or future limiting magnitudes. 
These agree well with the net CIB flux integrated directly from galaxy counts \citep[Table 5 of ][]{Kashlinsky:2005}; see also Fig. 4 of \citet{Driver:2016} for updated diffuse fluxes from counts \citep{Beckwith:2006,Windhorst:2011,Bouwens:2010}.

\citet{Driver:2016} derive from compiling counts survey data the net EBL of $24\pm4$ nW/m$^2$/sr between UV and 10 \mic\ and $26\pm5$ nW/m$^2$/sr in far-IR, 10--1000\mic. Integrating in the near-IR range of [1--5] \mic, would give $9^{+3}_{-1}$ nW/m$^2$/sr for the CIB contribution of known sources according to the HRK12 reconstruction.

The upshot of this discussion is that 1) galaxy counts from known populations produce finite CIB out to at least $m_{\rm AB}\gsim28$,  2) these counts are well approximated with the heuristic CIB/EBL reconstruction developed by \citet[][]{Helgason:2012a}/HRK12, and 3) any excess CIB, if found, must then arise in new populations, which are too faint or too distant to be detected, or both.

\subsection{Direct measurements of CIB}

Subtraction of solar system and Galactic foregrounds \citep[see review by][]{Leinert:1998} from space-based 
measurements of the absolute sky brightness, yield direct measurements of 
the mean CIB. Observations are usually obtained over multiple wavelengths,
and mean CIB estimates are derived by averaging data over large areas.
This approach was adopted by both the
{\it COBE}/DIRBE and the {\it IRTS}/NIRS instruments and its results were reviewed extensively before \cite{Kashlinsky:2005,Hauser:2001}. 
Here we briefly review only the new results on the mean near-IR CIB 
that appeared since the \citet{Kashlinsky:2005} review.

There have been several efforts to apply new modeling and analysis techniques
to existing data sets to make improved estimates of the mean CIB.
Updated results from the DIRBE measurements have been provided by 
\citet{Levenson:2008} and by \citet{Sano:2015, Sano:2016}. The {\it IRTS}
data were reexamined by \citet{Matsumoto:2015}. These reanalysis generally 
lead to smaller systematic uncertainties than earlier estimates, but like 
earlier work, they point to the presence of $\sim1-5\ \mu$m IR 
emission in excess of that expected from the integrated light of 
known galaxies (and zodiacal and Galactic foregrounds). New independent 
observations of the sky brightness by the {\it AKARI} spacecraft, 
also reinforce this picture \citep{Tsumura:2013}. New observations
from the CIBER suborbital mission, extend the CIB measurements to shorter
wavelengths, but indicate that the spectrum is flattening 
or falling from 1.25 to 0.8 $\mu$m \citep{Matsuura:2017}.
Deep {\it HST}/NICMOS observations have been analyzed by 
\citet{Thompson:2007,Thompson:2007a}, who report no near-IR CIB excess 
above the levels contributed by known galaxies to within a few 
nW/m$^2$/sr. However, the empirical method they apply is sensitive to 
structure in the CIB, but does not distinguish the mean CIB from 
the empirically subtracted zodiacal light.

\begin{figure}[t]
\includegraphics[width=3.5in, trim = 1cm 0 0 0 ]{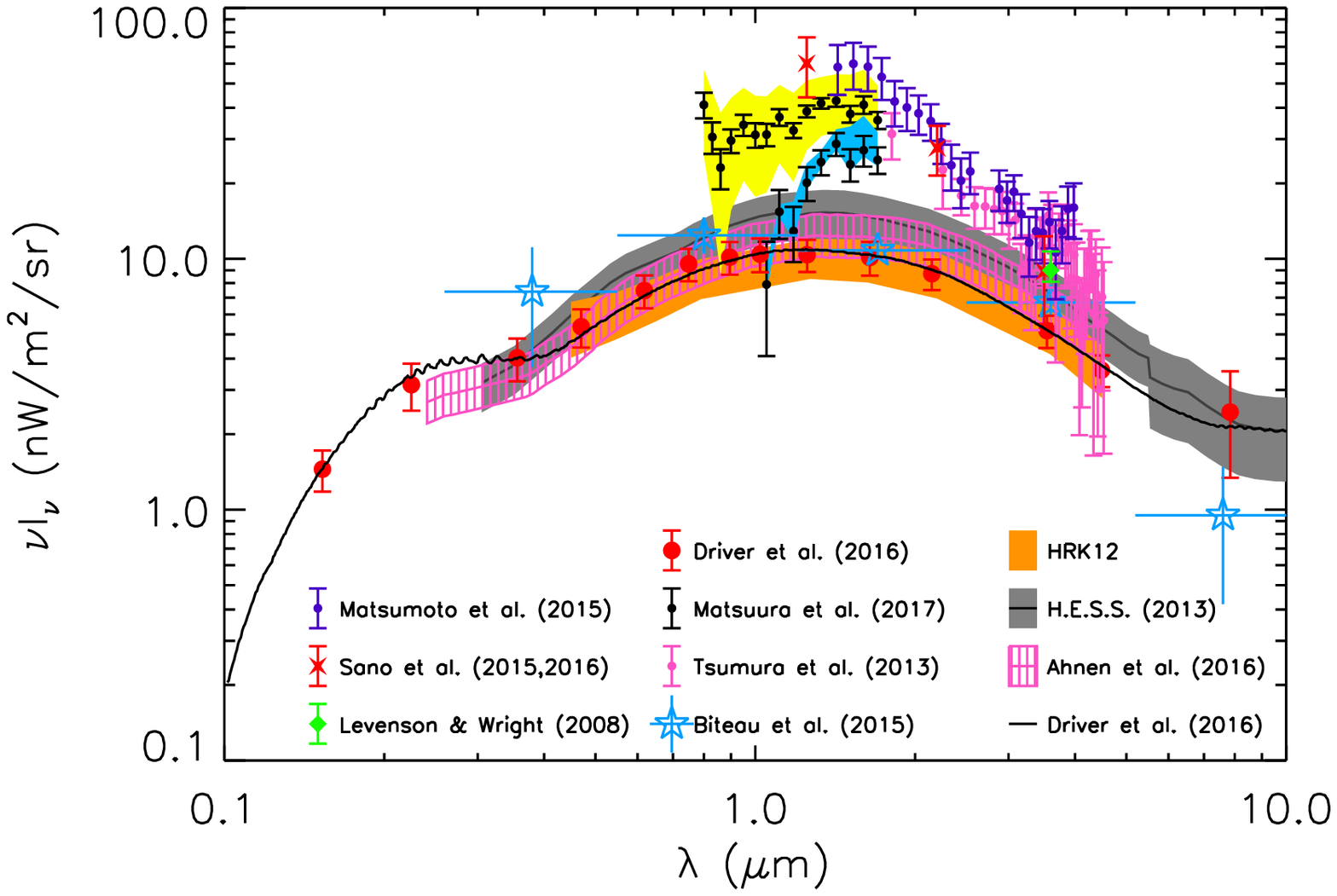}
\caption{\small  The SED of the mean CIB as derived from recent direct 
measurements and reanalyses: 
\citet{Levenson:2008} (green diamond), \citet{Sano:2015, Sano:2016} (red crosses),
\citet{Matsumoto:2015} (violet dots), \citet{Tsumura:2013} (pink dots),
\citet{Matsuura:2017} (black dots, points with yellow/blue (upper/lower) systematic 
uncertainty bands are derived with subtraction of 
the Kelsall et al./Wright zodiacal light models).
The \citet{Thompson:2007,Thompson:2007a} limit on the net CIB at 1.1 and 1.6 \mic\ is within 3-4 nW/m$^2$/sr of the level given by known galaxy counts. 
The orange (lower) band shows the HRK12 CIB, reconstructed from galaxy counts, 
bounded by its HFE to LFE uncertainties. Also shown are the 
modeled (black line) and measured (large red circles) CIB extrapolated 
from integrated galaxy counts by \citet{Driver:2016}.
Levels of CIB/EBL inferred from $\gamma$-ray absorption are shown from
\citet{H.E.S.S.Collaboration:2013} (gray band), \citet{Ahnen:2016} (striped pink band), 
\citet{Biteau:2015} (blue stars).
Adapted with modifications and additions from \citet{Driver:2016}.
}
\label{fig:cib_dc}
\end{figure}

Figure \ref{fig:cib_dc} displays the recent mean CIB measurements. 
Lower limits derived from the integration of the fluxes of resolved galaxies
are well below the mean CIB at 0.8 -- 4 $\mu$m. While the currently 
claimed direct mean CIB levels are in tension with constraints 
from $\gamma$-ray absorption (see Section \ref{sec:gray}), 
the CIB levels implied by the current 
fluctuation measurements, a few nW/m$^2$/sr, can be comfortably accommodated.

The primary difficulty with all direct measurement of the CIB
and the interpretation of these measurements is the large uncertainty associated
with the subtraction of bright foregrounds, particularly the zodiacal light. 
\citet{Dwek:2005} propose that the similarity of the energy spectra suggest that 
incompletely modeled zodiacal light could be responsible for the apparent CIB 
excess. However, with extension to shorter wavelengths, the similarity 
is less clear \citep{Tsumura:2010}.

It has
been noted \citep[e.g.][]{Cooray:2009} that a mission outside 
the interplanetary dust cloud can make greatly improved mean CIB and COB determination due to 
reduction of the zodiacal light. \citet{Greenhouse:2012} and \citet{Matsuura:2014} 
present concept studies for such a mission. Studies 
using {\it Pioneer} \cite{Toller:1983} and {\it New Horizons} \cite{Zemcov:2017} data 
set upper COB limits.

The uncertainties associated with foregrounds can be reduced with 
analysis of the CIB fluctuation rather than its mean intensity.
Fluctuation measurements were pioneered in the CIB context of DIRBE data by \citet{Kashlinsky:1996,Kashlinsky:1996a} and \citet{Kashlinsky:2000}. At optical wavelengths, such methodology has been explored earlier by \citet{Shectman:1973,Shectman:1974}.

\subsection{Limits from $\gamma$-ray absorption}
\label{sec:gray}

\begin{figure}[t]
\includegraphics[width=3in]{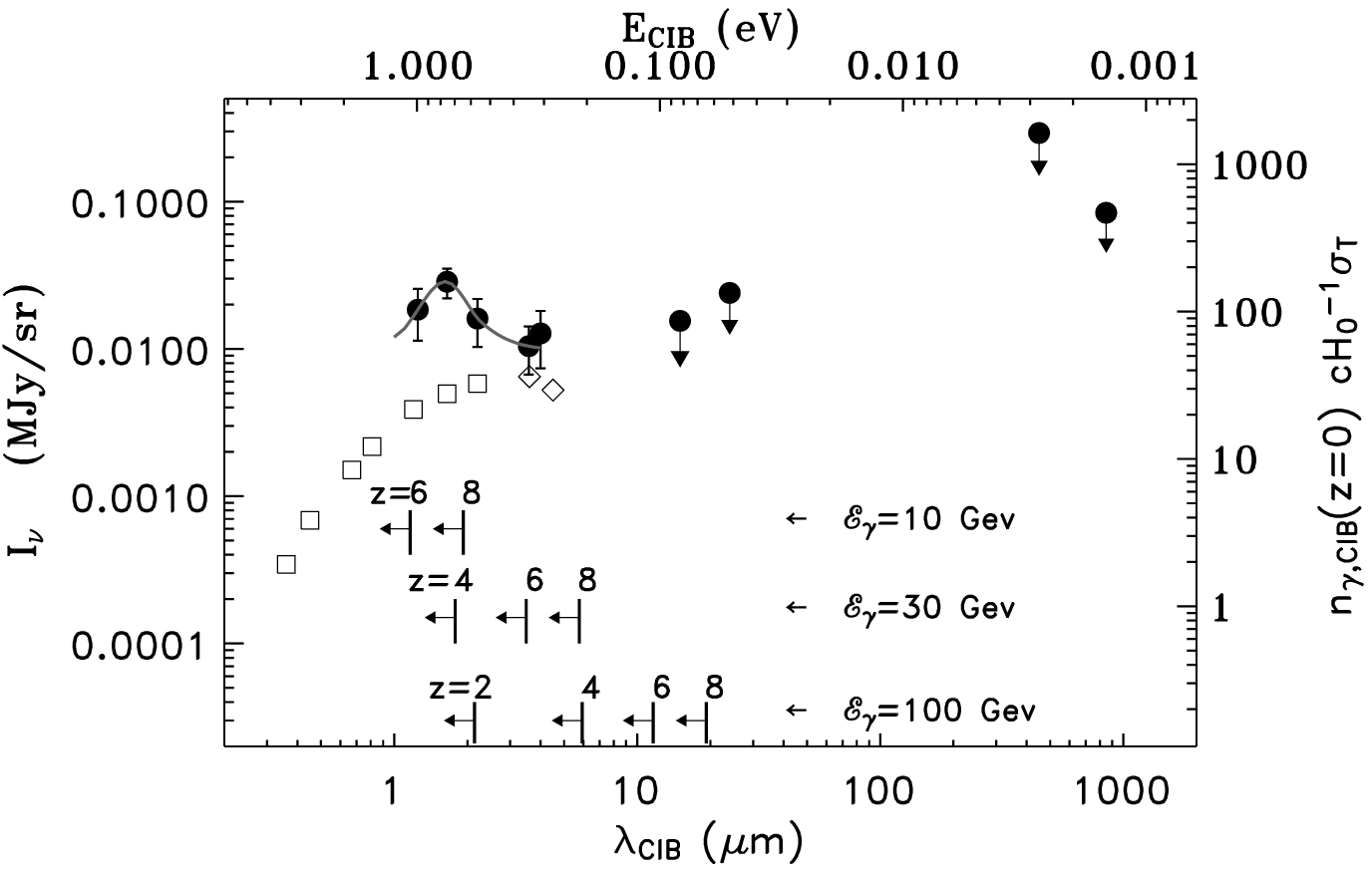}\\
\includegraphics[width=2.5in, trim=1.5cm 0 0 0]{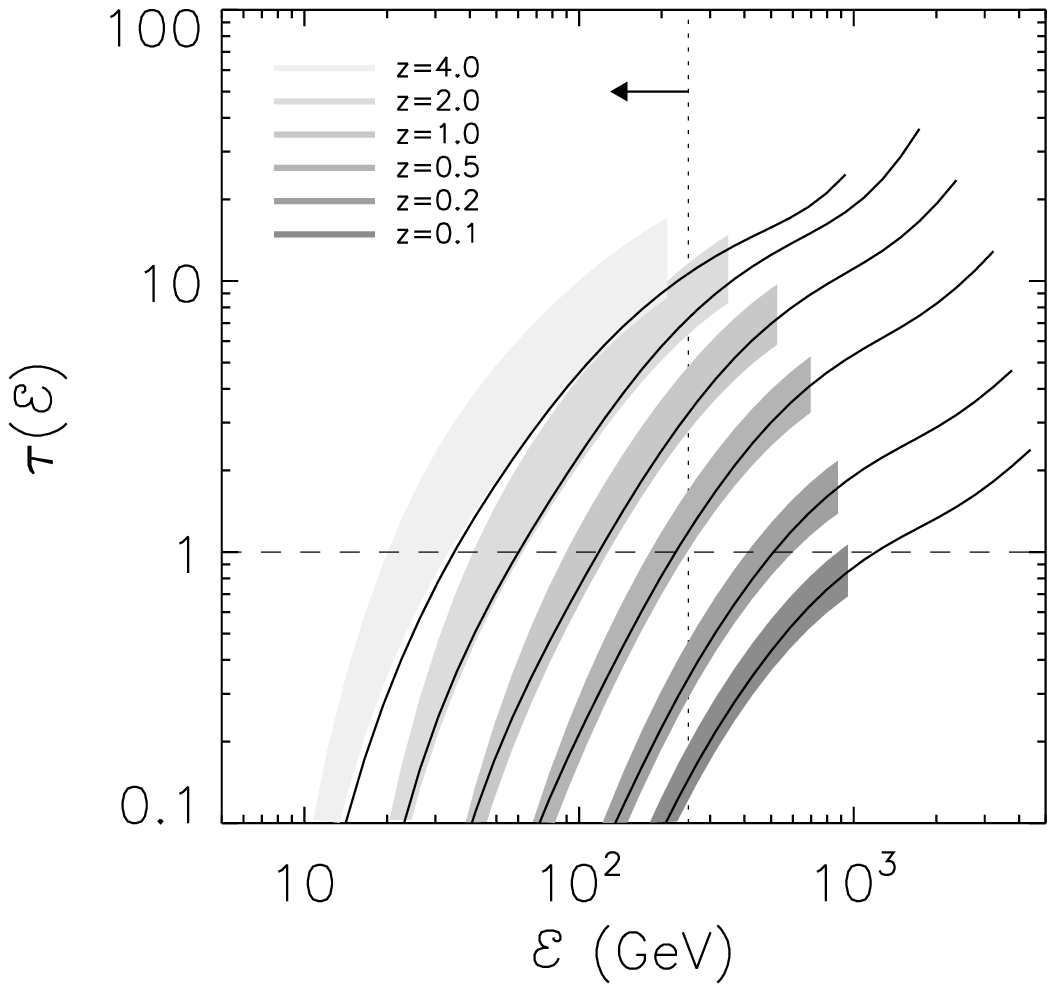}
\caption{\small {\bf Top}:  Adapted from \citet{Kashlinsky:2005b}. Filled circles show the CIB excess claimed by IRTS analysis at near-IR \cite{Matsumoto:2005} as derived in \citet{Kashlinsky:2005} and open squares show the integrated counts. Vertical bars with left-pointing arrows show the range where photon-photon absorption is possible for the redshifts and energies indicated {\bf Bottom}: Adapted from \citet{Helgason:2012}. Solid lines show the $\gamma\gamma$ optical depth out to the marked $z$ vs the observer $\gamma$-ray energy using the default reconstruction of \cite{Helgason:2012a} and the shaded regions show the boundaries of the HRK12 reconstruction. The dashed horizontal line marks $\tau=1$. The figure shows that the Universe is already optically thick to MeV photons at $z\gsim 0.1$.
}
\label{fig:cib_gray}
\end{figure}

CIB emissions may provide a source of abundant photons at
high $z$. The present-day value of $I_\nu$
corresponds to
a comoving number density of photons per logarithmic energy
interval, $d\ln E$, of $\frac{4\pi}{c}\frac{I_\nu}{h_{\rm
Planck}}=0.6(I_\nu/1\ {\rm MJy\ sr^{-1}})$ cm$^{-3}$ and if these photons come from high $z$
their number density would increase $\propto (1+z)^3$ at early
times. These photons with the present-day energies, $E$, would
also have higher energies in the past and they would thus provide absorbers
for sources of sufficiently energetic photons 
via $\gamma \gamma_{\rm CIB}\rightarrow e^+e^-$ when $E'_\gamma {\cal E'}_{\rm CIB} \geq (m_ec^2)^2$ \cite{Nikishov:1962,Gould:1967}.  
The $\gamma\gamma$
absorption, being electromagnetic in nature, has cross-section magnitude 
similar to that for the Thomson scattering, $\sigma_T$: it is given by 
$\sigma =\frac{3}{16}\sigma_T
(1-\beta^2)[2\beta
(\beta^2-2)+(3-\beta^4)\ln(\frac{1+\beta}{1-\beta})]
$ where $\beta\! = \![1-\frac{2m_e^2c^4}{E^\prime{\cal
E}^\prime(1-\cos\theta)}]^{1/2}$,  $E$ and ${\cal E}$ the present day energies of the
CIB and $\gamma$-ray photons respectively; the primes denote
rest-frame energies, e.g.\ $E^{\prime}$=$E(1+z)$. The cross-section has a
sharp cutoff as $\beta\rightarrow$1, peaks at $\simeq
\frac{1}{4}\sigma_T$ at $\beta\simeq 0.7$, and is $\sigma\propto
\beta$ for $\beta \lsim 0.6$. The mean free path of $\gamma$-ray photons in
the presence of CIB would be $(n_{\gamma_{CIB}}\sigma)^{-1} \sim
0.8 (\sigma_T/\sigma)(1\ {\rm MJy\ sr^{-1}}/I_\nu)(1+z)^{-3}$ Mpc. Fig.\ \ref{fig:cib_gray}, top shows
the CIB expressed as $I_\nu$ and as the comoving photon number density times
$\sigma_T cH_0^{-1}$, along with regions defined by the $\gamma\gamma$
absorption threshold.

This interaction generates absorption at sufficiently high $\gamma$-ray energies for a given IR/optical wavelength (marked in Fig.\ \ref{fig:cib_gray}, top).
Measuring this absorption provides independent constraint on the EBL/CIB and its evolution with $z$. 
However, the net diffuse flux probed in this way is not source-subtracted, and is not wavelength-specific.
If significant CIB comes from high-$z$ sources, it would 
have provided a far more abundant source of photons at high $z$
which interact with photons of present-day energy ${\cal E} \gsim
2M_e^2c^4/E^\prime \gsim 30(1+z)$GeV so that even a moderate CIB from first stars era could be identified in spectra of $\gamma$-ray sources at $z\gsim$3--5 \citep[][]{Kashlinsky:2005b}.
\citet{Helgason:2012} reconstruct $\tau$ from known sources with the multi-wavelength reconstruction of \citet{Helgason:2012a}. This gives the minimal absorption and shows that TeV photons are fully absorbed from nearby ($z\lsim 1$) sources (Fig.\ \ref{fig:cib_gray}), so to probe first stars era with this method more directly one needs GeV photons.

\citet{Dwek:2005a} and \citet{Aharonian:2006} examined the strong CIB
in the context of $\gamma$-ray absorption towards blazars and production from 
Population III systems. They conclude that Population III systems are unlikely 
to contribute much to the CIB excess claimed in the IRTS and DIRBE studies, although such statements quantitatively depend on the {\it assumed} SED of EBL \cite{Kashlinsky:2007d}. \citet{Ackermann:2012} detect attenuation from EBL in the combined sample of Fermi blazars out to $z\simeq 1.6$. Fig.\ \ref{fig:cib_dc} shows CIB/EBL levels from \citet{Ahnen:2016,H.E.S.S.Collaboration:2013,Biteau:2015}. Constraints from the observed $\gamma$-ray absorption \cite{H.E.S.S.Collaboration:2013} give upper limits of 17 and 14 nW m$^{-2}$ sr$^{-1}$ at 1.1 and 1.6 \mic. This is to be 
compared with the resolved CIB from faint galaxy counts estimated in Table \ref{tab:cib_counts} at these bands.  Thus $\lsim 8$ and 5 nW m$^{-2}$ sr$^{-1}$ currently appear feasible in CIB excess at these wavelengths. 

Constraints may however be less restrictive because of an alternative suggested explanation of secondary TeV photons produced by interaction of cosmic rays and EBL \cite{Essey:2010,Essey:2010a}. In that interpretation, the intrinsic spectra of blazars at TeV energies have absorption due to high CIB levels, but appear unabsorbed because cosmic rays (protons) from the blazar jets interact with lower energy EBL along the line of sight to produce pions and secondary $\gamma$-rays, when $E_{\rm p}E_{\rm EBL}\geq (m_\pi c^2)^2$. Those secondary $\gamma$-rays coincide with the blazar within the angular resolution of the Cherenkov telescopes because the intergalactic magnetic fields are weak ($\lsim 10^{-14}$G) and unable to deflect the cosmic ray protons from the line-of-sight. 

\subsection{Resolved cosmic X-ray background}
\label{sec:cxb}
\begin{figure}[t]
\includegraphics[width=3in]{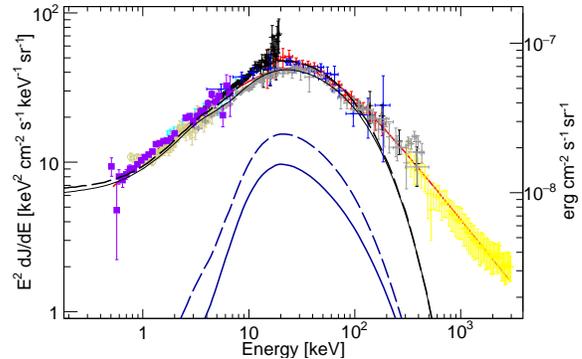}
\caption{ 
Compilation of independent measurements of the CXB spectrum from several different instruments in the 0.5-500 keV range \citep[adapted from][see references therein for the individual datasets]{Cappelluti:2017}. The magenta data points have been recently derived from the Chandra legacy data in the COSMOS field \citet{Cappelluti:2017}. The solid and dashed lines show population synthesis model curves using the \citet{Gilli:2007} model, adapted from \citet{Comastri:2015}. The thin solid and thick solid lines show the total AGN spectrum and the contribution of Compton-thick (mildly+heavily absorbed, see text) AGNs in the model, respectively. The dashed curves show the same information, but assume a 4 times larger abundance of the heavily absorbed Compton-thick AGNs. The flux of 1 keV$^2$cm$^{-2}$keV$^{-1}$sr$^{-1}=1.6\times10^{-9}$erg/cm$^2$/sr as shown in the right vertical axis.
\label{fig:fig_guenther1}}
\end{figure}

The CXB was discovered by \citet{Giacconi:1962} in a rocket flight originally designed to detect X-ray emission from the Moon; the CXB was the first cosmic background discovered. The shape of the CXB spectrum in the 3-50 keV range was first determined by HEAO-1 \citep{Marshall:1980} and shows a pronounced maximum emitted energy in the 20-30 keV range. Figure \ref{fig:fig_guenther1} \citep[from][]{Cappelluti:2017} summarizes the best measurements up-to-date. The measurement of the absolute level of the X-ray background is complicated, because of systematic uncertainties in the instrument responses, as well as the instrumental background and solid angle characteristics. Additionally, there are systematic differences in the contribution of relatively bright X-ray sources, which are present in wide-field collimated instruments and typically avoided in narrow-field imaging surveys. Thus, throughout the history of CXB measurements there have been systematic differences in the measured absolute CXB intensity, which are partially reflected in Figure \ref{fig:fig_guenther1}. There is a 10-20\% difference between the minimum and maximum flux measured in the energy range 1-20 keV, corresponding to systematic CXB flux uncertainty of $\sim$1--2 keV/cm$^2$/sec/sr around 1 keV. 

X-ray surveys are practically the most efficient means of finding active galactic nuclei (AGNs) over a wide range of luminosity and redshift. Deep surveys with focusing X-ray telescopes on {\it ROSAT}, {\it Chandra}, and {\it XMM-Newton} have resolved the majority of the extragalactic CXB into faint discrete X-ray sources. Enormous multi-wavelength photometric and spectroscopic follow-up efforts have identified optical and/or NIR counterparts to most of these sources, and have shown that the main contributors to the CXB are indeed AGN at redshifts up to $z\sim 5$ \citep[e.g.][]{Brandt:2005, Brandt:2015}. One of the key observational tools is the determination of the X-ray luminosity function of these AGN, and its cosmological evolution, which give strong constraints on the accretion history of the Universe. The best-fit model for the distribution of AGN as a function of luminosity and redshift is the so called ``luminosity-dependent density evolution'' (LDDE), which shows a strong dependence of the AGN space density evolution on X-ray luminosity, with a clear increase of the peak space density redshift with increasing X-ray luminosity. This ``AGN cosmic downsizing'' evolution is seen both in the soft X-ray (0.5-2 keV) and the hard X-ray (2-10 keV) bands \citep[e.g.][and references therein]{Ueda:2014,Miyaji:2015,Fotopoulou:2016}, as well as in other wavebands \citep[see discussion in][]{Hasinger:2008}.

The spectral shape of the CXB was a puzzle for some time, because it does not resemble typical AGN spectra. The resolution came from cosmological population synthesis models, where the evolving AGN luminosity function is folded with sophisticated AGN spectral model templates including the Compton reflection hump, and a wide distribution of neutral gas absorption column densities from unabsorbed heavily Compton-thick absorption \citep[e.g.]{Comastri:1995, Gilli:1999, Ueda:2003, Ballantyne:2006, Gilli:2007, Treister:2009, Ueda:2014}. In these models most of the AGN emission in the Universe is significantly absorbed by intervening gas and dust clouds, which is also the reason for the characteristic 20-30 keV peak of the X-ray background spectrum.
 
\begin{figure}[t]
\includegraphics[width=3in]{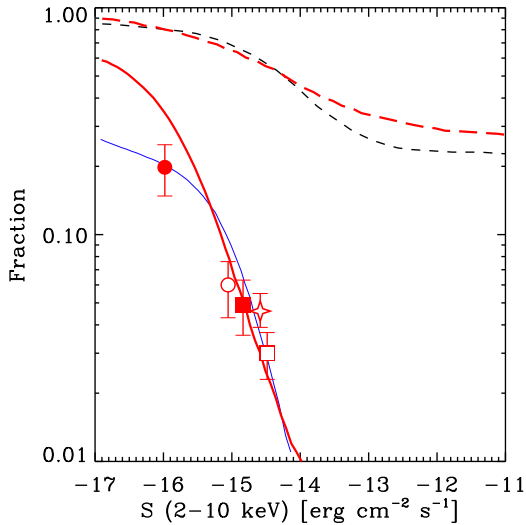}
\caption{Observed fractions of Compton-thick
AGN by Brightman and Ueda (2012, filled circle),
Brunner et al. (2008, open circle), Tozzi et al. (2006, filled
square), Kocevski et al. (2017, star) and Hasinger et al. 
(2007, open square), adapted from Ueda et al. (2014). 
Predictions from population synthesis
models of Gilli et al. (2007), adapted from Comastri et al. (2015)
and Ueda et al. (2014) for
Compton-thick AGN fractions ($\log N_{\rm H}$=24--26, thick red,
solid blue) and all obscured AGNs ($\log N_{\rm H}$=22--26, dashed red, 
dashed black) in the total AGNs are given as a function
of the observed 2-10 keV flux.}
\label{fig:fig_guenther2}
\end{figure}
 
Compton-thick AGN, where most primary X-ray emission is absorbed by a large column density of intervening material and only a very small reflected soft component escapes, are very hard to detect and therefore elusive in existing deep X-ray surveys. Fig.\ \ref{fig:fig_guenther2} summarizes our current knowledge of the relative fraction of observed Compton-thick AGN as function of X-ray flux, compared to the population synthesis models. There are still significant uncertainties in the actual contribution of Compton-thick to the luminosity function and cosmic evolution of AGN, and thus their contribution to the accretion history of the Universe.

Through the studies of the X-ray background and of large samples of black holes in nearby galaxies, it has become clear that supermassive black holes at the centers of galaxies must play an important role in the cosmic evolution of galaxies. AGN and galaxies in general undergo very similar evolution patterns, where the peaks of AGN activity and star formation occur in the same redshift range ($z$=1.5--2) and show a similar dramatic decline (downsizing) towards lower $z$. Strong correlations have been found between the BH mass and global properties of its host galaxy spheroid, like the bulge luminosity \citep{Kormendy:1995,Magorrian:1998} and the stellar velocity dispersion, i.e.\ the $M_{\rm BH}-\sigma$ relation \citep{Ferrarese:2000,Gebhardt:2000}. Using these correlations, the mass density of local dormant supermassive black holes in galaxy centers has been estimated, and is found largely consistent with the mass density accreted by AGN throughout the history of the Universe \citep{Marconi:2004,Merloni:2004}, yielding further evidence for a tight link between the growth of galaxy bulges and of their nuclear black holes through standard, high-efficiency accretion processes.

However, a recent comprehensive analysis of black hole mass measurements and scaling relations concluded that the canonical black-hole-to-bulge mass ratio, instead of being constant for all galaxies with values around $\sim0.1-0.23\%$ \citep[e.g.][]{Merritt:2001,Marconi:2003}, actually shows a mass dependence and varies from 0.1-0.2\% at $M_{\rm bulge}\sim 10^9$ $M_\odot$ to $\sim0.5$\% at $M_{\rm bulge} =10^{11}$ $M_\odot$ \citep{Graham:2013,Kormendy:2013}. However,  \citet{Shankar:2016}
argued that the above are overestimating black hole masses, and hence AGN counts.
The revised normalization would lead to a dramatically (a factor of 2 to 5) {\it larger} estimate of the local BH mass density, which is dominated by massive bulges. Conversely to the previous findings, this result means, that there must be other significant channels for BH growth, apart from those assumed in the standard population synthesis model for the CXB. Like others before, \citet{Comastri:2015} pointed out, that the systematic uncertainty in the normalization of the CXB spectrum allows significant contributions of so far undetected populations of heavily shrouded Compton-thick AGN without violating other observational constraints, e.g.\ in the mid-infrared. The dashed curve in Fig.\ \ref{fig:fig_guenther1} shows a variant of the population synthesis model, where the contribution of heavily obscured, i.e.\ reflection-dominated Compton-thick AGN has been increased by a factor of 4 
with respect to the standard model. This already goes some way towards augmenting the local BH mass density, but is not sufficient. \citet{Comastri:2015} therefore had to assume another, so far undetected component of BH mass growth, e.g.\ BHs which are completely shrouded by obscuring material and only radiate at mid-IR wavelengths. However, one major uncertainty in these estimates is the unknown cosmological evolution of the obscuration fraction \citep[e.g.][]{Treister:2006,Hasinger:2008}. There is also mounting evidence, that the fraction of galaxy mergers is significantly higher among Compton-thick AGN compared to the normal CXB population \citep{Kocevski:2015,Kocevski:2017}. In particular at high redshifts, where galaxy mergers are expected to be more common, obscured accretion can play a much larger role than locally.

\section{Theory behind CIB fluctuation studies}
\label{sec:theorycib}

\subsection{CIB fluctuations primer}
\label{sec:cibprimer}

\subsubsection{Theoretical basis}
\label{sec:basis}

CIB is a decisive tool when sources of interest are fainter than the sensitivity limits of the instrument or are too numerous to be individually resolved (i.e. are confused) at the instrument's angular resolution. 
The goal is to probe CIB levels from faint populations below the (ideally low) threshold defined by instrument noise and resolution, i.e.\  $F_{\rm CIB}(m\!>\!m_{\rm lim})$. For a sufficiently faint removal threshold, and suitable $\lambda$, the hope is that one would move sufficiently far along the redshift cone to probe the earliest sources.  

The rate of the net CIB flux production
probed in observer band at wavelength $\lambda$ is:
\begin{equation}
\frac{dF_\lambda}{dz} = \frac{c}{4\pi} {\cal L}_{\lambda^\prime}(z) \frac{1}{1+z} \frac{dt}{dz}
\label{eq:dfdz}
\end{equation}
where ${\cal L}(z)$ is the comoving luminosity density at the rest wavelength $\lambda^\prime$. Emissions from astrophysical sources in the rest frame UV are cut off at the Lyman break due to absorption by the intergalactic medium (IGM), which happens at rest $\lambda_{\rm Ly}$=0.0912\ \mic\ if the IGM is fully ionized or at the Ly$\alpha$  of $\lambda_{\rm Ly}$=0.1216\ \mic\ if it contains mainly \HI.

In the Cartesian limit (small angles), CIB fluctuations can be Fourier transformed, $\Delta(\vec{q})=\frac{1}{4\pi^2} \int \delta
F(\vec{x})
\exp(-i\vec{x}\cdot \vec{q}) d^2x$, and characterized by the 2-dimensional projected power spectrum, $P(q)=\langle |\Delta(\vec{q})|^2\rangle$, as a function of the angular frequency $q$ (or angular scale $2\pi/q$).  A
typical rms flux fluctuation is $\sqrt{q^2P(q)/2\pi}$ on the
angular scale of wavelength $2 \pi/q$.
Theoretically there are two types of contributions relevant for interpretation of the measured cosmological projected (2-D) power spectrum of source-subtracted CIB fluctuations: 1) shot noise from remaining sources occasionally entering the beam, and 2) the clustering component that reflects clustering of the remaining CIB sources. 

The shot-noise power is given by \cite{Kashlinsky:2005}:
\begin{equation}
P_{\rm SN} = \int_{m_{\rm lim}}^\infty S^2(m) \frac{dN}{dm} dm
\label{eq:sn}
\end{equation}
where $m_{\rm lim}$ is the limiting magnitude of sources remaining in the source-subtracted CIB map, $S(m)$
is the flux of a source of AB magnitude $m$, and $dN/dm$ is the number counts of the sources per $dm$. This component is intrinsically white, but convolved with the instrument beam.

When interpreting observations, it is useful to 
consider the
shot-noise as follows: source-subtracted CIB fluctuations are measured at a given shot noise level, which per eq.\ \ref{eq:sn} defines the equivalent  effective magnitude (or flux) of source removal. The net mean CIB flux from sources remaining in the data is then $F_{\rm CIB}(m\!\!>\!\!m_{\rm lim}) = \int_{m_{\rm lim}}^\infty S(m) \frac{dN}{dm} dm$. Hence the remaining shot noise is connected to the remaining CIB as $P_{\rm SN} \sim S(\bar{m}) F_{\rm CIB}$ with $\bar{m}$ being the effective magnitude of the remaining populations \citep{Kashlinsky:2007b}. 
When discussing observational results the shot noise power will be expressed in units of $[P_{\rm SN}]={\rm nJy}\cdot$nW/m$^2$/sr which is equivalent to $\frac{3}{\lambda(\mic)}\times10^{-12}$ nW$^2$/m$^4$/sr.
Measurements of the diffuse flux CIB fluctuations will be expressed in units of $[\sqrt{q^2P/(2\pi)}]= $ nW/m$^2$/sr. 

The clustering component is generally made up of two terms \cite{Cooray:2002}: 
the 1-halo term, and 
the 2-halo term. The 1-halo term is essentially a white noise term convolved with an ``average" halo profile of the remaining sources and so reflects an average halo profile below angular scales subtending a typical halo. It is unimportant for high-$z$ sources, but may be important for more local extended ones. The  projected 2-halo term is related to  the underlying 3-D power, $P_{3D}$, of the sources by the relativistic \citet{Limber:1953} equation: 
\begin{equation}
\frac{q^2P_{\lambda}(q)}{2\pi}= \int_0^{z_{\rm Ly}(\lambda)}
 \left(\frac{dF_{\lambda^\prime}}{dz}\right)^2 \Delta^2(qd_A^{-1}; z)dz
\label{eq:limber_1}
\end{equation}
where $\Delta^2(k,z)\equiv \frac{k^2P_{3D}(k,z)}{2\pi cH^{-1}(z)}$ is the mean square fluctuation in the source counts over a cylinder of diameter $k^{-1}$ and 
length $cH^{-1}(z)$ and $\frac{dF_{\lambda^\prime}}{dz}$ is the CIB flux production at rest $\lambda^\prime\equiv\lambda/(1+z)$ over the epochs spanned by the integration \citep{Kashlinsky:2004,Cooray:2004,Fernandez:2010,Kashlinsky:2015,Helgason:2016}. 
Because cosmological sources have a Lyman break due to IGM absorption by 
\HI\
at rest wavelength $\lambda_{\rm Ly}$,  {\it the integration  
stops at} $1+z_{\rm Ly}(\lambda)=\lambda/\lambda_{\rm Ly}$ because at larger redshifts sources emit only longward of the Lyman-break wavelength; the integration extending to redshift specified by the {\it far} edge of the filter of band $\lambda$. This will be used in the Lyman tomography in Sec.\ \ref{sec:ly-tomography} below.

The density field is today linear on scales $>r_8$ and the scale of non-linearity is smaller at higher $z$. It is reasonable to assume that on linear scales the density of luminous sources traces that of the underlying matter to within a scale-independent bias factor. For reference, angular scale of $1^\prime$ subtends $1.1, 1.4, 1.5, 1.6 h^{-1}$Mpc at $z=5,10,15,20$. As Fig.\ \ref{fig:p_lcdm} shows, the density field on these scales is in linear regime at $z\gsim$8--10. We assume a $\Lambda$CDM template for $P_{3D}$ in eq.\ \ref{eq:limber_1} for the high-$z$ contributions to CIB fluctuations on arcminute scales and beyond. If the range of $z$ spanned by the populations that are probed is narrow, as can arise if lower-$z$ sources are removed and very high-$z$ sources do not enter beyond $z_{\rm Ly}$, one can relate CIB fluctuations to the net CIB flux as $\delta F(2\pi/q) \sim F_{\rm CIB}\Delta(qd_A\bar{z})$, where $\bar{z}$ is a suitably averaged redshift of the sources.

Let us assume that a fraction $f_{\rm Halo}$ of all matter in the Universe collapses in halos capable of producing luminous sources at a given redshift, converting on average a fraction $f_*$ of the halo baryons into luminous sources. 
The bolometric diffuse flux produced by these populations, after they have converted their mass-energy into radiation with radiation efficiency $\epsilon$, is 
\begin{eqnarray}
F_{\rm tot} \simeq  f_{\rm Halo}f_*\left(\frac{c}{4\pi}\epsilon\rho_{\rm baryon} c^2\right) z_{\rm eff}^{-1}\simeq \nonumber \\
9.1\times 10^5 \epsilon f_{\rm Halo}f_*z_{\rm eff}^{-1} \;\frac{\Omega_{\rm baryon}h^2}{0.0227} \;\; \frac{{\rm nW}}{{\rm m^2 sr}}
\label{eq:f_cib_theor}
\end{eqnarray}
 where $z_{\rm eff}\equiv 1/\langle(1+z)^{-1}\rangle$ is a suitably averaged effective redshift factor which accounts for the radiation energy density decreasing with expansion as $\propto (1+z)^{-4}$ vs. the matter density $\propto (1+z)^{-3}$. 
\subsubsection{Observationally determined quantities and their uncertainties}
\label{sec:quantities}

Once CIB maps are produced, e.g.\ for a square field of width $\Theta$, the diffuse flux is Fourier transformed with pixels in the Fourier plane having a width of $\Delta q={2\pi}/{\Theta}$. Because flux is a real quantity, only half of the Fourier plane is independent.
The power spectrum is defined as $P(q)=\langle |\Delta(\vec{q})|^2\rangle$,
where $\Delta(\vec{q})$ is the 2-D Fourier transform (FT) of the
source-subtracted CIB. 
For ease of comparison with background intensities, we plot results as
the mean squared fluctuation at angular scale $2\pi/q$, defined as
$q^2P(q)/(2\pi)$. For spherical harmonic expansion, $\delta F(\theta,\phi)=\sum a_{\ell,m} Y_{\ell,m}(\theta,\phi)$, the power is $C_\ell=\langle |a_{\ell,m}|^2\rangle_m=\frac{1}{2\ell+1}\sum_{m=-\ell}^\ell |a_{\ell,m}|^2$. At small angular scales, the multipole in spherical harmonic expansion is related to the angular wavenumber via $\ell \simeq q($in rad$^{-1})$.

The cross-power describing the correlations between fluctuations at different wavelengths (1,2) is
$P_{\rm 1\times2} (\vec{q}) = \langle \Delta_{1}(\vec{q}) \Delta^*_{2}(\vec{q})\rangle = \langle[{\cal R}_{1}(\vec{q}) {\cal R}_{2}(\vec{q}) + {\cal I}_{1}(\vec{q}) {\cal I}_{2}(\vec{q})]\rangle$ with ${\cal R, I}$
standing for the real, imaginary parts of the Fourier transform, $\Delta(\vec{q})$.  The cross-power spectrum
is a real quantity which can be positive or negative. 

The correlation function, $C(\theta) = \langle \delta F(\vec{x}) \delta F(\vec{x}+\vec{\theta})\rangle$ and the 2-D power are interrelated via an integral transform, which in the limit of small angles $\theta\ll 1$ rad is: $C(\theta)=\frac{1}{2\pi} \int_0^\infty P(q) J_0(q\theta)q dq$ and $P(q)=2\pi \int_0^\infty C(\theta) J_0(q\theta)\theta d\theta$ with $J_0$ being cylindrical Bessel function of 0-th order. Any white noise power, such that $P$=const, results in $C(\theta)\!\!\propto\!\! \delta_D(\theta)$ and directly translates only into the zero-lag value of the correlation function (i.e.\ variance). The shot-noise component is white noise convolved with the beam and will be reflected in the correlation function values only up to roughly the beam scale. Non-zero values of the correlation function on scales much greater than the beam reflect a non-white power from clustering; e.g.\ a power-law $C\propto \theta^{(n-2)}$ corresponds to $P(q)\propto q^{-n}$.

The coherence between the two bands is defined ${\cal C}_{12}\equiv \frac{P_{12}^2}{P_1P_2}$. It should lie between 0 and 1 (no to full coherence).

We now turn to errors/uncertainties for the measured quantities: auto-powers, cross-powers and coherence. We assume that the underlying $\Delta(\vec{q})$ is Gaussian-distributed, but note that this may be affected by biasing \cite{Kaiser:1984,Jensen:1986,Bardeen:1986,Kashlinsky:1991,Kashlinsky:1998}.

The errors on the power measured from a finite size field are subject to the sampling (``cosmic") variance \citep{Abbott:1984}. Namely, if the power $\hat{P}$ at the central wavenumber $q$ is determined from a total of $N_q$ independent pixels in the Fourier plane the error on this measurement is $\sigma_P=\hat{P}/\sqrt{N_q}$. Because the auto-power is a quadratic quantity, and is  $\chi^2$-distributed, this approximation does not correspond to the standard $68\%$ confidence limit at the very largest scale, where $N_q \sim$(1-2), but at smaller scales it is a reasonable approximation. An additional issue is that masking of resolved sources in the maps generates coupling between various Fourier harmonics thereby biasing/distorting the measurement of the power from FT because of the convolution with mask. Thus one should proceed with caution and verify the power results from FFT with the much more CPU intensive computation of the correlations function \citep[e.g.][]{Kashlinsky:2005a,Matsumoto:2011}, which is immune to masking effects. In practice, when $\lsim$30-35\% of the maps are masked there is good consistency between the two approaches \citep{Kashlinsky:2005a}, but the two can at times diverge for much more aggressive masking with the correlation function being a more reliable estimate  \citep{Kashlinsky:2007c}.

The cross-power for uncorrelated quantities can be both positive and negative and would be distributed in a Gaussian manner if the underlying quantities are Gaussian-distributed. The cosmic variance error on its measurement from the same field at two different bands is $\sigma_{P_{12}}\simeq \sqrt{P_1P_2/N_q}$ \citep[see][]{Cappelluti:2013}.

For  errors on the coherence, or the square of the correlation coefficient ${\cal R}$, the situation is more complicated since statistical errors must be evaluated from the confidence contours of the quantity of interest (${\cal C}\equiv {\cal R}^2$ here), which must be derived from its underlying probability distribution function (PDF). Due to the highly non-linear structure of ${\cal R}$ with respect to the underlying quantities in both the numerator and the denominator, its PDF is not trivially derivable. 
However, once the errors on the power at each $q$ are determined, one can then propagate them via the Fisher transformation
\citep[][]{Fisher:1915} 
 to give the confidence contours of the resultant correlation coefficient. Because errors are always equivalent to confidence contours, one needs to evaluate the 68\% confidence limits of ${\cal R}$ from the errors on the power.  The Fisher transformation technique represents the standard way to evaluate the probability distribution of ${\cal R}$ and relate the uncertainties 
to those of the powers. The Fisher transformation works as follows: One evaluates the central value, ${\cal R}_0\equiv\sqrt{{\cal C}_0}$, of the correlation coefficient from the power data per above. 
 The Fisher transformation is to compute the quantity 
${\cal Z}=\frac{1}{2} \ln\left[\frac{(1+{\cal R})}{(1-{\cal R})}\right]
$, which is {\it normally} distributed in most practical cases \cite{Fisher:1915}.
This transformation, and its inverse
${\cal C}
={[\rm tanh}({\cal Z})]^2$, is then used to construct the corresponding  confidence interval for ${\cal C}$: one evaluates the 68\% contours of ${\cal Z}$ from the variances of the auto- and cross-powers, assumed to be equivalent to the 68\% confidence levels. The variance in ${\cal Z}$ is related to the errors on powers as $\sigma^2_{\cal Z}=\frac{{\cal C}_0}{(1-{\cal C}_0)^2} \left[ \frac{\sigma_{P_{12}}^2}{P_{12}^2} + \frac{1}{4}\frac{\sigma_{P_1}^2}{P_1^2}+\frac{1}{4}\frac{\sigma_{P_2}^2}{P_2^2}\right]$. 
The 68\% contours for ${\cal C}$ are derived from ${\cal Z}\pm1\sigma_{\cal Z}$, 95\% from ${\cal Z}\pm2\sigma_{\cal Z}$, etc. 
The confidence contours for ${\cal C}$ are thus constrained to 
the interval of $[0,1]$ (and $[-1,1]$ for ${\cal R}$).
\subsection{Contribution from remaining known galaxy populations}
\label{sec:cibfromhrk12}
\citet[][]{Helgason:2012a} developed a robust heuristic way of reconstructing CIB fluctuations from galaxy populations spanning wavelengths from UV to mid-IR out to $z\sim 6$. 
The assembled database for the reconstruction now covers over 340 luminosity function (LF) surveys from UV to mid-IR \citep[HRK12,][]{Helgason:2012,Helgason:2014}, and the methodology allows filling in the redshift cone with known galaxies across the required wavelengths. 

The HRK12 methodology works as follows: the LF in the optical and near-IR can be well
described by the \citet{Schechter:1976} function parameterized by $M^\star$,
$\phi^\star$ and $\alpha$. Table 1 in
HRK12 shows the
measured Schechter parameters from multiple surveys as a function of
both restframe wavelength and redshift. Whereas $M^\star$ and $\phi^\star$ are well measured out to
large distances, the faint-end slope $\alpha$ is poorly constrained
and often is simply kept fixed in fits. Deep
near-IR number counts provide the best constraints on the
faint-end LF slope as they are dominated by the faint end of
the LF at $z\sim 1-3$, where measuring $\alpha$ directly becomes
challenging. In other words, the faint galaxy counts at 1--5 \mic\
sample the faint end of the LF at different rest-frame wavelengths at
intermediate $z$ where the volume density of sources per solid angle is at maximum. 
More importantly, compared to the LF, the uncertainties in the counts are robust, i.e.\ they are not affected by systematic uncertainties associated with redshift determinations or degeneracy in the best-fit Schechter parameters. The only assumptions  in the reconstruction are 1) the LF is well described by the Schechter function, and 2) that the evolution and spectral behavior of $\alpha$ is smooth and does not exhibit sudden changes in a narrow interval. The uncertainties of the reconstruction, around the {\it default} model, are bracketed by the high-faint-end (HFE) and low-faint-end (LFE) limits from varying $\alpha$ within the limits allowed by the data.

The accuracy of the reconstruction is verified by the remarkably good fits to the subsequently measured, and much deeper than at the time, IRAC counts \cite{Ashby:2013,Ashby:2015} and Fig.\ \ref{fig:cib_counts} above.   
\subsection{Reionization limitations on first stars era}
\label{sec:reionization}

At recombination ($z_{\rm rec}\sim$1100) photons and baryons decouple 
and the Dark Ages begin lasting until the unknown redshift(s) when the first luminous 
sources formed, reionizing the Universe. Two opposite regimes govern the later 
evolution: even a small amount of neutral hydrogen (\HI) in the IGM would absorb 
any light emitted in rest 
UV
bands by resonant absorption in the Lyman 
lines of 0.1216, 0.1026 and 0.09725 \mic\ (Ly-$\alpha, \beta, \gamma$ respectively) 
with the largest cross-section being due to Ly-$\alpha$ \cite{Gunn:1965}. Conversely, the
ionized IGM affects the CMB in several ways, mainly: 1) the CMB angular power spectrum would be 
suppressed by Thomson scattering on sub-degree scales, 2) the Thomson scattering of CMB photons 
would also lead to linear 
polarization of the CMB \cite{Rees:1968}, and 3) peculiar motions generate new temperature aniosotropies (see \ref{sec:sz}). 
The probability of scattering is 
$\propto 1-\exp(-\tau_e)$ with $\tau_e$ being the Thomson optical depth and, 
since CMB angular structure is measured to have a clear peak structure 
at $\ell\gsim 100$, it follows that $\tau_e\ll 1$. The induced CMB polarization 
is fixed by the quadrupole anisotropy of the scattering IGM, 
so polarization on scales exceeding the horizon at $z_{\rm rec}$ 
(or $\sim 1^\circ$) provides evidence of Thomson scattering, or $\tau_e > 0$. 

Thus reionization encodes information about the nature of the first stars, 
first galaxies and the emergence of large-scale structure
\citep[see discussions in][ed.]{Mesinger:2016}. 
A nice overview of the underlying physics and measurements is 
found in \citet{Zaroubi:2013}.

\subsubsection{Gunn-Peterson absorption and neutral hydrogen at low $z$}
\label{sec:gp}

As pointed out by \citet[][hereafter GP]{Gunn:1965} 
the high value of the Ly-$\alpha$ cross-section,
$\sigma_\alpha = 4.88\times 10^{-18}$cm$^2$ (ignoring line-broadening effects), leads
to a very high optical depth:
\begin{equation}
\tau^{\rm GP}_\alpha(z)= 1.2\times 10^4 \int_0^z \frac{x_{\rm HI}(z') 
(1+z')^2}{\sqrt{\Omega_{\rm m}(1+z')^3+\Omega_\Lambda}} dz'
\label{eq:tau_gp}
\end{equation}
where we adopted a He mass fraction of $Y=0.24$.
This results in full absorption even for a very small fraction  
of the cosmologically distributed neutral hydrogen $x_{\rm HI}$.

The observed absence of the \HI\ trough
in quasar spectra at wavelengths shorter than the rest Ly-$\alpha$ line 
shows that by $z\simeq 6$ the intergalactic 
hydrogen has been reionized \citep[see review by][and references therein]{Becker:2015}.
Very broadly, eq.\ \ref{eq:tau_gp} and the observed lack of absorption of quasar 
spectra require $x_{\rm HI}\lsim10^{-4}(1+z)^{-3/2}$ out to $z\lsim 6$. Prior to 
that sources of UV radiation had to exist to ionize the surrounding gas.  
Similar limits on $x_{\rm HI}$ have been reached with probing the Ly-damping of GRBs \cite{Totani:2006}.

\begin{figure}[t]
\includegraphics[width=3in]{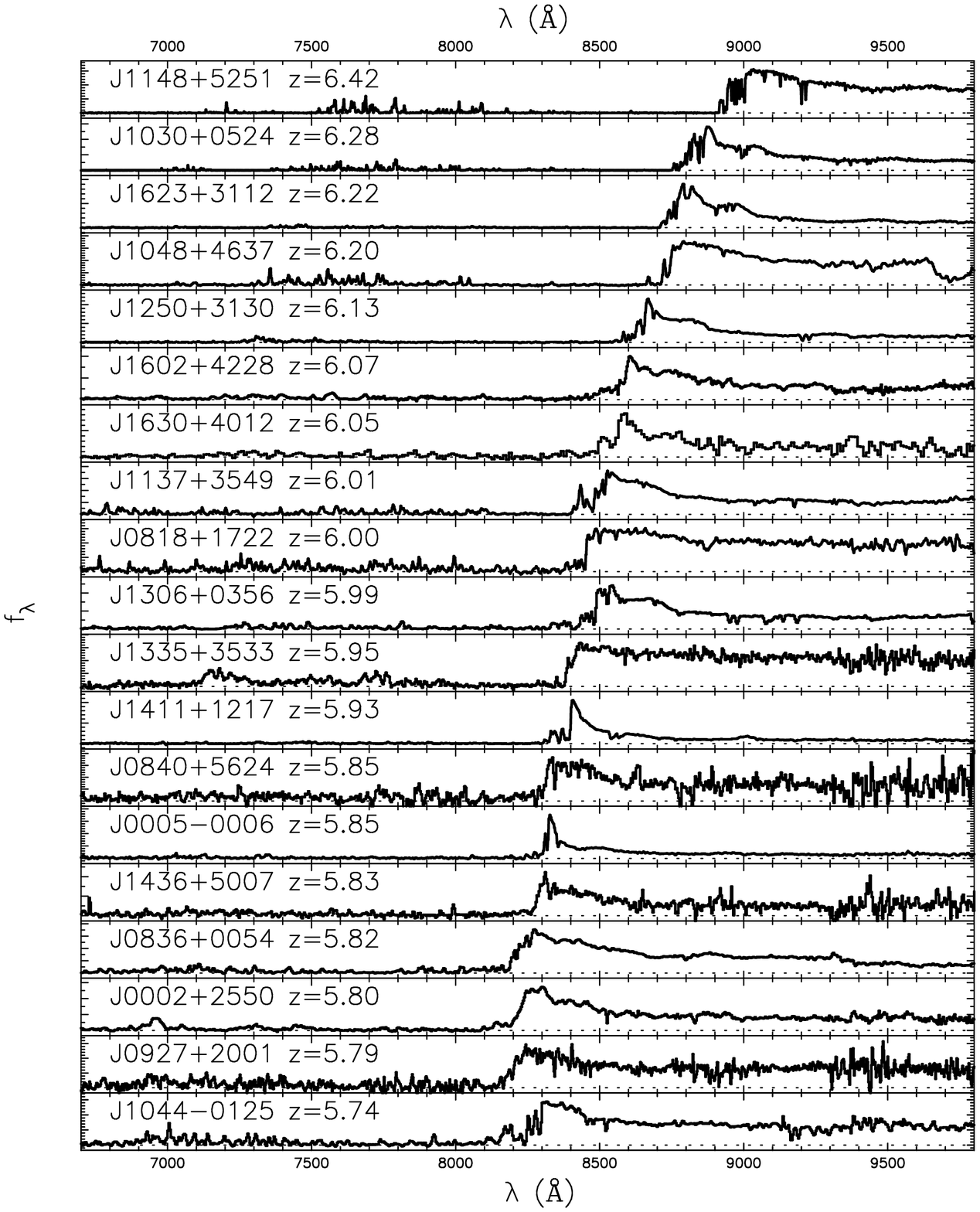}
\includegraphics[width=3.in]{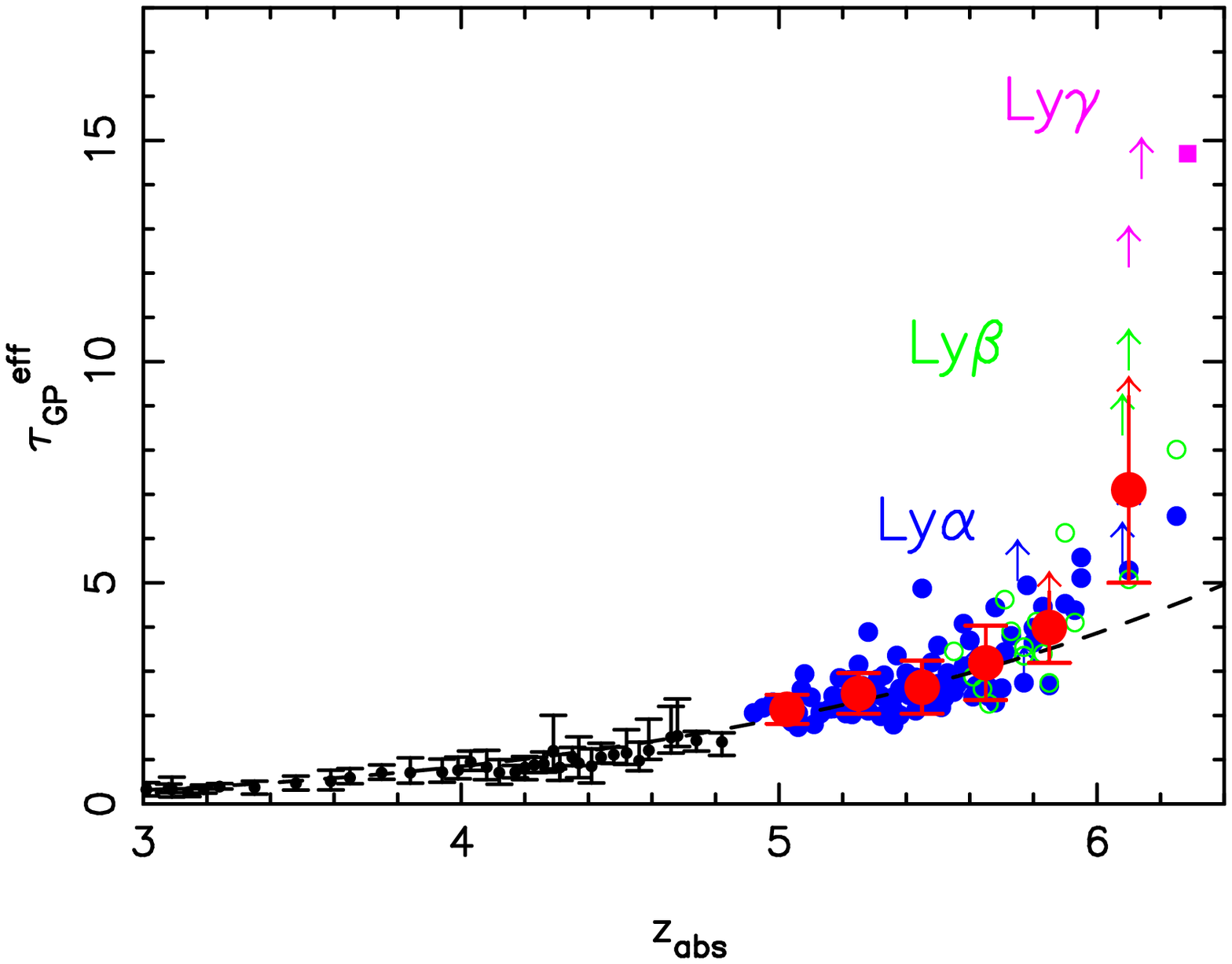}
\caption[]{\small{Adapted from \citet{Fan:2006}. {\bf Top}: Spectra of 19 SDSS quasars showing GP absorption. See also \citet{Banados:2016} for a more recent extensive compilation. {\bf Bottom}: Evolution of optical depth combined with the Ly-$\alpha,\beta,\gamma$ absorption measurements.
}}
\label{fig:tau_gp}
\end{figure}
Numerous observations suggest that reionization of 
hydrogen was complete by $z_{\rm ion, H}=6^{+0.3}_{-0.5}$ as summarized in 
\citet{Fan:2006b}. Follow-up high-resolution spectroscopy of the SDSS-discovered 
QSOs at $z\sim 6$ 
established that the Universe contained large amounts of neutral
IGM at $z>z_{\rm ion, H}$ as shown in Fig.~\ref{fig:tau_gp} \citep[from][]{Fan:2006}. 
The lower panel shows the reconstructed effective GP optical depth which 
is $\tau_{\rm GP} \simeq (1+z)^{4.3}$ out to $z\simeq 5.5$ rising 
exponentially at higher $z$. \citet{Bernardi:2003} find that the effective optical depth decreases suddenly after $z\sim 2.4$ by about 10\% and climbs back to the smooth scaling again by $z\sim2.9$.

These observations do not constrain when the hydrogen reionization
began or how it proceeded. 
\citet{Bolton:2007} argue for an extended reionization period. 
\citet{Bolton:2013} show with simulations that observations may not 
require a large change in $x_{\rm HI}$ between $z\simeq 6$ and 7, 
but ``may instead be indicative of the rapid decrease in the typical 
mean free path for ionizing photons expected during the final stages of reionization" from 
``the increasing incidence
of absorption systems which are optically thick to 
Lyman continuum photons''. More recent observations start to probe the $z\simeq 6-7$ range.
\citet{Ota:2017} found that comparison of models of Lyman-$\alpha$ 
emitters with the measured Lyman-$\alpha$ luminosity function  
suggested that the neutral fraction of H increased with redshift
at $z>6$. These observations are compatible with both fast (steep)
and extended reionization histories as described below.

\subsubsection{Thomson optical depth and high-$z$ ionization}
\label{sec:thomson}

Detailed transition modeling from neutral to ionized state of the IGM hydrogen is  
subject of intense current theoretical and observational investigations with the 
main observational constraints coming from CMB temperature anisotropies.
CMB photons are scattered off free electrons, damping the primary
anisotropies and generating a large-scale polarization signal
\cite{Mukhanov:2005} and secondary anisotropies \cite{Vishniac:1987,Atrio-Barandela:1994}. 
These effects are determined by the Thomson scattering (cross-section  
$\sigma_T=6.65 \times10^{-25}$cm$^2$) optical depth 
given by 
\begin{eqnarray}
\tau_e(0,z_{\rm reion})\! & = & \! 2\times10^{-3}\! \int_0^{z_{\rm reion}}\!\! 
\frac{x_e(z) (1+z)^2}{\sqrt{\Omega_{\rm m}(1+z)^3+\Omega_\Lambda}}dz\!   \nonumber \\ 
& \equiv &
0.038+\Delta \tau_e(z\!>\!6) 
\label{eq:dtaudz}
\end{eqnarray}
where $x_e$ is the fraction of free electrons at each redshift $z$.
In eq.~(\ref{eq:dtaudz}), the integration gives the total
optical depth. A small fraction of cold gas exists in the form of galaxies and Ly-$\alpha$
systems, that could be as large as 10\% \cite[see references in ][]{Salvador-Sole:2016}. 
Removing the contribution $\tau_e(0,6)\simeq0.038$ at $z\leq6$, as evidenced by the GP absorption probes, leaves
\begin{equation}
\Delta \tau_e(z\!>\!6)\simeq 0.003 \Omega_{\rm m}^{-1/2} \int_{z=6}^{z_{\rm reion}} 
x_e(z)\sqrt{z} dz
\label{eq:tau_hi-z}
\end{equation}
as the high-$z$ contribution to 
the net Thomson optical 
depth, which is constrainable by CMB and is of relevance here. 
It gives a weighted measure of the fraction of 
free electrons, $\langle x_e(z\!>\!6)\rangle$, since the
start of reionization at the unknown redshift $z_{\rm reion}$ until 
the epoch when the GP absorption is known to vanish, $z\simeq 6$. 
These epochs contain the first stars era.

After reionization a fraction $1$--$\exp[-\tau(0,z_{\rm reion})]\simeq 0.038+\Delta\tau$
of CMB photons is scattered off, so their contribution to the 
primary CMB fluctuations gets smeared out up to the reionization horizon scale
$\ell_{reion}\simeq \pi z_{\rm reion}^{1/2}\Omega_m^{0.09}$.
Due to the damping of the primary CMB radiation 
power spectrum $C_\ell$, CMB TT anisotropies constrain the 
amplitude of the matter power spectrum as $A_S \exp[-2\tau(0,z_{reion})]$ and, in combination with gravitational lensing measurements
that are sensitive to $A_S$, this can be used to place useful constraints on 
$\tau(0,z_{reion})$ \cite{Hu:2002,Planck-Collaboration:2016a}.

The large-scale E-mode polarization of the CMB is a very sensitive probe of reionization
\cite{Reichardt:2016}. Compton scattering produces polarization only when the incident field 
has a quadrupole moment \cite{Rees:1968,Hu:1997}. 
While photons and baryons are tightly coupled, only the dipole anisotropy is present. Thomson scattering  generates polarization causally from the 
quadrupole component of the underlying ionized matter distribution only 
up to the horizon scale at the time. E-polarization is only generated during 
recombination and reionization and so it reflects the horizon 
scale at reionization. Any such signal on super-degree scales directly indicates 
the epoch when reionization started. The amplitude of the polarization anisotropy 
is proportional to the duration of recombination/reionization and is maximal at
the scale of the horizon \cite{Mukhanov:2005} which corresponds
to $\ell\sim 100$/$\ell\sim 10$, respectively.
The angular scale and width of the reionization contribution to the
E-mode of CMB polarization power spectrum encodes
information about the reionization history.
At $\ell<10$ the amplitude of the E-mode
polarization power spectrum is two orders of magnitude smaller than the
temperature anisotropy power spectrum; the measurement requires not only
detector sensitivity to those low signals, but control of systematic
errors and foreground residuals down to those levels
\cite{Planck-Collaboration:2016b}.

\begin{table}
\caption{Thomson scattering optical depth from {\it WMAP} and {\it Planck} analyses.}
\begin{tabular}{|l|l|p{4cm}|}
\hline
Data & $\tau_e$ & Ref.\\
\hline
WMAP 1 yr  & $0.17\pm 0.04$ & \citet{Kogut:2003} \\
WMAP 3 yr & $0.10\pm 0.03$ & \citet{Page:2007} \\
WMAP 9 yr & $0.089\pm 0.014$ & \citet{Hinshaw:2013} \\
Planck 2013  & $0.089\pm 0.014$ & \citet{Planck-Collaboration:2014} \\
Planck 2015  & $0.075\pm 0.013$ & \citet{Planck-Collaboration:2016a} \\
TT+lens.+BAO & $0.067\pm 0.016$ & \citet{Planck-Collaboration:2016a} \\
Planck 2016 & $0.058\pm 0.012$ & \citet{Planck-Collaboration:2016b} \\
\hline
\end{tabular}
\label{table:tau}
\end{table}

In Table~\ref{table:tau} we list the values measured over time.
The large discrepancies and the constant decline of the central value
reflect the difficulty of the measurement.
WMAP 1yr value was obtained from the temperature E-mode of polarization cross
power. The quoted value of WMAP 3yr data was based on the E-mode of polarization;
subsequent WMAP data releases reduced the error bar. The 2013 Planck results
used the Planck based power spectra and WMAP polarization data and derived the
same result than the final WMAP 9yr data analysis. In Planck 2015, foreground
cleaning using Planck 33GHz and 353GHz maps further reduced the value of $\tau$.
Adding CMB lensing data, the optical depth decreased to $\tau=0.066\pm 0.016$
consistent with PlanckTT+Lensing+BAO result that uses no low-$\ell$ polarization
data. However,  since Planck measured a lensing power spectrum larger than the
amplitude expected from the $\Lambda$CDM model with Planck
measured parameters, lensing data tend to prefer lower values of $\tau$. 
The Planck 2016 result includes Planck temperature and 
High Frequency Instrument (HFI) polarization data.
These measurements, although
derived assuming reionization in a step-like transition, have important
implications for the physical processes driving the reionization of the IGM
\cite{Greig:2016,Mitra:2016}. 
The high value measured by WMAP 1yr supported models of early, $z\approx 15$
reionization \cite{Choudhury:2006}, driven by metal free Pop III stars.
The decreasing values of $\tau$ measured from subsequent observations
reduced the need for high-z galaxies as reionization sources 
\cite{Robertson:2015, Bouwens:2015, Mitra:2015}.  A steep reionization
favors models where quasar contributions were negligible
at $z\ge 6$ and the earlier reionization was driven by early galaxies.
Alternative sources such as dark matter annihilation and decay have
also been considered \cite{Liu:2016}.

\begin{figure}[t]
\includegraphics[trim=2.5cm 1cm 1cm 0, width=3in]{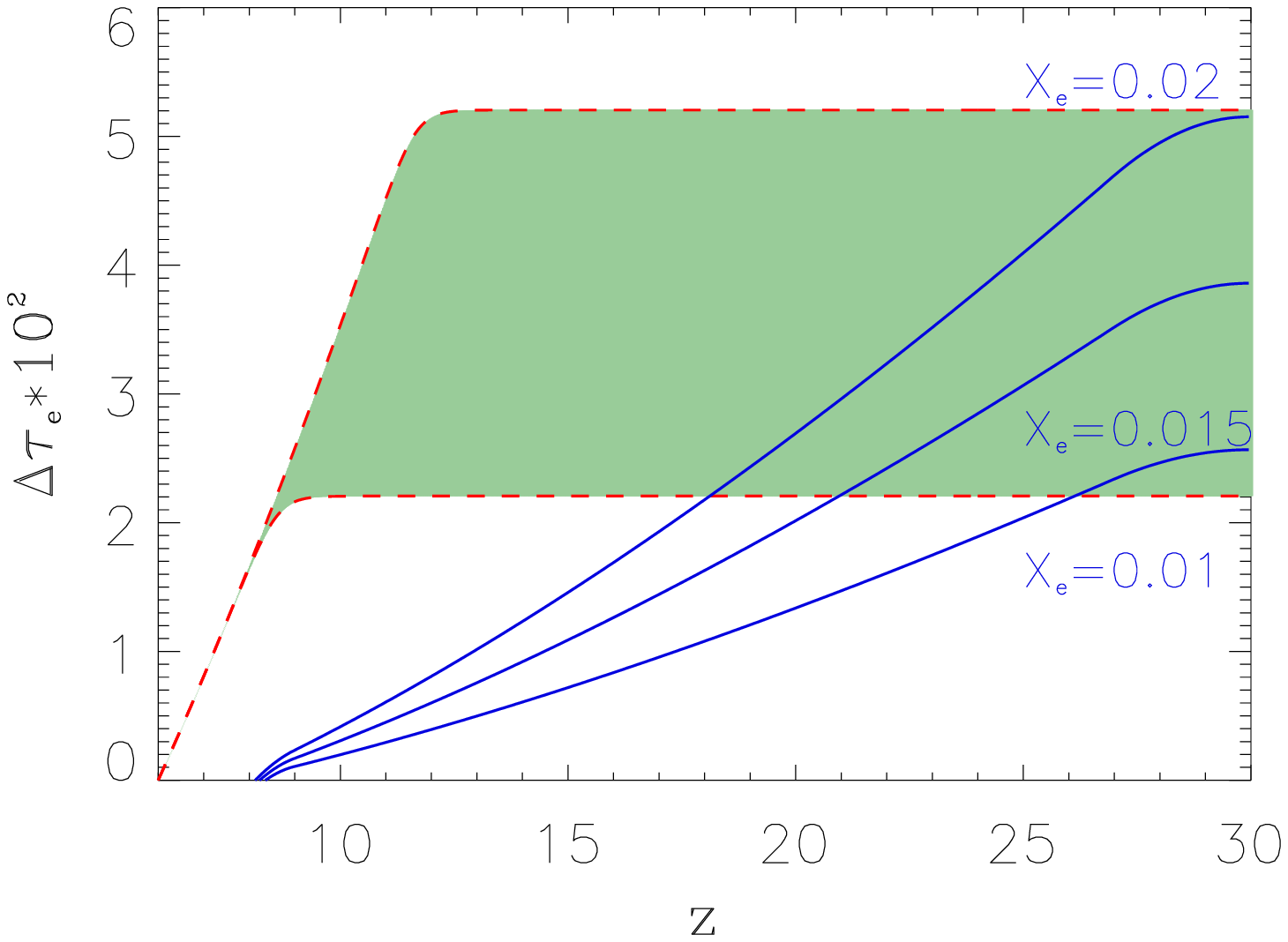}
\includegraphics[trim=2.5cm 1cm 1cm 0, width=3in]{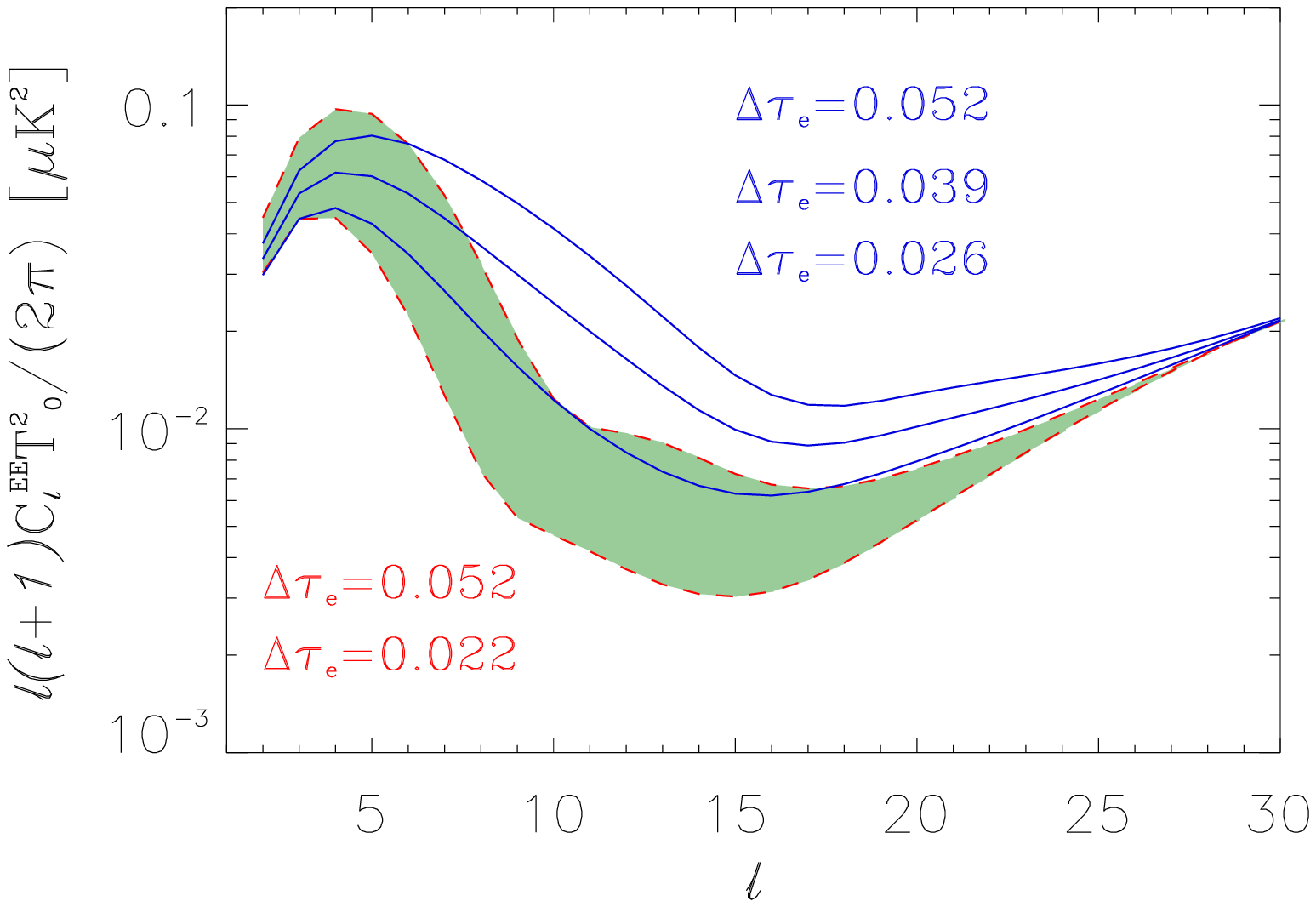}
\caption[]{\small{Top: $\Delta\tau_e$ from $z$=6 to 
start of reionization at $z_{\rm reion}$=30. Shaded area corresponds to a 
$z$-symmetric tanh transition of width $\delta z$=0.5. Dashed lines bound $\tau_e$ for $x_e$=1 at 
$z_{0.5}$=8.5/11.5 
corresponding to lower/upper
limits on net $\tau_e$ of \citet{Planck-Collaboration:2016b}. Blue solid lines 
correspond to ionization histories with $x_e$ growing linearly
from $z_{\rm reion}$ to $z$=27 until $x_e=0.01,0.015,0.02$, then remain 
constant to $z$=9, and then again grow linearly
until $z=6$ to $x_e$=1. The contribution to the optical 
depth from redshifts $z\le 6$ is taken to be $\tau_e(z<6)=0.038$ per eq.\ \ref{eq:dtaudz}.
Bottom: Power spectra of the E-polarization 
anisotropies. The shaded area bounded by the red dashed lines
corresponds to the symmetric hyperbolic tangent and the blue solid lines 
to the constant $x_e$ from $z$=9 to $z$=27 in the top panel. $T_0$ is the CMB temperature.
}}
\label{fig:tau}
\end{figure}
For a steep reionization the most recent value of  
Table~\ref{table:tau} implies that the average redshift at which reionization 
occurred was between $z=7.8$ and $8.8$ and lasted $\Delta z<2.8$.  
The Planck Collaboration used the reionization fraction
$x_e(z)=(f/2)[1+\tanh[(u(z_H)-u(z))/\Delta u)]$,
with $u(z)=(1+z)^{3/2}$, $\Delta u=(3/2)(1+z)^{1/2}\Delta z_H$ and $\Delta z_H=0.5$
\cite{Lewis:2008} and fit the value of $z_H$ to the measured value of $\tau$.
But the width and location of the polarization peak contains more 
information than the overall Thomson optical depth. 
Allowing arbitrary ionization histories shows a preference in the data for
more extended reionization processes out to $z\sim 30$  \cite{Heinrich:2016}.
\citet{Heinrich:2016} considered a fiducial model with 
a constant ionization $x_e=0.15$ in the range $6\le z\le 30$ although the
exact value fluctuates around this fiducial model to fit the 
$C_\ell^{EE}$ data. The excess of power in the E-mode of polarization in
the multipole range $10\leq\ell\leq 20$,
present in Planck 2015 LFI data, is compatible with $\sim 20\%$ of the
volume of the Universe being ionized by $z\sim 20$ \cite{Miranda:2016}. 

In Fig.~\ref{fig:tau} we plot the contribution to the Thomson optical depth for  
the two ionization histories discussed above as a function of redshift, $\tau(0,z)$.
When computing the fraction of free electrons, it is necessary to take into
account the contribution from He. Its first ionization happens in parallel to 
that of H but its second ionization, requiring 54eV photons, is assumed to have been
delayed until quasars, that can emit the necessary energetic photons are 
sufficiently abundant, at $z\sim 3-4$ \cite{Madau:1994, Miralda-Escude:2000,Becker:2011}.
Then, we take $f=1+f_{He}$ for singly ionized He and $f=1+2f_{He}$
for doubly ionized He. We model the He reionization by a tanh
function centered at $z_{He}=3.5$ and width $\Delta z_{He}=0.5$.
Since GP test shows the Universe is ionized by $z\simeq 6$,
the total contribution of the IGM to the Thomson optical depth 
up to $z=6$ is $\tau(0,6)=0.038$.
A small fraction of cold gas exists in the form of galaxies and Ly-$\alpha$
systems, that could be as large as 10\% \cite[see references in ][]{Salvador-Sole:2016}. 
Removing this contribution yields a conservative lower bound on the CMB optical depth
of $\tau(0,6)\simeq 0.035$. In Fig~\ref{fig:tau} the (blue) dashed and (red) 
dot-dashed lines correspond to the tanh-model with Planck 2015 and Planck LFI
2016 CMB optical depth values while the (black) solid line corresponds
to the extended reionization model. The contribution to the optical
depth at $z\le 6$ is $\tau=0.038$, common to all models. However, when
fitting Planck 2016 LFI data, the extended model allows for a much higher 
contribution from high-$z$ sources since
$\Delta\tau\simeq 0.045$ while this contribution is 
only $\Delta\tau\simeq 0.02$ for the tanh model. Note that the peak of 
the polarization power spectra of the symmetric hyperbolic tangent 
reionization model is narrower than the extended reionization model.

\subsection{New high-$z$ populations and their consequences}
\label{sec:highz-pops}

Here we discuss the various high-$z$ candidates that contribute to CIB, and 
the environments in which they form and subsequently influence.
These sources leave potentially detectable signatures of their redshifts
though the Lyman break that should truncate their UV emission.
The CIB that they leave behind is subject to reionization constraints discussed above in Sec.\ \ref{sec:reionization}. For more detailed information regarding emissions from the individual sources possible at high $z$ the reader is referred to excellent recent reviews by \citet{Bromm:2013a,Ferrara:2012b,Latif:2016}.

\subsubsection{First halo collapse}
\label{sec:firsthalos}

Given the underlying matter power spectrum, the number density
of available halos can be computed via the Press-Schecter formalism \citep{Press:1974}, assuming that any region that reached density
contrast $\delta_{\rm col}$=1.68 undergoes spherical collapse.
The emergence of the first luminous sources at the end of the cosmic dark
ages is largely governed by the ability of primordial gas to cool inside these halos \citep[e.g.][]{Bromm:2013}.
In the absence of any metal coolants, prior to the dispersal of the first
heavy elements from Pop~III supernovae, there are two principal cooling channels
in the early universe. At temperatures in excess of $\sim 10^4$\,K, line radiation
from atomic hydrogen, predominantly concentrated in the Ly$\alpha$ transition,
provides very strong cooling. For the $\Lambda$CDM power spectrum, the first DM halos are characterized by shallow gravitational potential
wells, with correspondingly low virial temperatures, $T_{\rm vir}$. 

\begin{figure}[t]
\includegraphics[width=3.in]{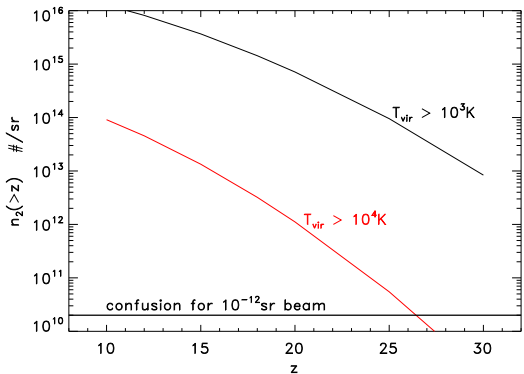}
\caption{\small  Adapted from \citet{Kashlinsky:2015a}. The projected angular density of early luminous halos for $P_{\rm PBH}=0$ at redshifts greater than $z$ assuming stars and accreting BHs form when $T_{\rm vir} \geq 10^3$K (black) and $10^4$K (red). If PBHs make up the DM, the number would be higher. Horizontal thick solid line shows the confusion limit for a beam of $10^{-12}$ sr (or 0.04 arcsec$^2$) in area.}
\label{fig:confusion}
\end{figure}
Halos with $T_{\rm vir}\lesssim 10^4$\,K will thus not be able to activate
atomic hydrogen cooling. In such low-$T_{\rm vir}$ systems, the so-called
minihalos, cooling has to rely on molecular hydrogen. The H$_2$ formation
chemistry in the absence of dust grains is catalyzed by free electrons 
left over from the epoch of recombination, with a rate that depends on
the gas temperature: ${\rm H}+e^-\rightarrow {\rm H}^{-} +\gamma$ followed by ${\rm H}^{-}+{\rm H} \rightarrow{\rm H}_2+e^{-}$ \cite{Yoneyama:1972}. (At higher densities 3-body reactions become important, $3{\rm H}\rightarrow {\rm H}_2+{\rm H}$ and $2{\rm H}+{\rm H}_2\rightarrow 2{\rm H}_2$ \cite{Palla:1983}). For sufficient H$_2$ production, temperatures
of $\sim 10^3$\,K are required. This effect selects DM halos with
$T_{\rm vir}\sim 10^3$\,K, minihalos, as the formation sites for the
first (Pop~III) stars. Molecular hydrogen, however, is fragile, and can
easily be destroyed by non-ionizing soft-UV photons in the Lyman-Werner (LW, 11.2--13.6 eV) bands.
Such a pervasive LW background is expected to rapidly emerge in the
aftermath of the initial Pop~III star formation.
It has therefore been argued that the first galaxies, defined as systems
that can sustain self-regulated star formation, will be hosted by more massive
DM halos. Indeed, ``atomic cooling halos'' with
$T_{\rm vir}\gtrsim 10^4$\,K are considered promising candidates for first-galaxy
hosts, as they would not have to rely on H$_2$ as a coolant, and could instead
tap into the much more efficient, and resilient, atomic hydrogen channel.
In summary, there are two characteristic scales for DM host halos, expressed
in terms of $T_{\rm vir}\sim 10^3$\,K and $\sim 10^4$\,K, where the former is
predicted to host the first stars, and the latter the first galaxies \citep[see review by][]{Bromm:2011}.

Fig. \ref{fig:confusion} shows the projected density of collapsed halos with parameters suitable for star formation assuming the power spectrum shown in Fig. \ref{fig:p_lcdm}.  The horizontal line shows the confusion limit for the {\it JWST}-type beam assuming confusion intervenes at $>0.02$ halos/beam \cite{Condon:1974}. There are various possibilities for boosting the small scale power in the spectrum of underlying matter fluctuations, from e.g.\ modified inflationary and early Universe physics \cite{Clesse:2015,Kashlinsky:2016,Kusenko:2017} increasing the levels of the CIB from first stars era.

\subsubsection{First stars}
\label{firststars}

Anthropic argument dictates that stars have typical mass of order solar, i.e.\ given the values of the fundamental constants a self-gravitating thermonuclear reactor must have mass $M_* \sim M_\odot$. This follows as: 1) pressure equilibrium of gas in stars gives their typical temperature $T_*\sim
m_p^{4/3}k_{\rm B}^{-1} GM_*^{2/3} n_p^{1/3}$, where $m_p$ and $n_p$ are proton mass and number density. 2) If all baryons participate in nuclear burning, $\gamma$'s would have number density similar to protons  $n_\gamma \sim n_p$. 3) Stars are optically thick, so photons are at the same thermodynamic temperature. Thus $P_{\rm radiation}\sim n_\gamma k_B T_* \sim P_{\rm gas}$. 4) Stars radiate as black-body $aT_*^4\sim n k_BT_*$, so $T_*\sim (k_B/a)^{1/3}n^{1/3}$. Combining 1)-4) leads to $M_*\sim M_{\rm Chandra} = (\frac{\hbar c}{G})^{3/2}m_p^{-2} \simeq 1.44M_\odot$.

So how do stars reach that mass when collapse starts in much more massive clouds and halos? In his seminal galaxy formation paper \citet{Hoyle:1953} proposed the so-called opacity-limited fragmentation theory for the origin of stars. He noted that initially, while the cooling time is much less than the collapse time protostellar clouds collapse isothermally. As a result the Jeans mass in the cloud, $M_{\rm Jeans}\propto T^{3/2}\rho^{-1/2}$, decreases as the density, $\rho$, increases during the collapse. The cloud becomes susceptible to fragmentation into progressively smaller clumps. This hierarchical fragmentation stops when density gets high enough to make the fragment opaque, and trap the radiation released, via shocks and such, during collapse. For absorption opacity this happens when the optical depth across the fragment reaches $\tau \sim 1$, the temperature starts rising adiabatically as $T\propto \rho^{2/3}$. The Jeans mass then stops decreasing and the final fragment forms. Hoyle showed that this occurs, for solar metallicity opacities, at a fraction of $M_\odot$; the rest presumably getting accreted after the fragmentation stops.

Prior to enrichment of gas with metals, atomic hydrogen can only cool gas to $T\sim 10^4$K and H$_2$, which is hard to form, could lower the temperature a bit as discussed below.  Given that the Jeans mass $\propto T^{3/2}$, it was thought early on that first stars would have to be massive. However, \citet{Rees:1976} showed very generally  that this does not have to be the case: The maximal achievable rate of cooling is the black body one, when emission is radiated by the surface of the fragment with radius $r$, at $\simeq aT^4 c (4\pi r^2)$. Collapse ceases being isothermal and fragmentation stops when the cooling is of order the free-fall rate of release of binding energy, $\simeq (GM^2/r) (G\rho)^{-1/2}$. Combining this with the Jeans criterion for fragment's mass, $M_F\simeq (\pi k_B T/m_pG)^{3/2}\rho^{-1/2}$ leads to the minimal fragment mass being $M_{F,{\rm min}}\simeq M_{\rm Chandra} (k_BT/m_pc^2)^{-1/4} \propto T^{-0.25}$. Note only the weak dependence on $T$ which arises because  while there are no coolants in the absence of metals to keep $T$ low, the same absence of metals makes the onset of $\tau\sim 1$ occur at higher $\rho$. Thus the absence of metals does not necessarily require high masses for forming stars. Efficiency of fragmentation is also affected by the angular momentum of the collapsing protogalactic clouds \cite{Kashlinsky:1982}.

The above assumes that stars formed in efficient fragmentation
of collapsing much more massive clouds, a condition that is not necessarily
applicable to the first objects forming out of smooth density field in
metal-free early Universe. 
Current models suggest that the first, metal-free (Pop III) stars formed at $z < $30 in dark-matter mini-halos with virial temperatures $T_\mathrm{vir} < 10^4$K cooling their gas via H$_2$ line emission. Detailed numerical work \citep[][]{Bromm:1999,Abel:2002} in the context of the standard $\Lambda$CDM model suggested that first stars are likely very massive forming out of high density clumps ($n\sim 10^8$cm$^{-3}$) inside the $\sim10^6M_\odot$ minihalos \citep[see review by][and refs therein]{Bromm:2004}. 

Once formed, Pop III stars affect their own evolution (feedback) in two ways: (a) by producing copious amounts of LW photons they photo-dissociate H$_2$ molecules in nearby objects \citep{Shang:2010, Sugimura:2014, Agarwal:2014, Regan:2014}, quenching their cooling and star formation; (b) by polluting the star forming gas with metals dispersed by Pop III SNe, thereby changing the gas fragmentation properties \citep{Schneider:2002, Schneider:2006}, and inducing a transition to a normal Pop II star formation mode \citep{Tornatore:2007, Xu:2013, Pallottini:2014}. Numerous evolution modes and stellar activity in pregalactic Universe have been discussed with varying constituents from quasi-normal stellar populations, massive stars, BHs forming in the course of stellar activity, massive binaries, etc \citep[e.g.][]{Kashlinsky:1983,Fernandez:2006,Fernandez:2010,Fernandez:2012,Santos:2002,Salvaterra:2003,Mirabel:2011,Cooray:2004a,Helgason:2016}.

Massive stars, 
such as hypothesized to dominate the first stars era, are radiation-pressure dominated, and emit nearly at the Eddington limit. In addition, they are close to fully convective with the entire stellar mass taking part in the hydrogen burning \citep{Bromm:2001,Schaerer:2002}. This leads to the  high radiative efficiency of $\epsilon\sim0.007$ and a correspondingly more efficient CIB production.  For normal Pop II stars, described by a Salpeter IMF, the effective efficiency is an order of magnitude lower since only a small core burns hydrogen. 

\subsubsection{First black holes}
\label{firstbhs}

For accreting BHs, the radiative efficiency can be as high as $\epsilon=0.4$ for maximally rotating Kerr-holes, reaching values much greater than that of H-burning. Thus, BHs can contribute significantly even with much smaller fraction than stars. Two type of stand-alone BHs appear relevant to discuss in this context: 
(a) Direct collapse BHs (DCBHs) forming during cosmogonic evolution during the first stars era. These BHs would be very massive, as discussed below, but of low abundance. And
(b) Primordial BHs (PBHs) which may have formed in the very early Universe \citep[e.g.][]{Carr:1975} with  much lower masses, comparable to the mass within the cosmological horizon at the time of their formation, but having much greater abundance.

\paragraph{DCBHs:}
The process by which astonishingly massive (BH mass $M_\bullet\approx 10^9 M_\odot$) BHs came into existence within 1 Gyr from the Big Bang is one of the most puzzling mysteries in cosmic evolution. The current paradigm stipulates that supermassive BHs (SMBHs) have grown from smaller seeds by gas accretion. This hypothesis, however, faces a number of difficulties.  
The most striking complication is connected with the short time available for the build-up of SMBH. Assuming that gas accretion occurs at the Eddington rate, assembling the SMBH mass ($M_\bullet = 2\times 10^9 M_\odot$) deduced for the most distant quasar ULAS J1120+0641 \citep{Mortlock:2011} at redshift $z = 7.085$ (or, cosmic age 0.77 Gyr) requires a seed mass $M_\bullet > 400 M_\odot$). Such value is about 10 times larger than the most recent estimates of the mass of first stars (and, consequently, of their remnant BHs). Serious concerns are raised also by the assumption that accretion occurs at the Eddington rate.
The most obvious route to form the early BHs during the first stars era is via the final collapse of sufficiently massive Pop III stars ($M > 30 M_\odot$, with the exception of the narrow pair-instability interval $150 < M/M_\odot < 260$ \citep{Woosley:2015, Aoki:2014}). A number of studies \citep{Alvarez:2009, Milosavljevic:2009, Johnson:2013, Jeon:2014} have shown, however, that stellar BHs accrete very inefficiently, although under some extreme conditions they might grow super-critically if accretion occurs through a slim disk \citep[e.g.][]{Madau:2014, Alexander:2014, Volonteri:2015} because they spend most of their lifetime in low-density regions. Thus, they appear unable to rapidly build the observed $z=7$ SMBH population. 
A compelling solution is to start with a significantly larger seed mass. The early proposals for the formation of massive BHs directly from the gas-phase by \citet{Loeb:1994,Eisenstein:1995} have now developed into more complete scenarios \citep{Begelman:2006, Regan:2009, Petri:2012, Johnson:2013, Latif:2013, Yue:2014, Ferrara:2014}. The direct collapse channel invokes the formation of  massive BHs in environments where gas gravitational collapse proceeds at sustained rates ($> 0.1 M_\odot$ yr$^{-1}$). 
The most promising candidates for these super-accreting environments are dark matter halos with virial temperature (a proxy for mass)  $T_\mathrm{vir} \sim 10^4$ K. In these halos primordial gas cools almost isothermally via collisional excitation of the hydrogen $1s-2p$ transition followed by a Ly$\alpha$ photon emission. As the accretion rate is $\propto T_\mathrm{vir}^{3/2}$, this mechanism guarantees extreme accretion rates, $\sim0.1-1 M_\odot$ yr$^{-1}$, feeding the central object, a central protostellar gas condensation. 
For efficient feeding, the accretion flow should remain smooth, i.e.\ it should not fragment. Fragmentation is in general induced by a softening of the equation of state below the isothermal value $\gamma = 1$, i.e.\ the gas cools as it gets denser. While Ly$\alpha$ cooling keeps the gas on the isothermal track, the presence of H$_2$ molecules, heavy elements or dust provides extra cooling, and induces fragmentation. In primordial gas, one has then to prevent only the formation of H$_2$. This can be achieved by irradiating the collapsing gas with a sufficiently strong external UV field that photo-dissociates H$_2$. Such UV radiation field is likely coming from a nearby star-forming galaxy and/or the general collective background radiation from all galaxies present at earlier redshifts.  
UV radiation effects on larger halos ($T_\mathrm{vir} > 10^4$ K) are spectacularly different \citep{Dijkstra:2014, Agarwal:2012, Visbal:2014}. If H$_2$ is photo-dissociated by a sufficiently strong LW intensity, $J > J_\mathrm{crit}$, hydrogen Ly$\alpha$ line emission and other processes sustain an almost isothermal collapse preventing gas fragmentation into stellar sub-units. Under these conditions, theoretical works \citep{Bromm:2003, Begelman:2006, Volonteri:2008, Regan:2009, Van-Borm:2014} show that the most likely outcome is a rapid ($\approx 1$ Myr) formation of a $M_\bullet =10^{4-6} M_\odot$ DCBH. However, this process can occur only as long as the gas is metal-free; otherwise fragmentation and star formation would take place \citep{Ferrara:2013}. As DCBH also emit LW radiation, they might stimulate additional DCBH formation \citep{Yue:2016}. Finally, X-rays from DCBH preheat the intergalactic medium, before galaxies reionize it. 
In conclusion, the key requirements of the mechanism are that the collapsing gas:  (a) can be cooled by Ly$\alpha$ line emission; (b) is metal-free, and (c) is exposed to a UV radiation field. Whether and for how long these conditions can be simultaneously met during cosmic evolution is unknown. In halos meeting conditions (a)-(c) the central gas condensation grows rapidly, turning into a Super Massive Star (SMS). The Kelvin-Helmholtz timescale of these objects is much longer than their accretion time, implying that they grow virtually without emitting light. 
If during the evolution smooth accretion can be maintained, the SMS grows until it finally encounters a General Relativity instability. This will induce a rapid, direct collapse into a massive BH, i.e.\ without passing through a genuine stellar phase. These objects, with masses up to $10^6 M_\odot$, are dubbed as DCBHs. If they exist, DCBH would represent the ancestors of SMBH and offer the ultimate solution of the problems plaguing the field. For a more thorough review of DCBH we refer the reader to \citet{Latif:2016}. 
Finally, in addition to direct collapse, \citet{Begelman:1978,Kashlinsky:1983} pointed out that massive seeds may also form as a result of star-star runaway collisions in young ultra-dense Nuclear Star Clusters \cite[for modern versions see, e.g.][]{Portegies-Zwart:2002, Lupi:2014}.  

\paragraph{PBHs:} \label{sec:PBH} The LIGO discovery of GWs from a pair of BHs of similar and unexpected mass ($\sim 30M_\odot$) \citep[][]{Abbott:2016} has rekindled suggestions that DM may be composed entirely or predominantly of PBHs \cite{Bird:2016,Clesse:2016,Kashlinsky:2016}. PBHs in the mass range of $\sim 10-100M_\odot$ appear allowed by the available observational data \citep[see discussion in e.g.][]{Carr:2016} and the required abundance would appear in broad agreement with recently claimed abundance of quiescent black hole X-ray binaries in our Galaxy \cite{Tetarenko:2016} and possibly also with the observations of high-velocity clouds near the Galactic Center driven by inactive BHs rapidly plunging into molecular clouds \cite{Takekawa:2017,Yamada:2017}.  The mass range also is within the cosmological horizon at $\sim 0.01-0.1$ GeV when various mechanisms for generating PBHs in the early Universe operate  \citep[e.g.][]{Jedamzik:1997}. The strongest constraint against them was claimed by \citet{Mack:2007,Ricotti:2007,Ricotti:2008} to arise from observations of the lack of distortions of CMB black-body spectrum from {\it COBE} FIRAS \citep{Mather:1990,Fixsen:2009}, but new recent reanalyses of the accretion efficiency onto PBHs during the pre-recombination era find significantly weaker constraints and argue against ruling out PBHs of $\lsim 100M_\odot$ as the dominant component of DM \citep[][]{Ali-Haimoud:2016,Aloni:2016,Horowitz:2016}. Numerous other observational tests of this proposal have been suggested \citep[e.g.][]{Brandt:2016,Hawkins:1993,Schutz:2017,Munoz:2016}. If DM is made up of PBHs, the latter would introduce a new Poissonian component to the underlying density field as pointed out first by \citet{Meszaros:1974,Meszaros:1975} prior to development of inflationary paradigm. This component would substantially accelerate the collapse of the first halos and potentially make substantial contribution to the CIB \cite{Kashlinsky:2016}. If PBHs are present they would also require theoretical modifications in the processes affecting first stars era objects. Gas at sound speed $c_s$ in halo of velocity dispersion $v_d$ is accreted within the radius $r_{\rm acc}=GM_{\rm PBH}/u^2$ with $u^2=v_d^2+c_s^2$. The total accretion mass is $M_{\rm acc} =2 (n_{\rm gas}/10^4 {\rm cm}^{-3}) (M_{\rm PBH}/30M_\odot)^3 (u/1{\rm km\;sec}^{-1})^{-6} M_\odot$. For typical parameters this may be a non-negligible fraction of the minihalo baryons at $\sim M_{\rm acc}/M_{\rm PBH}\times \Omega_{\rm CDM}/\Omega_{\rm bar}\propto M_{\rm PBH}^{2} u^{-6}$ up to a few percent, but will not increase the PBH mass dramatically.  
Radiation from accreting PBHs may inhibit H$_2$ formation and thus influence adjacent star formation and DCBH collapse and evolution as discussed e.g.\ in \citet{Bromm:2003,Agarwal:2012,Yue:2014}. 
At the same time, the increased fractional
ionization of the cosmic gas produced by PBHs increases the primordial
H$_2$ abundance by up to two orders of magnitude \citep{Ricotti:2008}. The
increase of the cosmic Jeans mass due to X-ray heating is negligible
for models consistent with the CMB data. Hence, the formation rate of
the first galaxies and stars would be enhanced by a population of
PBHs.  Furthermore, stellar dynamical evolution of PBH minihalo may play important cosmogonical role. The PBHs in minihalos will evolve via secular stellar dynamical effects similar to that discussed in \citet{Kashlinsky:1983} and by loss of energy to GW emissions.  Stellar evaporation will lead to a core-halo structure with the isothermal core of radius $r_{\rm c}$ and $N_{\rm PBH}$ PBHs evolving on Gyr timescales $t_{\rm evap}\sim N_{\rm PBH}/\ln N_{\rm PBH} \times r_{\rm c}/v_d$, at constant binding energy, or $v_d\propto N_{\rm PBH}^{-1/2}$, because evaporating PBHs carry zero energy. Formation of massive BHs may be accelerated here. A fraction of PBHs will become binary when GW emission exceeds their kinetic energy ($\sim v_d^2$)
\citep{Bird:2016}. The fraction of PBHs that will form binaries before evaporation is $f_{\rm PBH, binary} \sim \frac{N_{\rm PBH}^2}{\ln N_{\rm PBH}}\frac{10^{-8}{\rm pc}^2}{r_c^2} \left(\frac{M_{\rm PBH}}{30M_\odot}\right)^{-2} \left(\frac{v_d}{1{\rm km\; sec}^{-1}}\right)^{-18/7}$ \citep{Kashlinsky:2016}. Instead of evaporating the resultant binaries will spiral in to the center due to dynamical friction possibly forming a central large BH contributing to the massive BH formation in early Universe. 

\subsubsection{Impact on/from thermal history of IGM}
\label{sec:igm}

The IGM is heated by the radiative and, to a smaller extent, mechanical 
energy deposition by stars, black holes, and possibly dark matter 
annihilation/decay along cosmic evolution. Heating from these sources 
occurs primarily in the form of photoionization heating. As photons 
with energies $h\nu > 13.6\, \mathrm{eV}$
ionize neutral 
hydrogen atoms, the energy of the photoelectron is gradually thermalized 
in the gas resulting in a temperature increase.  The same process controls 
the ionization of He atoms, requiring 24.6 eV and 54.4 eV to produce 
singly- or doubly-ionized ions of He. Contributions from heavier 
elements are negligible due to their low abundances. 

Photoionization largely dominates the thermal budget of the IGM. Shock-heating 
of the gas produced, e.g.\ by supernova-driven galactic outflows, is confined 
in small volumes \citep{Ciardi:2000, Ciardi:2005} around galaxies, 
in the so-called circum-galactic medium extending up to about the virial 
radius of the galaxy host halo. This conclusion is supported by the low 
level of intergalactic turbulence measured in the IGM \citep{Rauch:2001, Evoli:2011}. 
Supernova-heated gas might have marginally ($<10$\%) contributed to 
reionization by up-scattering of CMB photons inside hot bubbles 
\citep{Oh:2001, Johnson:2011}. Additional heating to the IGM can be 
provided by cosmic rays \citep{Sazonov:2015}. However, both the production 
and diffusion of these energetic particles is quite uncertain in the 
Epoch of Reionization (EoR).  
  
For this reason, the thermal history of the IGM is intimately connected 
to the process of cosmic reionization. As ionized (\HII) regions grow 
around the sources and merge, progressively filling the intergalactic 
space, the gas within them is heated to a temperature typical of ionized 
regions, i.e.\ in the range $(1-30) \times 10^4$ K. The characteristic 
volume of these ``bubbles'' is given by the classical Str\"omgren 
formula $V_I = \dot N_\gamma /\alpha_B \bar n_e^2$, where $\dot N_\gamma$ 
is the source ionizing photon rate, $\alpha_B = 2.6\times 10^{-13} T_4^{-1/2}$ 
is the Case B recombination rate of hydrogen, and $\bar n_e(z)$ is the mean 
IGM electron density at the relevant redshift. If the sources have hard spectra, 
containing significant amounts of helium ionizing radiation, analogous \HeI\ and 
\HeII\ spheres (typically embedded in the H one) will be produced. These bubbles 
form the typical ``patchy'' structure characterizing the reionization process, 
in which \HI\ is progressively destroyed.

Outside \HII\ regions the IGM remains largely neutral. However, if some X-ray 
emitting sources like accreting BHs, high-mass X-ray binaries, 
annihilating/decaying dark matter, or hot emitting plasma do exist, the 
IGM can be heated well outside \HII\ regions. This 
is possible because the comoving mean free path, $\lambda_X$, of an X-ray photon 
of energy $E_X$ is 
$\approx (E_X/13.6\ \mathrm{ eV})^{2.6}$ 
times longer than that of UV photons
\begin{equation}
\lambda_X = 20 \bar x_{HI}^{-1} \left({E_X\over 0.3\ \mathrm{keV}}\right)^{2.6}
\left({1+z\over 10}\right)^{-2} \mathrm{Mpc},
\label{Xmfp}
\end{equation}
where $\bar x_{HI}$ is the IGM mean neutral fraction. Such long mean free 
path reduces the patchiness of the ionized gas, and results in a more uniform 
ionization field. 

In addition, X-ray ionization is often incomplete, with 
$\bar x_{HI}
\approx 10-30$\%, as when $\bar x_{HI}$ exceeds a few percent, most of the 
photon energy is deposited by secondary electrons \citep{Shull:1985, Valdes:2008} 
in the form of heat. The X-ray illumination therefore produce very extended patches 
of mostly neutral gas heated to temperatures of $\approx 1000$K in which much 
smaller \HII\ regions are embedded. 

The temperature evolution of the neutral IGM component is of great interest 
for the \HI\ 21 cm redshifted tomography of the IGM \citep{Mesinger:2013}. Lacking 
so far sensitive measurements of such signal (this situation is bound to change 
with the advent of the Square Kilometer Array), only weak lower bounds to the 
IGM spin temperature -- a good proxy of the kinetic temperature as the two are 
efficiently coupled by the Wouthuysen-Field effect -- can be derived: 
$T_s > 6$K at $z=8.4$ \cite{Greig:2016}. See review by \citet{Furlanetto:2006}. Hence, here we 
will concentrate on the temperature of the ionized gas during the reionization process. 

The equation describing the IGM kinetic temperature $T$ at mean density 
(where temperature measurements are available) is \cite{Theuns:2002}:   
\begin{equation}
\label{Tev}
\frac{1}{T}\frac{dT}{dt} = -{2 H} + \frac{1}{\mu}\frac{d\mu}{dt} 
+ \frac{2\mu}{3 k_B T } \frac{dQ}{dt},
\label{eq:t_igm}
\end{equation}
where $H$ is the Hubble parameter, $k_B$ Boltzmann’s constant, $\mu$ the mean 
molecular weight, and $Q$ is the effective (heating - cooling) radiative 
cooling rate. The latter includes photo-electric heating, and cooling via 
recombination, excitation, inverse Compton scattering, collisional ionization, 
and bremsstrahlung. The second term on the right hand side of the above is 
relatively unimportant, accounting for the change in the number of particles 
in the thermal bath; it becomes marginally important (few percent level)  
only during He II reionization.

Note that in the absence of heating the IGM temperature evolution 
would follow a purely adiabatic evolution imposed by the Hubble 
expansion (first term on the left side of eq. \ref{eq:t_igm}), corresponding to $T\propto (1+z)^2$. 
Adiabatic expansion remains the dominant cooling mechanism for gas around 
the cosmic mean, $\bar n
 \approx 2.3 
\times 10^{-7} \mathrm{cm}^{-3}(1+z)^3$; however, at $z>7$ the contribution 
of inverse Compton cooling off CMB electrons cannot be neglected. 

As the above equation depends on the number of particles, and hence on the 
ionization state of the gas, its solution requires a derivation of $x_e$. 
This is usually done by balancing the ionization rate from all the sources, 
$\Gamma$, with the recombination rate. For hydrogen, such equation in 
equilibrium is simply  $n_{HI} \Gamma =  n_e^2 \alpha_B(T)$; similar 
equations hold for the different ionization stages of He. As the 
photoionization timescale ($\approx \Gamma^{-1}$) is much shorter 
than the cooling timescale ($H^{-1}$), the implicit assumption of 
ionization equilibrium is well justified. 

A general expression for $\Gamma$ valid for both H and He is:
\begin{equation}
\label{Gamma}
\Gamma_i = c \int_{\nu^i_T}^\infty d\nu \frac{u_\nu}{h\nu} a_\nu^i,
\end{equation}
where $a_\nu^i$ is the photoelectric cross section of the species 
$i=$H, He, ${\nu^i_T}$ is the photoionization threshold frequency of 
species $i$, and $u_\nu$ is the specific energy density of the UV 
background. The specific energy density is related to the specific 
intensity of the radiation by $u_\nu = 4\pi J_\nu/c$.

The most standard approach is to adopt the new \cite{Haardt:2012} 
prescription for the UV background intensity evolution, and a spectral 
shape of the form $J_\nu \propto \nu^{-\alpha}$. The power index depends 
on the spectra of sources considered ($\alpha_s$) and the filtering due 
to radiative transfer effects in the Ly$\alpha$ forest whose logarithmic 
slope of the column-density distribution is $\beta$=$1.3 \pm 0.2$. Then, 
$\alpha$=$\alpha_s + 3(\beta-1)$, with $\alpha_s$=0.5--1.0 for stellar 
sources, and $\alpha_s$=$1.5 \pm 0.2$ for quasars. The considerable 
uncertainty in $\alpha$ does not represent a major problem as the IGM 
temperature sensitivity to this parameter is limited by the optically 
thin conditions prevailing in the IGM. Then the photoheating rate 
$dQ/dt \propto {(2+\alpha)}^{-1}$ varies at most by a factor of 2, 
corresponding to an even smaller temperature change as $T\approx (dQ/dt)^{0.6}$.

Thus, once the initial temperature of the gas, $T_i$, at some fiducial 
 $z_i$ is assigned, the thermal history can be computed 
straightforwardly. Most models take $z_i$ as the redshift of reionization, 
thus postulating that reionization is instantaneous. 
While this is known not to be the case, such assumption is justified 
if eq.\ \ref{Tev} is thought to describe the evolution of 
a Lagrangian fluid element that has been engulfed by an expanding \HII\ region at $z_i$. 

The outcomes of such models are used to interpret the temperature measurements 
obtained from the Ly$\alpha$ forest data in quasar absorption line experiments. 
These measurements became available around the beginning of this century 
\citep{Schaye:2000, Ricotti:2000, McDonald:2003}. Although uncertain, such 
measurements allowed to conclude that the IGM temperature at the mean density, 
$T_0$, at $z \approx 3$ was too high to be consistent with the heat input 
produced by hydrogen reionization alone. The most popular solution to this 
problem involved extra-heating due to \HeII\ reionization occurring around 
$z=3$ \citep{Hui:2003}, an idea also supported by the tentative (and debated) 
detection of a bump in the $T_0$ evolution located at that epoch. 

The situation has become now clearer with the renewed interest in the IGM 
thermal history sparkled by the new observations by \citet{Becker:2011} 
\citep[see also][]{Rudie:2012}. The key advance has been to move away 
from the uncertain determination of $T$ at $\rho = \bar \rho$, and measure 
it at a critical density $\rho_*$ at which the Ly$\alpha$ Gunn-Peterson 
optical depth $\tau \approx 1$. In this regime, the Ly$\alpha$ forest lines 
are most sensitive to temperature. The 
difficulty
is that to transform 
$T_*\equiv T(\rho_*)$ into the usually quoted $T_0$ one needs to know 
the adiabatic index, $\gamma$, entering the equation of state 
$T(\rho)=T_0(\rho/\bar \rho)^{\gamma-1}$. These data, complemented by 
more recent ones by \citet{Boera:2014}, have allowed to put together a 
high-quality sample extending in the range $1.6 < z < 4.8$.  
The new data confirm the peak at $z \!\approx$3.1, where $T_0\simeq$2$\times10^4$K. Such feature is most straightforwardly interpreted 
with the extra heating provided by \HeII\ reionization, although models involving 
intergalactic absorption of TeV blazars \citep{Broderick:2012, Puchwein:2012} 
or heating from cosmic rays \citep{Lacki:2015} have been suggested as alternative 
explanations. Note that in the absence of \HeII\ reionization heating, hydrogen 
reionization would have left the IGM at a much lower temperature ($\approx 5000$K) 
at $z=3$. 

The above data can also constrain the temperature of the IGM in the EoR 
\citep{Furlanetto:2009, Bolton:2012, Lidz:2014}. The models must 
be anchored to the highest $z$ \cite{Becker:2011} datapoints, which imply 
that at $z=4.8$ the IGM temperature was $(6.5 \pm 1.5) \times 10^3$K. Then 
one varies the value of $z_i$ and $T_i$ and selects models that predict values 
within the error bars. As already mentioned, UV background spectral index variation 
have only minor effects on the thermal evolution.  One can also explore slightly 
more sophisticated models in which different parcels of gas are heated at 
different temperature in a given reionization redshift span ($\Delta z_i, 
\Delta T_i$), and then average over the results. Additional models 
\cite{Bolton:2014} calibrate their predictions on numerical simulations 
to derive the evolution of $\gamma$.

In brief, due to the rapid cooling imposed by adiabatic expansion which 
forces the temperature to set onto an asymptotic value, a large degeneracy 
exists among many reionization models with different ($\Delta z_i, 
\Delta T_i$). However, some extreme models in which hydrogen reionization 
is either (i) very short $\Delta z_i \le 3$ and ends at $z=6$, or (ii) 
produces too high ($> 25000$K) temperatures in the ionized gas are excluded 
as the gas cannot timely cool to the measured temperatures at $z=4.8$. It 
appears that the temperature value which is consistent with the largest 
number of thermal histories, also including those in which reionization 
can start earlier than $z=9$, is $T=2\times 10^4$K.   

\subsubsection{Sunyaev-Zeldovich contributions and imprints}
\label{sec:sz}

Free electrons in the IGM resulting from reionization produce CMB
temperature anisotropies via the Sunyaev-Zeldovich (SZ) effect. The 
two main contributions to this effect are the anisotropies
produced by the thermal motion of electrons, known as thermal SZ (TSZ) \cite{Sunyaev:1972}
and those produced by their peculiar motion, termed kinematic SZ (KSZ) \cite{Sunyaev:1980}. 
The CMB temperature anisotropies generated by the ionized
gas in the direction $\hat{n}$ are
\begin{equation}
\frac{\Delta T_{\rm SZ}(\hat{n}) }{T_{\rm CMB}}=\int \left[G(x)\frac{k_BT_e}{m_ec^2}
-\frac{\vec{v}_e\cdot\hat{n}}{c}\right]\frac{d\tau_e}{dz} dz
\label{eq:sz}
\end{equation}
where $m_e$ $T_e$, $\hat{v}_e$ are the electron 
mass, temperature and peculiar velocity. The TSZ has a characteristic
frequency dependence $G(x)$=$x{\rm coth}(x/2)-4$ with 
$x$=$h\nu/k_BT_{CMB}$; ignoring relativistic corrections $G(x)\simeq -2$ below $\sim 217$ GHz, vanishes at 217 GHz and goes positive at higher $\nu$. 

At the physical conditions
expected to hold during reionization, the temperature would be 
$T_e \le 10^4$K and the KSZ effect would be about
three orders of magnitude larger than the TSZ contribution. Nevertheless,
the TSZ effect offers a direct probe to the physical conditions of the ionized gas. 
Cross-correlation of CMB temperature anisotropies with CIB fluctuations
could provide a direct measurement of the temperature of the IGM during reionization
\cite{Atrio-Barandela:2014}: the subdominant TSZ component can be isolated in the presence of multi-frequency CMB maps, when frequency differencing remove the primary CMB and any KSZ components. The potentially measurable TSZ component carries information on the condition of the IGM at the pre-reionization epochs being proportional to the product of  $\tau_e$ and $T_e$ integrated along the line-of-sight.

In the KSZ effect we can distinguish ``homogeneous''
linear \cite{Vishniac:1987} and non-linear \cite{Hu:2000} contributions 
due to the peculiar motion of baryons in a completely ionized IGM and the ``patchy''
anisotropies generated by peculiar motions when the ionization fraction varies 
in space \cite{Aghanim:1996,Gruzinov:1998,Knox:1998,Mesinger:2012}. 
As the first stars and BHs
start producing UV photons, they generate ionization spheres around them.
Before those spheres merge, the Universe would be ionized in patches
generating KSZ anisotropies of 
$(\Delta T/T_{\rm CMB})\sim \tau_e (v_{\rm rms}/c)
\theta (1+z_{\rm reion})^{3/4}(\Delta z_{\rm reion})^{1/2}$, where $v_{\rm rms}$ 
is the rms 
peculiar velocity, $\theta$ the angular scale subtended by the ionized patches,
and $\Delta z_{\rm reion}$ is the redshift duration of the patchy phase. 
Thus, the patchy component 
can be used to set an upper limit on the duration of reionization \cite{Zahn:2012}. \citet{Munshi:2016} discuss how to separate it from the
homogeneous contribution.

Numerical radiation-hydrodynamical simulations estimate 
an amplitude
$[\ell(\ell+1)C_\ell^{\rm KSZ}/2\pi]|_{\ell=3000}\!\!\!\sim$0.6--2.8$\mu$K$^2$  at $\ell$=3000 
\cite{Battaglia:2013}. Data from {\it Planck},
ACT and SPT-SZ provide consistent constraints;
\citet{George:2015} found  $[\ell(\ell+1)C_\ell^{\rm KSZ}/2\pi]|_{\ell=3000}<
3.3\mu$K$^2$ at the 95\% confidence level and translated this upper 
limit into a constraint on the duration of the period when the 
electron fraction grows from 20\% to 99\%
of $\Delta z<5.4$ also at the 95\% confidence level.
Lower ionization fractions are largely made up of ionized regions
too small to be probed by the SPT data  \cite{George:2015}.
Furthermore, if star formation is suppressed in
low mass dwarf galaxies and minihalos located in ionized
or LW-dissociated halos, then more extended reionization histories
are compatible with upper limits on the KSZ power spectrum \cite{Park:2013}.
A similar constrain was derived from the UV luminosity functions 
of star forming galaxies at $z\sim 6-10$. \citet{Ishigaki:2017}
concluded that the redshift interval where the ionization
fraction grows from 0.1 to 0.99 was $\Delta z=4.1\pm 1.7$. 

\subsubsection{Sub-mm first dust emission}
\label{submm}
Dust grains are a fundamental constituent of the interstellar medium (ISM) of galaxies. A large fraction ($\approx $ 50\%  in the Milky Way) of the heavy elements produced by nucleosynthetic processes in stellar interiors can be locked into these solid particles. Most relevant here, they efficiently absorb optical/ultraviolet (UV) stellar light, by which they are heated, and re-emit this energy as longer (FIR/sub-mm) wavelength radiation that can freely escape from the galaxy. It is then natural to expect a tight relation between the UV ``deficit'' and the IR excess produced by this process.

The presence of dust at high ($z\gsim 6$) redshift implies that conventional dust sources (AGB and evolved stars) are not the dominant contributors. This is because their evolutionary timescales are close to or exceed the Hubble time at that epoch ($\approx 1$ Gyr). Following the original proposal by \citet{Todini:2001}, it is now believed that the first cosmic dust were formed in the ejecta of supernovae ending the evolution of much more fast-evolving massive stars \citep{Hirashita:2002, Nozawa:2007,  Bianchi:2007, Gall:2011}. For similar reasons the standard grain growth acting on grains during their residence time in molecular clouds of contemporary galaxies cannot increase the amount of dust by considerable amounts \citep{Ferrara:2016}. Thus, albeit quasar host galaxies show remarkably high dust masses \citep{Beelen:2006, Michalowski:2010}, in general the dust/gas ratio towards high-$z$ rapidly decreases  \citep{Dunlop:2013} as also witnessed by the observed steepening of early galaxies UV spectra. This does not come as a complete surprise given that the average metallicity of the Universe increases with time. 

\citet{Ferrara:1999} \citep[for a recent calculation see][]{da-Cunha:2013} noticed another important feature of high-$z$ dust. Due to the redshift increase of CMB temperature, $T_{\rm CMB}= 2.725(1+z)$ K, the FIR signal from dust becomes increasingly swamped by the CMB. At $z=6$, for example, $T_{\rm CMB}$=19K; as usually dust temperatures in the \textit{diffuse}   
ISM of galaxies are in the range $20-40$ K, the effect cannot be neglected. Even more dramatic, if not complete, might be the suppression of the signal from dust in dense regions (e.g.\ molecular clouds) where the dust is in thermal equilibrium with the CMB.

The superb sensitivity of the ALMA interferometry has allowed detection of the FIR signal of a handful of Lyman Break Galaxies (LBGs) for which HST rest-frame UV photometry (and hence the UV slope $\propto \lambda^\beta$ determination) by \citet{Capak:2015}, reporting a puzzling deviation of detected LBGs from the more local infrared excess (IRX) vs $\beta$ relation \citep{Meurer:1999}. In practice, these galaxies, although characterized by relatively flat $\beta \approx -1$ values, indicative of non-negligible dust attenuation, show a noticeable FIR deficit, i.e.\ they are relatively ``FIR-dark''.   

Such suggested deficit has been strongly reinforced by an even more recent report by the ASPECS survey \citep{Bouwens:2016}. The authors have performed deep 1.2 mm-continuum observations of the Hubble Ultra Deep Field (HUDF) to probe dust-enshrouded star formation from 330 LBGs spanning the redshift range $z = 2-10$. The striking result is that the expectation from the Meurer IRX-$\beta$ relation at $z=4$ was to detect at least 35 galaxies. Instead, the experiment only provided 6 tentative detections (in the most massive galaxies of the sample). Clearly, redshift evolution either of the dust temperature and/or mass must play a key role. 

An exception to the above scenario is the puzzling case of A1689-zD1 \citep{Watson:2015, Knudsen:2016}, a $z=7.5\pm 0.2$ gravitationally-lensed LBG where the thermal dust emission has been detected by ALMA. The large FIR flux $L_\mathrm{FIR}=(6.2\pm 0.8)\times 10^{10}$ L$_\odot$ indicates considerable dust amounts, consistent with a Milky Way dust-to-gas ratio. A similar result has been obtained by \citep{Laporte:2017}  for the $z\approx 8$ Y-band dropout galaxy, A2744 YD4. The ALMA 1 mm detection can be interpreted to arise from dust thermal emission, with an estimated dust mass of $6\times 10^6 M_\odot$. How this large dust amount formed so quickly is a challenging question for the future. 

Overall, the CIB level produced by the dust components from early times is generally expected to be small \cite{De-Rossi:2017} in comparison to the mean CIB detected at these wavelengths \cite{Puget:1996,Fixsen:1998}. It is unlikely to be detectable in direct measurements, but with enough dust may be isolated in some suitably constructed cross-correlation studies.

\subsubsection{Reconstructing emission history via Lyman tomography}
\label{sec:ly-tomography}

It is important also to isolate the CIB production as a function of redshift. Different cosmogonical models predict different modes of evolution at various high $z$ including the range of epochs that cannot be probed even after the advent of the JWST.   \citet{Kashlinsky:2015,Kashlinsky:2015a} proposed a methodology to reconstruct CIB contributions by $z$ using the Lyman tomography in the presence of two adjacent, non-overlapping filters at wavelengths $\lambda_2>\lambda_1$.  The discussion assumes that there are no emissions below some Lyman-{\it cutoff} wavelength $\lambda_{\rm Ly}$, which corresponds to Ly$\alpha$ at rest 0.1216\mic\ when reprocessing is done by the halo \HI, or Ly-continuum (0.0912\mic) otherwise \citep[][]{Santos:2002}. Such cutoff is fundamentally different from situations such as the Balmer break, where emissions, albeit of different amplitudes, exist on both sides of the wavelength and which gets washed out in the CIB integrations over different $z$.

The projected CIB auto-power 
is related to the underlying $P_{3D}$ of the sources by the relativistic Limber equation \ref{eq:limber_1}.
 {\it The integration range 
stops at} $z_{\rm Ly}(\lambda)$ because at larger redshifts sources emit only longward of the $\lambda_{\rm Ly}$; the integration extending to $z$ specified by the 
long wavelength edge of the filter bandpass.
The cross-power between two bands, $\lambda_2>\lambda_1$, extends only to
$z_{\rm Ly}(\lambda_1)$:
\begin{equation}
\frac{q^2P_{12}}{2\pi} = \int_0^{z_{\rm Ly}(\lambda_1)}
 \frac{dF_{\lambda_1^\prime}}{dz} \frac{dF_{\lambda_2^\prime}}{dz} \Delta^2(qd_A^{-1}; z) dz
\label{eq:cross-power}
\end{equation}
At $\lambda_2>\lambda_1$, we can write eq.\ \ref{eq:limber_1} as:
\begin{eqnarray}
\label{eq:limber_2}
\hspace{-1cm}\frac{q^2P_{2}(q,<\!\!z_{\rm Ly}(\lambda_2))}{2\pi} \!\!=\!\!\int_{z_{\rm Ly}(\lambda_1)}^{z_{\rm Ly}(\lambda_2)} 
\left(\frac{dF_{\lambda_2^\prime}}{dz}\right)^2\!\!\Delta^2(qd_A^{-1}; z) dz
 \\
\hspace{-1cm}
+\frac{q^2P_{2}(q, <\!\!z_{\rm Ly}(\lambda_1))}{2\pi}\!\!=\!\!\frac{q^2}{2\pi}\left[P_{\Delta z} + \frac{1}{{\cal C}_{12}(<\!\!z_{\rm Ly}(\lambda_1))}\frac{P_{12}^2}{P_1}\right]\nonumber
\end{eqnarray}
$P_{\Delta z}$ above probes emissions spanning $\Delta z$ at $z_{\rm Ly}(\lambda_1)<z<z_{\rm Ly}(\lambda_2)$ and arises from populations inaccessible to $\lambda_1$, but present at $\lambda_2$. Here $P_1,P_2$ are auto-power spectra at the adjacent bands $\lambda_1,\lambda_2$ with coherence ${\cal C}_{12}=\frac{P_{12}^2}{P_1P_2}$.

One would like to isolate the power, $P_{\Delta z}$, arising from luminous sources between
 $z_{\rm Ly}(\lambda_1)$ and $z_{\rm Ly}(\lambda_2)$. 
Rewriting (\ref{eq:limber_2}) leads to CIB fluctuation generated over $z_{\rm Ly}(\lambda_1)<z<z_{\rm Ly}(\lambda_2)$ as follows:
\begin{equation}
\frac{q^2P_{\Delta z}(q)}{2\pi}= \left[\frac{q^2}{2\pi}(P_2-\frac{P_{12}^2}{P_1})\right]_{\rm data} + \frac{q^2}{2\pi}P_{\rm sys}
\label{eq:lyman_tomography_p_df}
\end{equation}
where the first rhs term is fully given by the data and the last term is driven by incoherence of the sources at the two adjacent bands which occupy the {\it same} span of redshifts $z<z_{\rm Ly}(\lambda_1)$:
\begin{equation}
\frac{q^2}{2\pi}P_{\rm sys}\! =\! \left[\frac{{\cal C}_{12}(q, z<z_{\rm Ly}(\lambda_1))-1}{{\cal C}_{12}(q,z<z_{\rm Ly}(\lambda_1))}\right]\! \times\! \left[\frac{q^2}{2\pi}\frac{P_{12}^2}{P_1}\right]_{\rm data}\leq0
\label{eq:systematic}
\end{equation}
Subscript ``data" refers to directly measurable quantities. $P_{\rm sys}\leq 0$ because ${\cal C}\leq 1$ and the measurable quantity $(P_2-\frac{P_{12}^2}{P_1})$ sets a strict {\it upper} limit on the CIB fluctuations arising at $z_{\rm Ly}(\lambda_1)<z<z_{\rm Ly}(\lambda_2)$. 

\subsection{New diffuse sources at intermediate and low $z$}
\label{sec:ihl-theory}

The possibility of non-negligible CIB fluctuations arising at low to intermediate $z$ from a ``missing light"  associated with galaxy populations but distributed in diffuse structures around masked sources has been proposed by \citet{Cooray:2012}. This ``missing light" is termed the intrahalo light (IHL) which would permeate the Universe. IHL is to be distinguished from ``intracluster light" \citep[ICL;][]{Mihos:2005,Mihos:2016,Lin:2004a}, associated with clusters of galaxies, which in turn are removed/isolated in CIB fluctuations studies; much of the ICL
is further linked to extended halos of brightest cluster galaxies. The mean luminosity of an IHL contributing halo of mass $M$ is assumed to be modeled at rest $\lambda$ as $l_\lambda(M,z)=f_{\rm IHL}(M)[  F_\lambda L_{2.2\mic}(M)](1+z)^\alpha$ with $F_\lambda$ being the SED of the IHL component, normalized to unity at 2.2 \mic\ and assumed to be the same as that of old red stellar populations of elliptical galaxies. The fraction of the halo light stripped away as IHL is modeled as $f_{\rm IHL}=A_f (M/10^{12}M_\odot)^\beta$ and the free parameters $(\alpha,\beta, A_f)$ are adjusted to fit CIB observations. The parental halo luminosity is normalized per ICL observations of \citet{Lin:2004} to be $L_{2.2\mic}(M)=5.6\times 10^{12}(M/2.7\times10^{14}M_\odot)^{0.7}L_\odot$.
The angular power spectrum of the IHL is then calculated from the 1-halo term associated with the halo assumed to follow the NFW profile \citep[][NFW]{Navarro:1997}, and a 2-halo term reflecting the underlying clustering. The halo number density is derived from the underlying $\Lambda$CDM hierarchy via a \citet{Sheth:2002} variant of the Press-Schechter prescription. 

A very generic prediction of the IHL model is that the CIB excess there is produced by 1) the same types of populations as in known galaxies, that 2) are located at $z\ll 10$ and hence their CIB component is coherent with the diffuse light at visible wavelengths, and 3) have no enhanced BH activity of populations. Because the IHL-producing stellar populations have normal Salpeter-type IMF with emissions dominated by normal stars, to produce the same CIB levels one would need to convert more baryons than from very massive stars radiating close to the maximal efficiency of H-burning. Additionally, as discussed in \citet[][Sec.4.3]{Helgason:2014}  the light-to-mass ratio of the IHL is calibrated based on 
intracluster light (at $2.7 \times 10^{14}M_\odot$), and extrapolated as a power-law down to much lower mass scales. The bulk of the IHL is thus associated with 
low-mass systems so that it requires low-mass systems to host IHL exceeding their own stellar light. This results in IHL comparable to the integrated energy produced 
by the entire galaxy populations.

A possibility also exists of CIB contributions at low $z$ from a new particle decay \cite{Bond:1986,Gong:2015,Kohri:2017}.

\section{Current measurements and datasets}
\label{sec:measurements}

While mean levels of CIB are generally overwhelmed by foregrounds, \citet{Kashlinsky:1996a} noted that the same is not true for fluctuations and have pioneered the use of fluctuations to study CIB, applying it to DIRBE \citep{Kashlinsky:1996,Kashlinsky:2000}. The low angular resolution ($0.7^\circ$) of the DIRBE beam did not allow
removal of many sources and restricted the probing of the net CIB
fluctuation levels at the DIRBE bands from 1.25 to 240 \mic. Further development came with deep 2MASS study of higher angular resolution, but ground-based instrument, where \citet{Kashlinsky:2002,Odenwald:2003} developed studies of {\it source-subtracted} CIB to isolate CIB fluctuations at 1.1, 1.6, and 2.2 \mic\ from galaxies fainter than $m_{\rm AB} \simeq$20--21. The next significant step was made using the IRAC instrument \cite{Fazio:2004a} onboard {\it Spitzer}, where \citet{Kashlinsky:2005a} identified significant source-subtracted CIB fluctuations at 3.6 and 4.5 \mic, after subtracting known sources to deeper levels, which exceed the contribution from remaining known galaxies. This signal was confirmed with numerous follow-up studies. {\it AKARI}-based analysis by \citet{Matsumoto:2011} showed consistency with the IRAC measurements, but also identified significant source-subtracted CIB fluctuations at 2.4 \mic. NICMOS-based study \cite{Thompson:2007,Thompson:2007a} reached well beyond the depth of the 2MASS CIB results, but the shallower CIBER results at 1.1, 1.6 \mic\ \cite{Zemcov:2014} conflict both 2MASS and NICMOS.

\begin{figure}[t]
\includegraphics[width=3.2in]{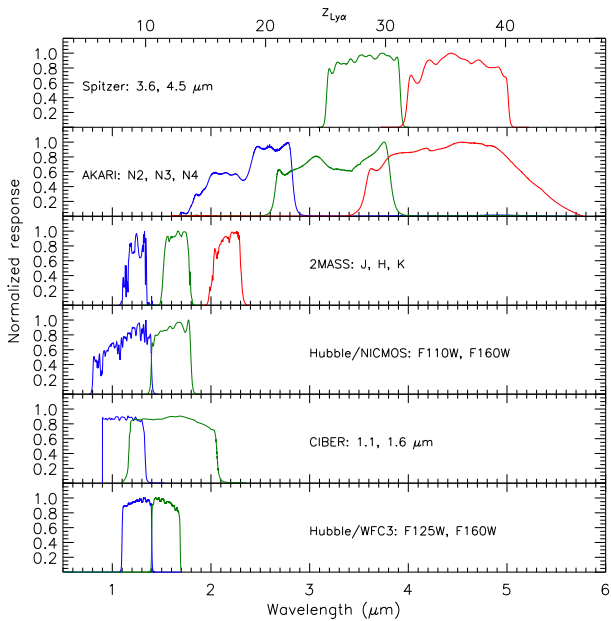}
\caption[]{\small{Filters used in CIB measurements to date. }}
\label{fig:filters}
\end{figure}

The currently available results are discussed below after summarizing the requirements for probing CIB fluctuation component as faint as that expected from early sources. While there is currently overall agreement about the source-subtracted CIB signal identified at 2--5 \mic, the various measurements, discussed below, at 1--2 \um led to currently conflicting and mutually exclusive measurements; hence the division of the discussion in this section.
Fig.\ \ref{fig:filters} shows the wavelength range of the filters employed in the data analyses discussed below.

\subsection{Requirements for probing source-subtracted CIB fluctuations from new populations}
\label{sec:requirements}
CIB science goals are driven by the need to reliably uncover, via CIB fluctuations, populations which cannot be resolved because they are fainter than the confusion limit of the present-day instruments. \citet{Arendt:2010} discuss in detail the exhaustive search for systematics in the analysis of the deep {\it Spitzer}
 data placing particular emphasis on the map-making algorithms and understanding of the instruments. 
There are three main requirements, broken down into detail below:
I) Maps must be prepared that isolate the source-subtracted CIB fluctuations down to 
the (faint) levels such as expected from first stars era; 
II) tools used to analyze the processed (and clipped) data must properly
evaluate random and systematic contributions to large scale fluctuations; and 
III) a robust cosmological interpretation of the results must demonstrate
certain characteristics and rule out others. 

$\bullet$ I. {\it Map assembly}:
{\bf 1}. Maps of diffuse emission should be constructed carefully,
removing artifacts well {\it below} the
expected cosmological signal. In practice, this means that the
maps should not have any structure at levels above $\delta F \sim
0.01$ nW m$^{-2}$ sr$^{-1}$ at 
arcminute scales at the IRAC
wavelengths.
{\bf 2}. No correlations should be introduced when constructing
the maps.
{\bf 3}. In constructing the maps, one should avoid spatial
filters that may remove the very populations which are in the
confusion noise and whose signal is to be identified.
{\bf 4}. Because of temporal variations of the zodiacal light,
data should be collected in as short time intervals as possible.

$\bullet$ II. {\it Analysis tools}:
{\bf 5}. Both the amplitude and the power spectrum of the
instrument noise must be estimated from the data (e.g., time
differenced A--B maps). This is particularly necessary for
shot noise estimates.
{\bf 6}. If sources are removed from the images via modeling 
and subtraction, the source model should accurately account 
for extended low surface brightness emission, i.e.\ the low level wings 
of the PSF and the intrinsic brightness profile of extended galaxies.
The modeling should not be pushed so deep that it 
alters the random noise distribution of the measurements.
{\bf 7}. The effects on the power spectrum caused by masking
sources need to be considered carefully. If the fraction of removed
pixels is small (typically
$\lesssim 30\%$) one can apply FTs;
otherwise the correlation function must be evaluated to demonstrate explicitly that the power spectra recovered
are consistent with the computed correlation functions.
{\bf 8}. The beam must be reconstructed and its large and
small-scale properties understood.

$\bullet$ III. {\it Interpretation}:
{\bf 9}. A true cosmological signal must be demonstrated to be
isotropic on the sky.
{\bf 10}. End-to-end simulations must be done to test that no
artifacts mimic the signal found.
{\bf 11}. Foreground contributions must be evaluated: Galactic ISM
(cirrus) can be extrapolated from locations and wavelengths where
it is brighter; zodiacal emission can be measured via its temporal
changes at different epochs.

\subsection{Measurements at 2--5 \um}
\label{sec:data2-5}

First measurements here were motivated by theoretical suggestions of \citet{Kashlinsky:2004,Cooray:2004} of a measurable CIB fluctuation signal, in certain configurations, that arises from first stars era. While \citet{Cooray:2004} proposed a configuration of wide fields with relatively shallow depth, \citet{Kashlinsky:2004} suggested analyzing deep relatively small regions, where more galaxies can be removed but the angular scales are more limited. 
\citet{Kashlinsky:2005} identified a first suitable dataset from early {\it Spitzer} IRAC observations \citep{Barmby:2004} and laid the ground
for future work by establishing the required machinery and identifying for the first time a source-subtracted CIB fluctuation component.
This component exceeded that from remaining known (``ordinary") galaxies and was proposed to originate in sources from the first stars era.
\subsubsection{Spitzer}
\label{sec:spitzer}
NASA's {\it Spitzer} Space Telescope is a 0.85m diameter infrared telescope launched in 2003 on an Earth-trailing orbit \citep[][]{Werner:2004}. Its CIB results have been obtained with the InfraRed Array Camera (IRAC) \cite{Fazio:2004a}, which covered 4 channels at 3.6, 4.5, 5.8, 8 \mic\ when it operated in the cryogenic regime until mid-2009. After its cryogen was exhausted it continued operating in a warm phase at 3.6 and 4.5 \mic.\\[-10mm]
\paragraph{Self-calibration and map processing} were established as described in \citet{Arendt:2010}. A method for self-calibration intended to make optimal use of 
IR imaging data with minimal (or no) need for separate calibration data
was outlined by \citet{Fixsen:2000}. The procedure essentially calculates
a least-squares fit between the data and a model of the data. The 
data model includes parameters describing the astronomical sky
(e.g.\ the intensity at each pixel in an image of the observed field), 
and various detector parameters (e.g.\ gain factors and offsets for
each pixel of the detector). An example of the use of a more complex 
data model is provided by \citet{Arendt:2002}.
The self-calibration procedure relies on use of an 
observing strategy that allows the determination of the model
parameters for the sky, without degeneracies \citep{Arendt:2000}.
Standard {\it Spitzer}/IRAC observations are designed with this in mind, through 
the use of relatively large scale and highly varied dither 
patterns.

\begin{figure}[t]
\includegraphics[width=3.375in]{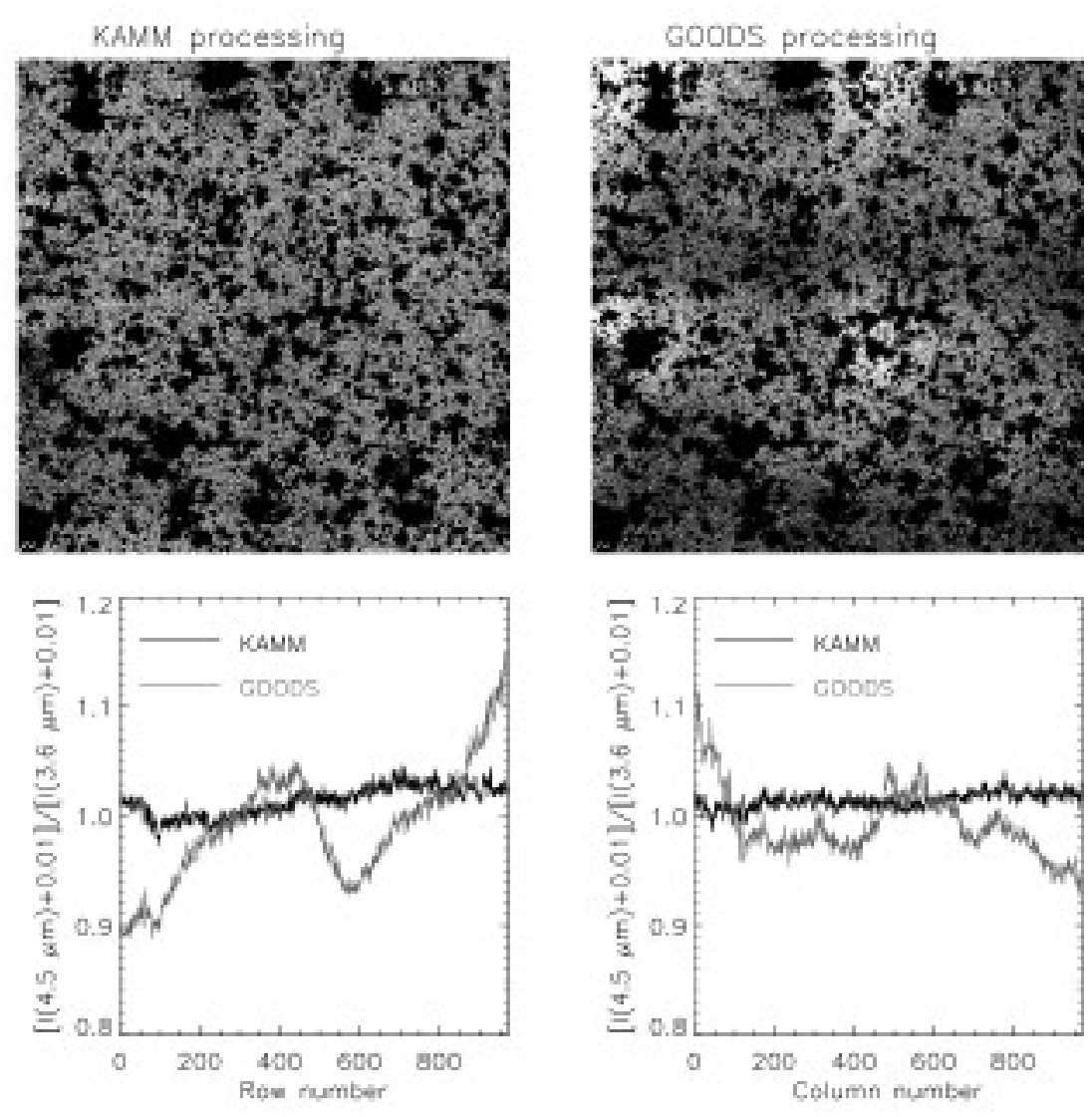}
\caption[]{\small{From \citet{Arendt:2010}. The effectiveness of self-calibration is 
illustrated by maps of the ratio of 4.5 $\mu$m/3.6 $\mu$m 
images of the CDFS field for self-calibrated (left) and 
more conventional (right)\footnote{\tiny\url{http://irsa.ipac.caltech.edu/data/SPITZER/GOODS/docs/goods_dr3.html}} processing; see full discussion in \citet{Arendt:2010}.
Bright sources in the images have been masked identically. 
Any color variations in the intrinsic background should {\it not}
be correlated with the observing mapping pattern. Lower
panels compare median intensities across each ratio image as a
function of row and column with small offsets added to the ratios
so that they are always positive with a mean near 1. The
pattern seen in alternative processing is related to the calibration
of the detector offsets.}}
\label{fig:akmm}
\end{figure}

Self-calibration has proved beneficial for the analysis of IRAC data,
because it can identify and remove instrumental artifacts that are 
not fully corrected in IRAC's standard basic calibrated data (BCD).
The extensive verification of the self-calibration processing
is provided by \citet{Arendt:2010}. 
Figure \ref{fig:akmm} (from that work) illustrates
the improvement that self-calibration can make in the removal of 
large-scale background variations induced by the observing strategy 
and mosaicking procedures.

\paragraph{Results} obtained with {\it Spitzer} IRAC instrument are discussed below in chronological order.

The analysis of early {\it Spitzer} observations of the QSO 1700+6416 field by \citet{Kashlinsky:2005a} established the methodology later used for such CIB studies and provided first indication of significant source-subtracted CIB fluctuations. The QSO1700 field encompassed $5'\times 10'$ ($1\times2$ IRAC FoVs) integrated over $\sim 8$ hrs/pix. After applying self-calibration to assemble maps from individual AORs, the assembled maps at 3.5, 4.5, 5.8 and 8 \mic\ were processed by iteratively 1) clipping off resolved sources and 2) removing outer parts of sources via a suitably modified CLEAN algorithm \cite{Hogbom:1974,Arendt:2010} out to a given level of the remaining shot-noise (and a given factor above the noise power) to produce the diffuse flux maps suitable for CIB study. The clipped pixels were filled with $\delta F=0$ as originally done in \citet{Kashlinsky:2002,Odenwald:2003} so as not to add power per Parseval theorem. 
The clipped fraction of the map was $\lsim 25\%$. The power spectrum of the noise was computed using FFTs from the time-differenced data, $(A-B)$, and the CIB power was evaluated as $P=P_{A+B}-P_{A-B}$. 
The clipping was then allowed to run deeper, removing up to $\sim 75\%$ of the map. Then the {\it correlation function} was evaluated instead 
and was found to remain consistent in amplitude and shape with the CIB power computed from FFTs at the $\lsim 25\%$ clipping as shown in the Supplementary Information of \citet{Kashlinsky:2005a}. 
Foreground emission contributions to the measured power were found to be well below the identified fluctuation except at 8\mic, where the signal was consistent with being dominated by Galactic cirrus emissions. However, the diffuse maps at 3.6 and 4.5 \mic, on the one hand, and 8 \mic\ on the other were found correlated at a weak, but statistically significant level suggesting that the populations contributing to the diffuse power at the former wavelengths are also present at the latter. There was no correlation between the removed sources and the residual diffuse flux maps. The power from remaining known galaxies reproduced well the shot-noise at small angular scales, but was shown to be well below the identified CIB power from clustering at $\gsim 20"$. It was suggested there that the clustering arises in new populations, posited to be at the first stars era.

After a set of new deeper {\it Spitzer} measurements became available through the significantly deeper GOODS observing program at $\simeq 24$ hr/pix \cite{Dickinson:2008}, \citet{Kashlinsky:2007a} analyzed the data in four parts of sky probing source-subtracted CIB fluctuations from maps of 
$10'\times10'$.
Despite the deeper shot-noise levels 
(see Table \ref{tab:spitzer}),
they identified a similar CIB clustering component extending the measurements to larger angular scales. Importantly, that study allowed to probe potential systematics better: the deeper observations were from two distinct epochs separated by 6 months, when each epoch was still sufficiently deep ($\sim 12$ hr/pix) but when the IRAC detectors were rotated by 180$^\circ$ with respect to the previous epoch. Reanalysis used in \citet{Kashlinsky:2007} used finer pixelization (0.6$''$ instead of 1.2$''$) resulting in a larger fraction of the sky, $f_{\rm sky}$, left for the power spectrum computation. If the diffuse fluctuation signal originated from the detectors, it would have been different at the two different orientations contrary to what was observed. Using the new observations at much lower shot-noise \citet{Kashlinsky:2007b} have further refined the high-$z$ interpretation of the CIB signal quantifying the high-$z$ luminosity density and the typical source fluxes in that case.

Using an alternative scheme for map production and analysis, \citet[][also \citet{Chary:2008}]{Cooray:2007} claimed that the power spectrum drops with additional masking from shorter band data and the signal originates in faint blue galaxies located at the peak of star formation and not associated with first stars era. However, their maps had only $20-30\%$ of the sky remain for FT and \citet{Kashlinsky:2007c} showed that - {\it in their data} - when the correlation function is computed for such heavily masked maps instead, the signal remains the same within the statistical uncertainties. After adopting the self-calibration scheme of \citet{Arendt:2010} that claim appears to have been abandoned \cite{Cooray:2012}. \citet{Kashlinsky:2007} found no correlations between the source-subtracted CIB fluctuations from clustering identified in {\it Spitzer} data and very faint galaxies found in visible with HST/ACS out to 0.9 \mic\ and $m_{\rm AB}\gsim 28$.

The above analyses were all done for the {\it Spitzer} cryogenic mission, which ended in 2009 after the telescope's supply of cryogen was exhausted. In the warm {\it Spitzer} mission only IRAC's channels at 3.6 and 4.5 \mic\ remained operational. During the warm mission observation, the SEDS observing program 
\citep{Ashby:2013} 
supplied new data that, while at depth of 12-13 hrs/pix were intermediate between the QSO1700 and GOODS observations, covered substantially larger areas of the sky. \citet{Kashlinsky:2012} have processed the suitably covered areas extending, for the time, the measurements to $\sim 1^\circ$ where the signal remained consistent with a high-$z$ population of sources as posited in the original analysis of \citet{Kashlinsky:2005a}. It was shown that, correcting a high-$z$ $\Lambda$CDM power template for masking effects iteratively, the measured power is well reproduced by a population clustered in that manner.

\begin{figure}[t]
\includegraphics[width=3.5in, trim=1.25cm 0 0 0]{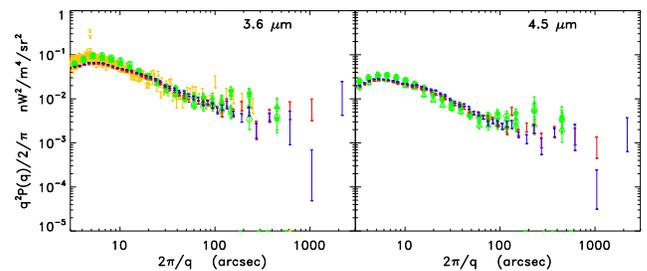}
\caption{\small From \citet[][Fig. 11]{Kashlinsky:2012} showing {\it Spitzer}-based CIB fluctuation data from 7 different fields at similar shot-noise levels.
}
\label{fig:cib_spitzer}
\end{figure}
Fig.\ \ref{fig:cib_spitzer} shows the CIB fluctuation data from seven different regions in the sky analyzed at the shot-noise level of the QSO1700 field. At similar shot-noise levels the signal appears isotropically distributed on the sky, consistent with its cosmological origin. 

\begin{figure}[t]
\includegraphics[width=2.5in, trim=1cm 0 0 0]{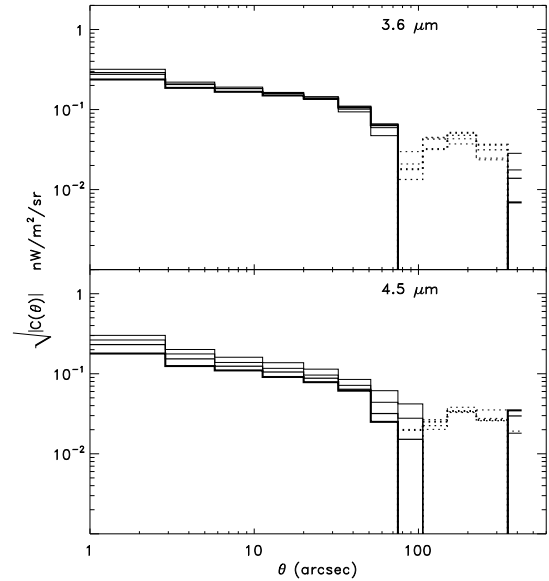}
\caption{\small Adapted from \citet{Kashlinsky:2005a}. Correlation function derived by \citet{Kashlinsky:2005a} for deep clipping of the {\it Spitzer} data out to $f_{\rm sky}=30\%$ of the pixels remaining for analysis. Deeper clipping is specified by lower value of the cutting parameter, $N_{\rm cut}$, while masking area around each pixel is increased by increased value of $N_{\rm mask}$. Solid lines show positive $C(\theta)$, dotted correspond to $C<0$.
}
\label{fig:cib_spitzer_corrfun}
\end{figure}
The dependence of the {\it Spitzer} first results on the clipping was addressed by \citet{Kashlinsky:2005a}. Fig.\ \ref{fig:cib_spitzer_corrfun} shows the CIB correlation function from that analysis for the various clipping and demonstrates robustness of the measured signal in the presence of more aggressive masking, specified by deeper cutting parameter, $N_{\rm cut}$, and wide mask size for each pixel applied, $N_{\rm mask}$. 

\begin{figure}[t]
\includegraphics[width=3.5in, trim=1cm 0 0 0]{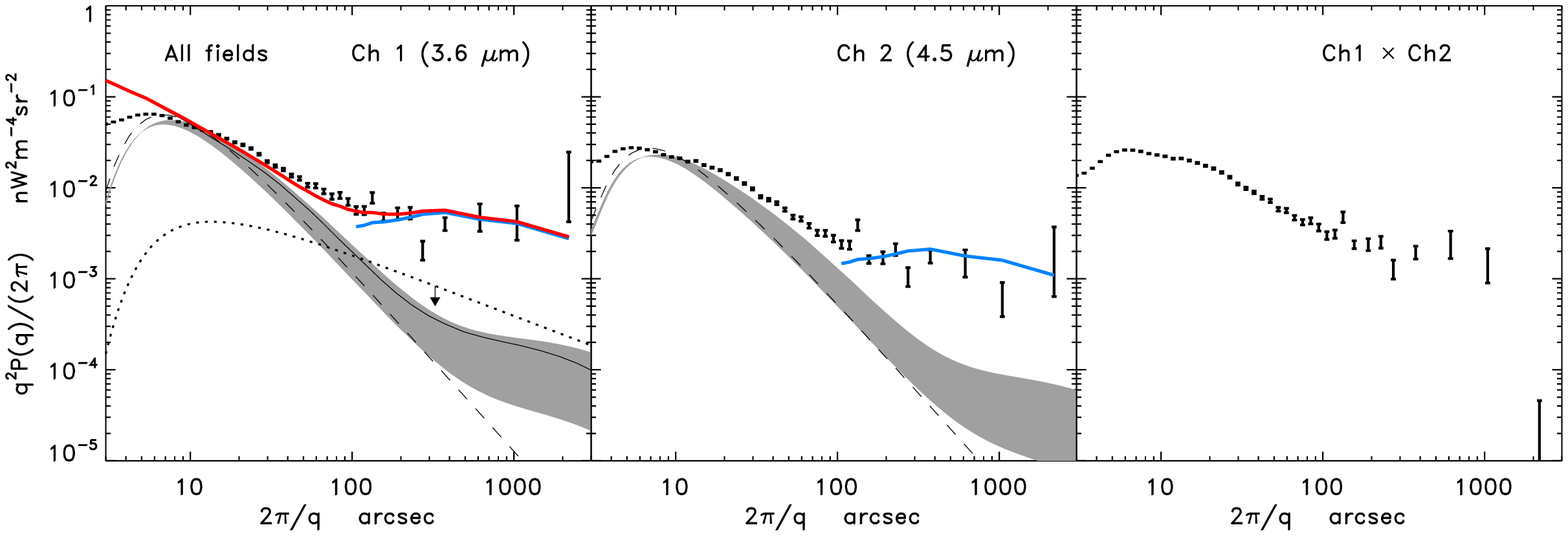} 
\caption[]{ \small{From \citet{Kashlinsky:2012}. Field-averaged CIB fluctuations at 3.6, 4.5 \um\ and the cross-power spectrum. Dotted line shows the upper limit on CIB fluctuations from remaining known galaxies derived in \citet{Kashlinsky:2005a}. 
Black solid line is the contribution of the remaining ordinary galaxies per \citet{Sullivan:2007} which clearly lies beneath the data contrary to the text there. Shaded areas show the reconstructed residual fluctuations from \citet{Helgason:2012a} due to ordinary galaxies. The dashed line shows the shot-noise contribution: $P_{\rm SN} = 57.5$ nJy$\cdot$nW/m$^2$/sr (or $4.8 \times 10^{-11}$ nW$^2$/m$^4$/sr) at 3.6 \um\ and $P_{\rm SN} = 31.5$ nJy$\cdot$nW/m$^2$/sr ($2.2 \times 10^{-11}$ nW$^2$/m$^4$/sr) at 4.5 \mic. Blue solid line corresponds to the high-$z$ $\Lambda$CDM (toy)-model processed through the mask of each field and then averaged. The "toy" model here refers to the tempate of $k^2P_{\rm 3D}$ evaluated at $k=qd_A$ with $d_A\sim 7$ Gpc corresponding to $z\sim 10$ and with the amplitude fitted to the data. Thick solid red line shows the sum of the three components. } } 
\label{fig:cib_final}
\end{figure}

Fig.\ \ref{fig:cib_final} shows the CIB fluctuation, including the cross-power, to $\sim$1$^\circ$ averaged over two SEDS field at the shot-noise levels corresponding to $\simeq 13$ hrs/pix. The cross-power, that could be identified because the region of overlap between the two IRAC channels was large for SEDS observations, appears consistent between the two IRAC channels despite their separate optical paths. The shape of the clustering component to $\sim$1$^\circ$ is consistent with sources distributed according high-$z$ $\Lambda$CDM model.

\citet{Cooray:2012} reproduced the source-subtracted CIB fluctuations from much shallower, but wider Bootes field. The {\it Spitzer} integrations were only 6 min/pix over a net area of 8 deg$^2$ and additional optical data were used to remove sources to sufficiently low shot noise level, but leaving less than 50\% of the assembled map for power spectrum computation. They identified the signal to $\sim 1^\circ$, consistent with an earlier analysis of \citet{Kashlinsky:2012}, and suggested an alternative origin for the fluctuations arising in the IHL from {\it new} populations stripped of their paternal galaxy halo at intermediate redshifts $z\sim 2-3$.

\begin{table}
\caption{Analyzed {\it Spitzer} CIB data. 
}
\begin{tabular} {|p{0.525in} | p{0.575in} | p{0.385in} |  p{0.25in} | p{0.45in} | p{0.65in} | p{0.125in} | p{0.125in}|}
 \hline 
{ \footnotesize (1)} & { \footnotesize (2) } & \footnotesize{(3)} & { \footnotesize (4)} & { \footnotesize (5)} & { \footnotesize (6)} & { \footnotesize (7)} & { \footnotesize (8)} \\
 \hline
{ \footnotesize QSO1700} & { \footnotesize 94.4, 36.1} & {\footnotesize $5\times10$} & {\footnotesize 7.8} & {\footnotesize (70, 40)} & {\footnotesize (24.5,24.5)} & {\footnotesize 77} & { \footnotesize [1]} \\
{ \footnotesize HDFN-E1} & {\footnotesize 125.9, 54.8} & {\footnotesize $10\!\times\!10$} & {\footnotesize 20.9} & {\footnotesize (26, 14)} & {\footnotesize (25.1,25.1)} & {\footnotesize 77} & { \footnotesize [2]} \\
{ \footnotesize HDFN-E2} & { \footnotesize 125.8, 54.8} & {\footnotesize $10\!\times\!10$} & {\footnotesize 20.7} & { \footnotesize (26, 14)} & {\footnotesize ibid} & {\footnotesize 77} & { \footnotesize [2]} \\
{ \footnotesize CDFS-E1} & { \footnotesize 223.7,--54.4} & {\footnotesize $10\!\times\!10$} & {\footnotesize 23.7} & { \footnotesize (26, 14)} & { \footnotesize ibid} & {\footnotesize 76} & { \footnotesize [2]} \\
{ \footnotesize CDFS-E2} & { \footnotesize 223.5,--54.4} & {\footnotesize $10\!\times\!10$} & {\footnotesize 22.4} & { \footnotesize (26, 14)} & { \footnotesize ibid} & {\footnotesize 77} & { \footnotesize [2]} \\
{ \footnotesize EGS} & { \footnotesize 96, 59.8} & {\footnotesize $8\!\times\!62$} & {\footnotesize 12.5} & { \footnotesize (50, 30)} & { \footnotesize (24.75,24.75)} & {\footnotesize 73} & { \footnotesize [3] } \\
{ \footnotesize UDS} & { \footnotesize 170,--59.9} & {\footnotesize $21\!\times\!\!21$} & {\footnotesize 13.6}  & { \footnotesize (50, 30)} & { \footnotesize ibid} & {\footnotesize 73} & { \footnotesize [3]} \\
{ \footnotesize Bootes} & { \footnotesize 57.5, 67.3} & {\footnotesize $3.5^\circ\!\times\!3^\circ$} & {\footnotesize 0.1} & { \footnotesize (80, 100)} & {\footnotesize (24.3,23.8) } & {\footnotesize 47} & { \footnotesize [4]} \\
 \hline
\end{tabular}
\begin{flushleft}
Columns: (1) Name of the field. (2) Galactic coordinates ($l_{\rm Gal},b_{\rm Gal})^\circ$. (3) Size ($'$). (5) $\bar{t}_{\rm exp}$ (hrs) (6) $P_{\rm SN}$ at (3.6, 4.5 \mic) in nJy$\cdot$nW/m$^2$/sr. (6) Limiting $m_{\rm AB}$ at (3.6, 4.5 \mic). (7) Sky fraction, $f_{\rm sky}$ in \%, remaining for $P(q)$ computation. (8) References: [1] \citet{Kashlinsky:2005a}, [2] \citet{Kashlinsky:2007a,Kashlinsky:2007}, [3] \citet{Kashlinsky:2012}, [4] \citet{Cooray:2012a}.
\end{flushleft}
\label{tab:spitzer}
\end{table}

Table \ref{tab:spitzer} sums up all the {\it Spitzer}-base measurements in the various sky configurations discussed above. The signal appears the same in different location and has now been measured to $\sim 1^\circ$. Its origin is now agreed upon to arise in new sources with two competing theories of a high-$z$ origin or IHL at low to intermediate $z$.

\paragraph{Foreground contributions} are important to evaluate:

Zodiacal light is the strongest foreground in terms of total intensity. 
However, the zodiacal light is very smooth on a wide range of angular scales.
Apart from distinct orbital structures: the asteroidal dust bands, 
the earth-resonant ring, and comet dust trails, only upper limits 
have been set on the structure of the zodiacal light at mid-IR wavelengths 
where zodiacal light is brightest \citep{Abraham:1997,Pyo:2012}.
Extrapolating these limits to {\it Spitzer's} near-IR wavelengths 
indicates that spatial fluctuations of the zodiacal light must be 
comparable or less than the observed fluctuations.

More direct and restrictive estimates of the possible contribution of 
zodiacal light to large scale fluctuations has been made by 
examining the power spectra in A-B difference maps, where A and B 
represent observations of the same field collected $\sim6$ or 12 months 
apart \citep{Kashlinsky:2005}. 
These power spectra isolate the contribution the zodiacal light,
because the structure must vary with time, and thus does not cancel out 
as do the Galactic and extragalactic signals. Similarly, the presence of
significant cross correlation between the structure at different epochs, 
indicates that zodiacal light cannot be the dominant signal \citep{Kashlinsky:2012}.

Most recently, \citet{Arendt:2016} examined the 3.6 and 4.5 $\mu$m 
power spectra for a $10'\times10'$ region in the COSMOS field for 
5 epochs. The epochs were chosen to span the widest possible range of 
solar elongation and brightness variation of the zodiacal light. 
They found that the large scale power showed no correlation with the
zodiacal light intensity, but noted that roughly 50\% of the 
white noise (best characterized at the smallest scales) correlates with the zodiacal light intensity, presumably due to the 
photon shot noise of the zodiacal light.

A different foreground to consider is stellar emission from our Galaxy.
The unresolved starlight is a significant contributor to the mean IR
background in low resolution studies, such as {\it COBE}/DIRBE 
\citep{Hauser:1998,Arendt:1998}. However, with higher angular resolution
and sensitivity, resolved sources can be identified and subtracted
from the data at 
a level 
well below the point where
extragalactic sources outnumber Galactic stars \citep[e.g.][]{Ashby:2013}. 

Emission from the Galactic ISM (i.e.\ cirrus) is perhaps the most 
difficult foreground to address. Estimates of the contribution of cirrus 
emission to the fluctuations are generally made by extrapolation measurements
made at other wavelengths and locations where the ISM is more easily detected.
In some fields, there is evident cirrus at 8 $\mu$m where PAH 
features yield relatively bright emission. In these cases, an upper limit 
on the cirrus contribution at shorter wavelengths can be made by assuming that 
the 8 $\mu$m power spectrum is dominated by cirrus, and rescaling the power 
spectrum to shorter wavelength using a spectral energy distribution established
from observations in low-latitude Galactic studies 
\citep{Arendt:2010,Kashlinsky:2005,Kashlinsky:2012}.

\paragraph{Contribution from remaining known galaxies} needs to be robustly estimated in the balance of the CIB. In the original study \citet{Kashlinsky:2005} have already shown, by calculating an {\it upper} limit on the CIB power assuming the power law of galaxy clustering observed on small scales to extent throughout, that known galaxies below the removal threshold do not reproduce the CIB clustering signal. This is shown with the dotted line as an upper limit in Fig.\ \ref{fig:cib_final}. In other words, if populations at lower redshifts and spanning longer cosmic periods with less biasing were to explain the measurement, they would require 
production of much larger CIB, which would be comparable to the net CIB flux at 3.6 and 4.5 \mic\ from all the known galaxies out to $m_{\rm AB} \gtrsim 
26$ \citep{Fazio:2004,Ashby:2013,Ashby:2015}. In their calculations \citet{Sullivan:2007} confirmed this, 
as shown in their Figure, 8, although the text of the paper contradictorily states throughout that the clustering can be produced by normal galaxies at $22.5 < m_{\rm Vega} < 26$.
Their estimate, shown as the solid line in Fig.\ \ref{fig:cib_final}, is clearly below the upper limit worked out earlier in \citet{Kashlinsky:2005}. \citet[][HRK12]{Helgason:2012a} have done a very sophisticated analysis described in Sec.\ \ref{sec:cibfromhrk12} and confirmed this, further lowering the possible contributions from known galaxy populations at $z\lsim 6$. 

\paragraph{Coherence with unresolved CXB}\label{CXO} has been identified by \citet{Cappelluti:2013} in the {\it Chandra}-based cross-correlation analysis of the source-subtracted CIB and CXB maps.
The {\it Chandra X-ray Observatory} is 
sensitive to X-rays in the [0.1-10] keV band
with an energy resolution of 150 eV and a FoV of $16.9'\times16.9'$. 
 The sharp imaging capabilities of $Chandra$ and its highly elliptical orbit allow observations with a very low background. In 
 X-ray observations there are two main background components, one which is purely astrophysical produced by blending  
 of all sources below the detection limit (CXB) and diffuse emission from the Galaxy and the Local Hot Bubble. The 
source-subtracted
CXB flux depends on the observation depth since deeper exposures 
 yield a larger fraction of resolved CXB.
The second component, which we call particle internal background (PIB), arises from charged solar wind or cosmic rays particles interacting with the spacecraft and/or producing secondary X-ray photons by fluorescence.
The local (Galactic) components of the X-ray diffuse emission are dominant at low energy ($E<1.5-2$ keV) while, at higher energies,
the PIB and the extragalactic CXB dominate the signal.

Because of the 
grazing incidence design
of X-ray telescopes (in contrast to optical and IR telescopes), 
regions of the detector close to the optical axis 
have better point-source sensitivity
than outer parts of the field of view: the PSF FWHM varies from $0.5''$ on axis to $>8''$ off-axis. Also the effective area varies with the off-axis angle. As a result, on-axis sources would produce a detection while 
off axis ones would instead contribute to CXB. This means the mean background level is a function of the off-axis angle. These properties are important to consider when studying unresolved CXB fluctuations. 
An advantage of X-ray data, compared to optical/near-IR, is that for every photon the CCD records 
time of arrival, energy, and position. In this way, by sorting the events in time of arrival one can produce
maps of odd (A) and even (B) events to be used later for time-differenced (A-B) evaluation of the noise floor
in power spectra. 

\citet{Cappelluti:2013} used 1.8 Ms data from the AEGIS-XD survey to produce CXB fluctuation maps, $\delta F_X(\vec{x})$, after removing 
X-ray detected sources \citep{Goulding:2012}. 
The AEGIS-XD survey consists of 66 {\em Chandra} ACIS-I pointings. The common area between Chandra and Spitzer 
is a narrow strip of 8$'\times45'$. 
A subsequent analysis of CIB--CXB crosspower by \citet{Mitchell-Wynne:2016} used 4 Ms of {\em Chandra} and Spitzer data in the CDFS area covering $\sim$110 arcmin$^2$.
Both used the approach developed by \citet{Cappelluti:2012} and improved in \citet{Cappelluti:2013} for producing X-ray fluctuation maps:
 resolved point sources like AGN, star-forming galaxies, X-ray binaries and diffuse galaxy clusters are removed as of first step of maps production. X-ray sources have a low source surface density compared to optical/NIR sources. \citet{Lehmer:2012} showed that
 at the depth of {\em Chandra} deep fields, the source density is of the order of $\sim3\times10^4$ sources/deg$^{2}$. This means that masking 
 X-ray sources removes less than 10\% of the pixels. 
 However, when producing X-ray fluctuation maps one must take into account also peculiarities 
of X-ray telescopes, such as position-dependent amplitude and nature of the two components of the background (i.e.\ 
 the CXB and the PIB).
To model the PIB they took advantage of the observations of ACIS-I
in stowed mode. The instrument is exposed but is stowed out of the focal plane and far from the onboard calibration source, when only PIB signal is present. 
Since the mean PIB level in this ``dark frame'' differs from the observation but its spectral shape is constant within 1-2\%, these maps are then scaled to match the actual background level. After masking and subtracting the background they evaluate
the position-dependence of the astrophysical background using an exposure map. 
Another effect to account for is the low pixel occupation number of X-ray photons (e.g.\ in the full
[0.5-7] keV band $\sim$1.1 photons/Ms/pix). This means that the Poisson noise dominates the noise on small scales. 

To measure the X-ray spectrum of the CIB--CXB crosspower, while still minimizing the Poisson noise, 
\citet{Cappelluti:2013} divided the total counts into 3 X-ray bands, [0.5-2] keV, [2-4.5] keV and [4.5-7] keV, each with $\sim$130,000 X-ray photons. 
In a later analysis, \citet{Mitchell-Wynne:2016} used deeper exposures of a much smaller field, collecting about 1/4 of the photons used in the earlier study but with a similar occupation number. 

\citet{Cappelluti:2013} also studied dependence on the masking by subjecting the data to 1) the 
IR mask from \citet{Kashlinsky:2012} and 2) an X-ray mask that specifically removes X-ray groups and cluster down to $\sim10^{13}M_\odot$. CIB power spectra with or without the additional X-ray masking agree to better than 5\% on all scales, consistent with the populations responsible for the CIB fluctuation signal being unrelated to the remaining known galaxy or galaxy cluster populations in the field.

After verifying that the CIB and CXB maps noise were uncorrelated, the cross-correlation analysis
between [0.5-2] keV and [2-4.5] keV and [4.5-7] keV 
was performed with the IRAC source-subtracted CIB maps at 3.6 and 4.5 \mic. 
Their analysis identified the cross-power between 4$''$-1200$''$ and 
evaluated the significance by characterizing the actual dispersion of the cross-power on scales of 10$''$-1000$''$. In \citet{Cappelluti:2013} the overall CXB-CIB cross-power was significant at 3.6$\sigma$ and 5.6$\sigma$ for 3.6 \mic\ 
and 4.5 \mic\ vs. [0.5-2] keV, respectively, while no significant correlation between any IRAC maps and harder X-ray channels was identified. \citet{Mitchell-Wynne:2016} find 3.7$\sigma$ and 4.7$\sigma$ significant cross-powers above 20$''$, respectively. Unlike \citet{Cappelluti:2013} they also find a marginally significant signal in 3.6 \mic\ vs. [2-7] keV (2.7$\sigma$) and 4.5 \mic\ vs. [2-7] keV (3.7$\sigma$), respectively. 

The new study by \citet{Cappelluti:2017a} uses data from the Chandra Deep Field South (CDFS), Hubble Deep Field North (HDFN), EGS/AEGIS field and UDS/SXDF surveys comprising 1,160 Spitzer hours and $\simeq$12 Ms of Chandra data collected over a total area of 0.3 deg$^2$. They show the consistency between the measured cross-powers in each of the regions and, after combining/stacking, report a highly significant detection of a cross-power signal from clustering on large angular scales $>20''$ between the 3.6\mic, 4.5\mic\ and [0.5-2] keV bands. The total significance of the detected clustering component of the cross-power is $\simeq 5\sigma$ and $\gsim6\sigma$, respectively. The level of coherence between the two background fluctuations from clustering is at least ${\cal C} \sim 0.15-0.2$, this being a {\it lower} limit with the CXB power of the new sources being unknown and limited from above observationally. At the same time they find no significant correlation with harder X-ray bands. Accounting for the contribution of known unmasked source population at $z<7$, this excess appears about an order of magnitude at $5\sigma$ level. 

\begin{figure}[t]
 \centering
 \includegraphics[width=3.25in]{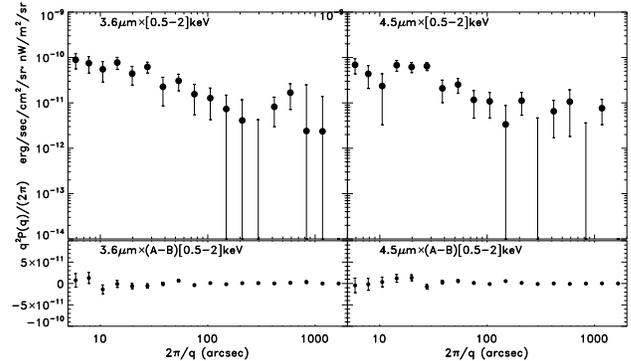}
 \caption{\label{fig:cps} Based on results from \citet{Cappelluti:2017a}.
{\bf Top}: the fluctuations' cross-power spectrum between IRAC 3.6 (left) and 4.5 \mic\ CIB and Chandra soft CXB. {\bf Bottom}: same but for the noise of Chandra time-differenced data.} 
\label{fig:fig1_c17}
\end{figure}

Fig. \ref{fig:cps} presents overall results for the 3.6, 4.5 \mic\ cross-powers with the soft [0.5-2]keV unresolved CXB from \citet{Cappelluti:2017a}.

\paragraph{Spitzer CIB correlations with diffuse light at other wavelengths} were probed in several studies as a tool to provide insights in addition to cross-correlating source-subtracted CIB from {\it Spitzer} with unresolved CXB fluctuations. Of particular relevance to interpretation would be whether the CIB correlates with visible light since any high-$z$ CIB component should not exhibit a visible counterpart because of the Lyman break in the sources around $\simeq 0.12/(1+z)\mic$. Soon after the discovery of the source-subtracted CIB fluctuations with {\it Spitzer}, \citet{Kashlinsky:2007} demonstrated that there are no correlations between the
source-subtracted IRAC maps and the {\it HST}/ACS data to $m_{\rm AB}\simeq28$.
This result implies that the Lyman break wavelength is red-shifted
beyond the longest ACS wavelength at 0.9 \um unless the CIB anisotropies
come from more local but extremely faint ($L<2\times 10^7L_\odot$) and so far unobserved
galaxies.
This likely
requires that the detected CIB fluctuations arise from objects
within the first Gyr of the Universe's evolution 
\citep[cf.][]{Mitchell-Wynne:2015}.

At longer wavelengths, \citet{Kashlinsky:2012} find only marginal correlations between their 3.6, and 4.5 \mic\ source-subtracted diffuse maps with those at 8 \mic, consistent with either cirrus, remaining known galaxies or new populations contributing to diffuse light at both wavelengths. \citet{Matsumoto:2011} find no correlations of the {\it AKARI} source-subtracted diffuse maps with the {\it AKARI} far-IR data at 100 \mic. \citet{Thacker:2015} claimed a cross-correlation of {\it Spitzer} data with diffuse maps from {\it Herschel} at 250, 350 and 500 \mic. The {\it Spitzer} data at 3.6 \mic, 
with a 0.1 hr net integration depth, was
repixelized at the common resolution of 6$''$, leading to both significantly larger removed sky and greater shot-noise (i.e.\ shallower depth) in the combined images. The resultant images had less than 40\% of the map pixels available for Fourier analysis, yet the correlation function has {\it not} been evaluated in the paper to substantiate the robustness of the strong cross-power on sub-degree scales, which was interpreted as coming mostly from IHL. However, a close look shows that the adopted contribution from remaining known galaxies, shown in \citet[][green dashes in Fig. 9]{Thacker:2015} without uncertainties, corresponds to the low-faint-end of the HRK12 reconstruction. Our evaluation of the high-faint-end limit of the HRK12 reconstruction for the appropriate parameters, which is as plausible, increases the CIB power from remaining galaxies by up to an order of magnitude on sub-degree scales. Consequently, the power from remaining known galaxies could be revised upwards by a high enough factor to largely explain the claimed levels of coherence with remaining known galaxies. The uncertainties in the contributions from remaining known galaxies are thus sufficient to account for the claimed coherence at the levels of ${\cal C} \sim (1-3)\%$: the cross-power can be explained if about $\sqrt{{\cal C}}\sim 10\%$ of the sources are common to both the near- and far-IR channels. Galactic cirrus further increases the cross-power, particularly at the largest angular scales probed. The claimed necessity of the IHL in explaining the reported cross-power advanced in \citet{Thacker:2015} thus appears unsubstantiated.
 
\subsubsection{AKARI}
\label{sec:akari}
AKARI is the most recent Japanese IR satellite 
\cite{Murakami:2007}. Its 0.68 m telescope is 20\% smaller {\it Spitzer's},
leading to a lower angular resolution, but the field of view of AKARI's
Infrared Camera \citep[IRC;][]{Onaka:2007}
is $10'$, which is twice as wide as {\it Spitzer's} IRAC.
The IRC obtained images in three near-IR bands at 2.4, 3.2, and 4.1 $\mu$m,
with $1.46''$ pixels. It also included mid-IR bands at 7, 9, 
and 11 $\mu$m with $2.34''$ pixels and 15, 18, and 24 $\mu$m 
with $\sim2.45''$ pixels. 

\citet{Matsumoto:2011} used IRC data from the North Ecliptic Pole (NEP) 
Monitor Field to investigate fluctuations in the CIB in the three near-IR bands.
The NEP Monitor field \citep{Wada:2007} was observed 
regularly throughout the mission. 
Observations were dithered over a single $10'$ field of view, but the field
rotation throughout the mission yields a uniformly covered circular region 
of $10'$ in diameter. The shot noise level after source subtraction and masking 
corresponds to limiting AB magnitudes of 22.9, 23.2, 23.8 at 
wavelengths of 2.4, 3.2, 4.1 \mic, respectively \citep{Matsumoto:2011}.
 
\citet{Matsumoto:2011} describe the procedures used for data reduction, including 
flat fielding, dark subtraction, and corrections for instrumental artifacts.
Source subtraction on the final stacked images was very conservative/aggressive. 
Pixels in the maps exceeding $2\sigma$ were masked, and the procedure was
iterated, until no further pixels exceeded $2\sigma$. To remove lower surface 
brightness portions of sources, the IRAF DAOPHOT package 
was used to find and subtract sources as seen in the unmasked images 
to the $2\sigma$ level, and higher resolution
ground-based $K_s$ band images were convolved with the IRC PSF and used to 
subtract the emission of extended sources. These source-subtracted images 
were then masked with the original $2\sigma$ clipping mask, plus an 
additional margin of 1 pixel ($1.5''$) around all clipped regions. This 
left $\sim47\%$ of the circular field available for analysis.

\begin{figure}[t]
\includegraphics[width=3.5in]{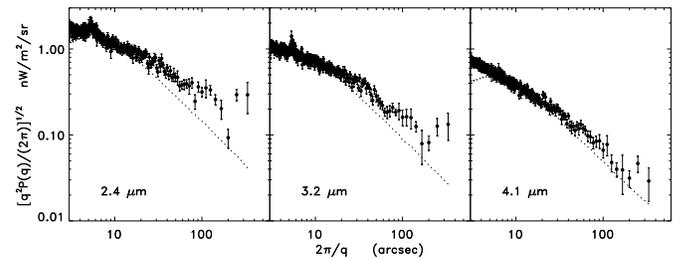}
\caption[]{\footnotesize{Adapted from \citet{Matsumoto:2011}. Power spectra in the AKARI Monitor field after
subtraction of the noise power measured from dark maps \citep{Matsumoto:2011}.
Dotted line shows the fitted shot (white) noise level of remaining
sources that were too faint to subtract/mask. Convolution with the beam 
attenuates the power in this component at the smallest angular scales. 
}}
\label{fig:fig_akari1} 
\end{figure}
Power spectrum analysis, using FFT, of the source-subtracted and clipped images
revealed spatial fluctuations in the data in excess of the power 
shown in dark maps, generated from an equivalent number of concurrent dark 
frames. The dark maps are found to be very similar to time-differenced A--B maps, which 
are generated by inverting the sign of 
half the data, so that any fixed signal from the cancels out and only 
noise (and systematic errors) remain. The power that is measured in 
excess of the A-B noise, appears to be dominated by shot noise (white) 
components at small angular scales, but with a non-white excess increasing 
at large scales ($\gtrsim50''$), especially at 2.4 and 
3.2 $\mu$m (Figure \ref{fig:fig_akari1}). 

\begin{figure}[t]
\includegraphics[width=3.25in]{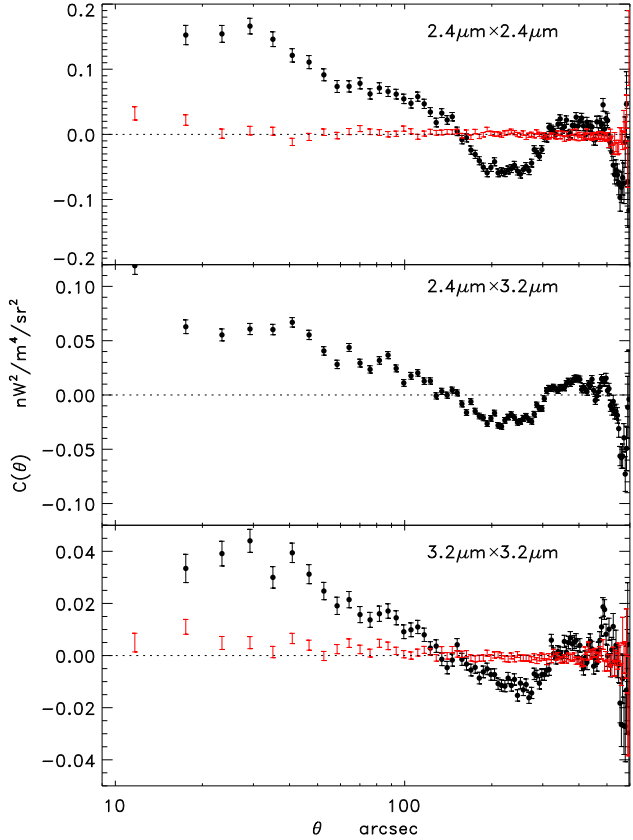}
 \caption[]{\footnotesize{Correlation and cross correlation functions 
for the source-subtracted backgrounds in the AKARI Monitor field 
\citep[adapted with modifications from][]{Matsumoto:2011}. The pixel scale of the maps was 
modified from $1.5''$ to 6$''$ to speed up these calculations, at the expense of 
finer resolution. The errors bars without symbols (red) are derived from the dark maps.
The black symbols are derived from the actual sky maps, and clearly show 
significant large scale structure.
}}
\label{fig:fig_akari2} 
\end{figure}
Because of the large clipping fraction, \citet{Matsumoto:2011} have also verified their FFT-based power results by computing the correlation 
function. 
The correlation function, reproduced in Figure \ref{fig:fig_akari2}, 
also indicates the presence of large scale 
structure, and cross correlations show that the structure is similar in all
three near-IR bands. Importantly, the cross-power evaluated there confirms that the same populations are present at all wavelengths and cluster on similarly large scales. Furthermore, since the shot-noise contribution to $C(\theta)$ 
is contained within $C(0)$,
the correlation function directly isolates the clustering component of the underlying populations.

\begin{figure}[t]
\includegraphics[width=2in]{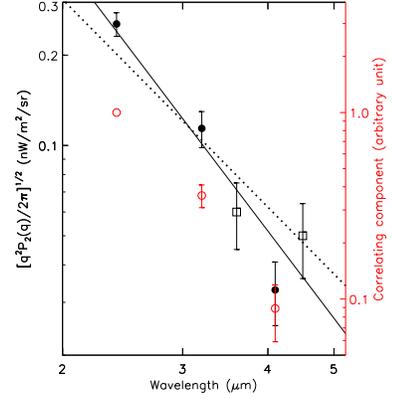}
 \caption[]{\footnotesize{Adapted from \citet{Matsumoto:2011}. The spectral energy distribution of the 
source-subtracted background fluctuations averaged over scales of 
$100''-350''$ from
AKARI measurement (filled circles) is compared to that of {\it Spitzer} (open squares). The open red circles are the 
SED derived from the slopes of pixel-to-pixel correlation of the AKARI 3.2 and 
4.1 $\mu$m data with its 2.4 $\mu$m data (scaled on the red right-hand vertical axis); the 2.4\mic\ open red circle hence has no error bar 
as its correlation is 1.0 by definition.
Solid and dashed lines are 
a Rayleigh-Jeans ($\nu I_{\nu} \sim \lambda^{-3}$) fit to the AKARI data,
and a model of the expected SED of high $z$ Population III sources 
\citep[Figure 20 of][]{Fernandez:2010}.
}}
\label{fig:fig_akari4} 
\end{figure}

The amplitudes of the power spectra at large angular scales are indicative of 
the mean spectral energy distribution (SED) of the sources that produce the 
fluctuations. The SED is found to rise towards shorter wavelength with 
a slope similar to the Rayleigh-Jeans tail of a blackbody 
(Figure \ref{fig:fig_akari4}). This is consistent 
with the rising SED implied by the {\it Spitzer}/IRAC data, but extends the 
trend to shorter wavelengths (2.4 $\mu$m). 

\begin{figure}[t]
\includegraphics[width=2.5in]{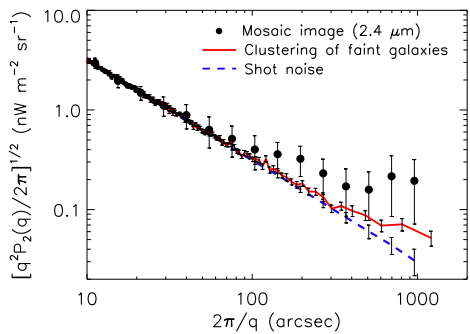}
 \caption[]{\footnotesize{Adapted from \citet{Seo:2015}. The 2.4 $\mu$m power spectrum of the 
source-subtracted AKARI NEP Deep field \citep{Seo:2015}, compared to 
the shot noise level, and the power expected from faint remaining galaxies
as derived from simulations. At scales $>100''$ the measured power exceeds 
that expected from the remaining galaxies that are too faint to 
subtract or mask.
}}
\label{fig:fig_akari_seo} 
\end{figure}

\citet{Seo:2015} have extended the \citet{Matsumoto:2011} analysis to 
larger angular scales by using data from NEP deep survey \citep{Wada:2008}. 
These data 
are not as deep as the Monitor Field, but span angular scales up to $1000''$
(Figure \ref{fig:fig_akari_seo}).
\citet{Seo:2015} use only 2.4 and 3.2 \mic\ data from this data set, 
because the 4.1 \mic\ band had insufficient depth for 
source-subtracted CIB studies. The data reduction used by \citet{Seo:2015}
is similar to that used by \citet{Matsumoto:2011}. However, the analysis 
of the source-subtracted images differs as power spectra are 
corrected for mode coupling due to masking, the map-making transfer function,
and the beam via the same procedure as in \citet{Cooray:2012}. Although the clipping fraction approaches 70\% in the maps, no correlation function has been presented. 

Both \citet{Matsumoto:2011} and \citet{Seo:2015} considered, tested, and
rejected the possibility that zodiacal light or Galactic foregrounds 
can contribute to the reported power at large angular scales ($\gtrsim 100''$).
\citet{Matsumoto:2011} ruled out zodiacal light because of: the lack of signal 
in A--B tests using data collected at different times, the discrepancy 
between the SED of the residual large scale power and 
the zodiacal light, and the extrapolation from limits on the zodiacal light 
fluctuations established in the mid-IR \citep{Pyo:2012}. The extrapolation 
is also used by \citet{Seo:2015} to rule out zodiacal light.
The contribution of faint Galactic stars is dismissed in both papers,
as both resolve sources to depths where source counts are strongly 
dominated by galaxies rather than Galactic stars.
Both papers rule out cirrus contributions, based on lack of 
correlation with far-IR emission at 90 $\mu$m \citep{Matsuura:2011}, which 
does show evidence of cirrus emission at the NEP. In contrast, at mid-IR
wavelengths (7-11 $\mu$m) \citet{Pyo:2012} find that 
the large-scale mid-IR power can be accounted for by rescaling 
the large-scale power at 90 $\mu$m according to a typical cirrus spectrum.
At 15-24 $\mu$m \citet{Pyo:2012} report the photon shot noise as the 
dominant component of the power spectrum, even at large angular scales.

The upshot of the {\it AKARI}-based analysis is 1) consistency with the {\it Spitzer} results at 3.6 and 4.5 \mic, 2) identification of the source-subtracted CIB at 2.4 \mic\ and demonstrating, via cross-correlation, that it arises from the same populations as at the {\it AKARI} longer IRC channels, and 3) identifying the energy spectrum of the sources-subtracted CIB which approximates $\nu I_\nu \propto \lambda^{-3}$.

\citet{Helgason:2016a} suggest possible systematics in the interpretation due to the beam modeling uncertainties of \citet{Matsumoto:2011}: the deduced shot-noise level is sensitive to the beam and with the beam from \citet{Seo:2015} they recover a larger shot-noise power. This decreases the effective limiting magnitude by $\Delta m\sim 0.5$ resulting in larger contributions from remaining known galaxies, which they suggest are enough to explain the bulk (although not all) of the detected CIB {\it power spectrum}. They point out however, that ``the same is not true for {\it Spitzer}/IRAC measurements at similar wavelengths, which still show fluctuations in excess of what can be attributed to faint galaxies" and which are consistent with the {\it AKARI} results as discussed below. While this is indeed an important point, we note that the power they attribute is predominantly shot-noise, is flat and will not contribute to the measured by {\it AKARI} correlation function beyond the beam scale ($\theta \sim 3''$), which is shown in Fig. \ref{fig:fig_akari2} and which they do not attempt to model. The correlation function shows the same populations at all of the {\it AKARI} wavelengths, which are highly coherent and with a distribution distinct from white/shot-noise.

\subsubsection{Currently established CIB fluctuation properties at 2--5 \um}
\label{sec:properties2-5}

We now sum up the properties of the source-subtracted CIB fluctuations that currently appear established in this wavelength range.

\begin{figure}[t]
\includegraphics[width=3.5in, trim= 1cm 0 0 0]{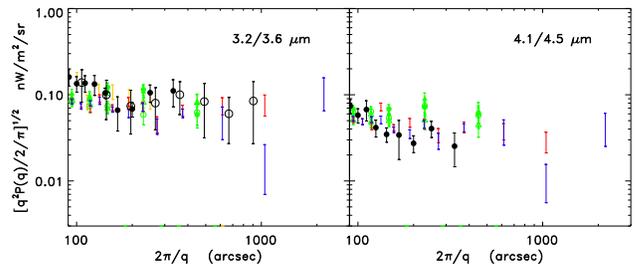}
\caption{\small Comparison of CIB fluctuations from {\it AKARI} with {\it Spitzer}. {\it Spitzer} results are shown in same color scheme as Fig.\ \ref{fig:cib_spitzer}. {\it AKARI} results are scaled to {\it Spitzer} wavelengths with the $\propto \lambda^{-3}$ SED - black filled circles are from \citet{Matsumoto:2011} and black open circles in the left panel are the shallower measurements from \citet{Seo:2015} at 3.2 \mic\ with a higher shot noise levels dominating small scales. All measurements trace the same populations at angular scales $\gsim 30''$, although, because of their shallower depth and smaller area, the {\it AKARI} data are more polluted by the remaining known sources. 
}
\label{fig:cib_akari_vs_spitzer}
\end{figure}

\paragraph{Cosmological origin of fluctuations in new populations:}
Fig.\ \ref{fig:cib_spitzer} illustrates that the CIB fluctuation signal from clustering detected by {\it Spitzer} is consistent with being isotropic on the sky as required by its cosmological origin. Fig.\ \ref{fig:cib_akari_vs_spitzer} shows that the same signal is present in the {\it AKARI} measurements at the adjacent wavelengths.
The fields analyzed using {\it Spitzer} data and shown in Table \ref{tab:spitzer} span a factor of $\sim 3$ in cirrus intensity, yet exhibit a consistently similar large scale component out to $\sim 1^\circ$. Likewise, the signal appears temporarily invariant suggesting small, if any, contribution from zodiacal emission. 
The detected CIB fluctuation thus appears to arise from clustering of new 
extragalactic
populations. The CIB fluctuations contain two components: small scales arise from the shot noise from remaining galaxies and large scales arise from the clustering of contributing sources.

\begin{figure}[t]
\includegraphics[width=3.4in, trim = 2cm 0 0 0]{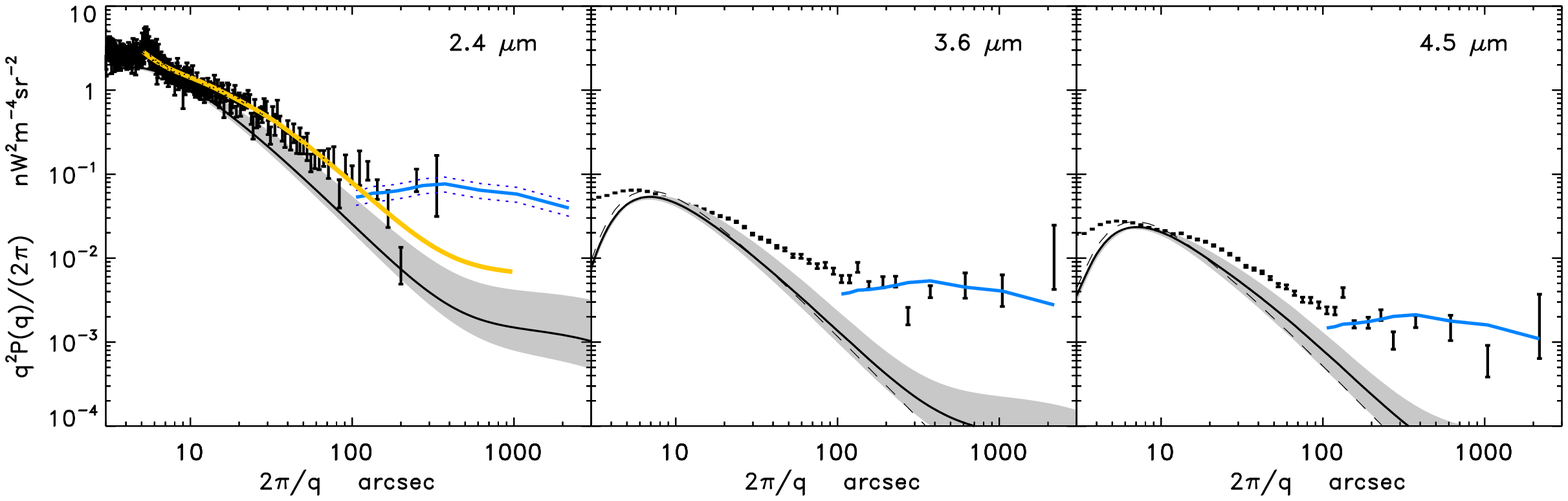}
\caption{\small Adapted from \citet{Kashlinsky:2015a}. Mean squared source-subtracted CIB spatial fluctuations at 2.4, 3.6, 4.5 $\mu$m. Black dashes show the shot-noise component remaining in the IRAC maps. Black solid line shows the ``default'' reconstruction of the CIB from remaining known galaxy populations with uncertainty shown by shaded area from \citet{Helgason:2012a}; the yellow line shows the contribution from \citet{Helgason:2016} at the revised AKARI shot noise. Blue solid line shows the template of the high-$z$ $\Lambda$CDM model; it is extrapolated to the 2.4 \um\ data from the IRAC channels using the $\lambda^{-3}$ energy spectrum with the uncertainty marked with blue dots.  {\bf Left}: AKARI results from \cite{Matsumoto:2011}. {\bf Middle and right} panels show the IRAC-based measurements from \citet{Kashlinsky:2012}. 
}
\label{fig:cib}
\end{figure}

\paragraph{Contribution from remaining known galaxies} appears negligible at large scales from {\it Spitzer} measurements, but the situation may be less clear for {\it AKARI}, given the somewhat small deep field \citep{Matsumoto:2011} which leads to larger statistical uncertainties, or the shallower wider field \citep{Seo:2015} which leaves pollution from remaining known galaxies at larger levels. Nonetheless, the {\it AKARI} data at 2.4 \mic\ exhibit excess at larger scales that are consistent with large-scale fluctuations from {\it Spitzer} at 3.6 and 4.5 \mic, and the correlation functions evaluated there show clear deviations from shot-noise dominating remaining known galaxy contributions at all wavelengths. Fig.\ \ref{fig:cib} sums up the contributions from remaining known galaxies to the measured powers.

\paragraph{Spectral Energy Distribution:}
{\it AKARI}-based analysis extended the CIB fluctuation measurement to 2.4 \mic\ and suggested an approximately Rayleigh-Jeans type spectral energy distribution of the sources producing them, $\nu I_\nu \propto \nu^{-\alpha}$ with $\alpha\sim 3$. Fig.\ \ref{fig:cib} shows the combined {\it AKARI+Spitzer} results with a fit of a high-$z$ $\Lambda$CDM template extrapolated from {\it Spitzer} band to the {\it AKARI} 2.4 \mic\ channel shown in blue; solid line shows the least-squares amplitude derived at 2.4 \mic\ and the blue dotted denote the 1-$\sigma$ error span.

\begin{figure*}
\includegraphics[width=6.5in]{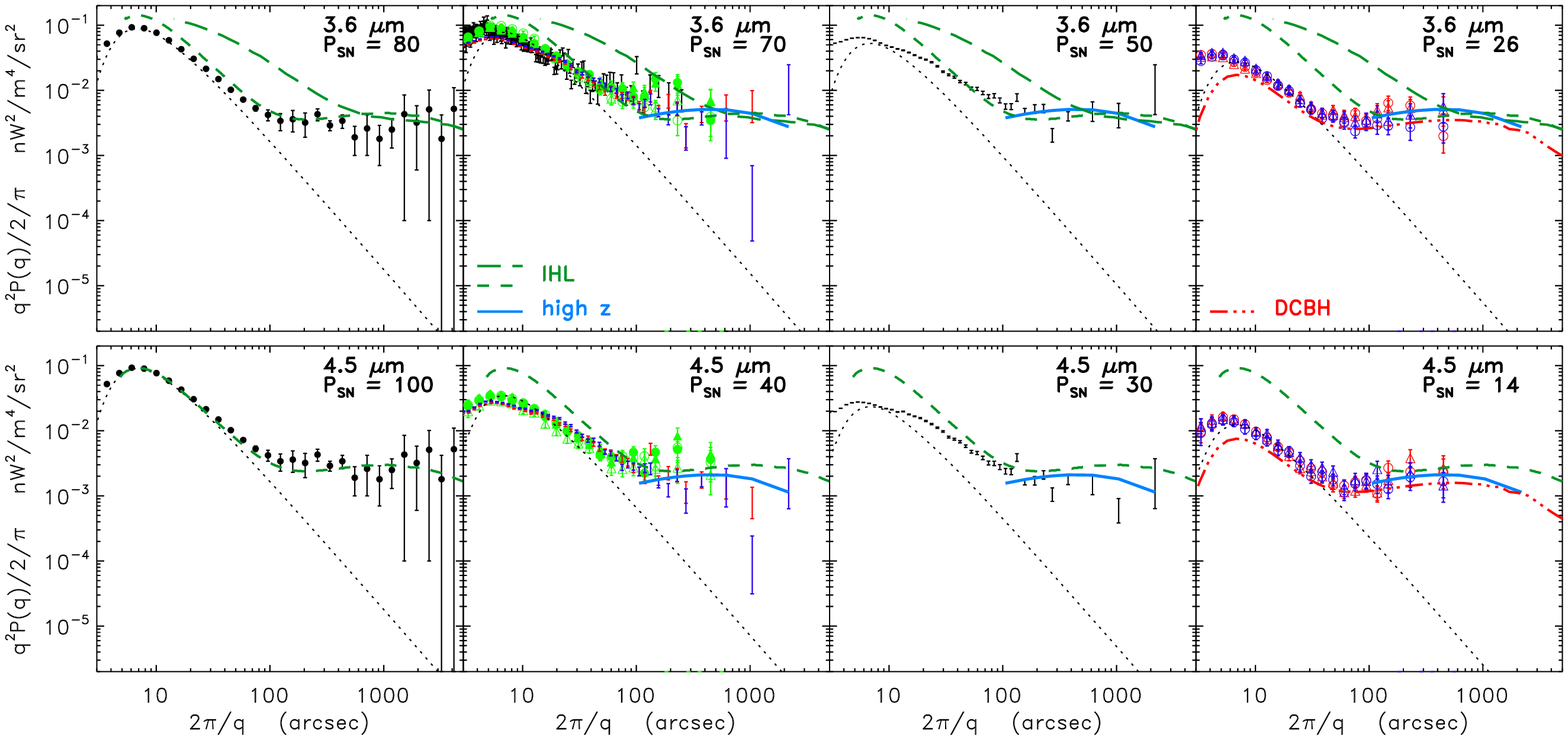}
\caption{\small Adapted from \citet{Kashlinsky:2015a}. Current {\it Spitzer}/IRAC-based measurements at {\it different shot-noise levels} (decreasing in amplitude from left to right) from \citet{Kashlinsky:2005,Kashlinsky:2007,Kashlinsky:2012,Cooray:2012}. The remaining shot noise power, $P_{\rm SN}$, is shown in each panel in units of nJy$\cdot$nW/m$^2$/sr. Upper panels correspond to 3.6\mic, lower to 4.5 \mic. Dotted lines show the remaining shot noise fluctuation in the {\it Spitzer}/IRAC maps convolved with the IRAC beam. {\it No decrease of the large-scale clustering component is yet apparent at the lower shot-nose levels}. This appears to conflict with the currently developed
IHL models, shown in green from \citet{Cooray:2012} (short dashes) and \citet{Zemcov:2014} (long dashes), where the 1-halo component contributes an effective shot noise, which may be related to the large scale amplitude driven by the 2-halo term.
Solid blue line shows a high-$z$ $\Lambda$CDM template, $k^2P_{\rm 3D}(k)$ at $k=qd_A$ with $d_A\sim 7$Gpc, normalized to the CIB fluctuation from {\it Spitzer} and corrected for the mask as described in \citet{Kashlinsky:2012}. The PBH model of \citet{Kashlinsky:2016} naturally produces the required CIB, has sources located at these distances and is effectively represented by the solid blue line. The DCBH model of \citet{Yue:2013} is plotted with red triple-dot-dashed lines. Both BH models appear to match the current data since the shot noise amplitude, being fixed by the abundance of the individual sources and their fluxes, is below the levels reached in these measurements.
}
\label{fig:clustering_sn_irac}
\end{figure*}

\paragraph{Clustering component vs shot-noise power:}
The source-subtracted CIB fluctuations measured with {\it Spitzer} data appear with low shot noise, while exhibiting a substantial clustering component, which indicates the origin of the clustering component in very faint populations (currently $S\lsim$20 nJy at 3.6 and 4.5 \mic). The source-subtracted CIB fluctuations measured from {\it Spitzer} data at progressively lower shot noise levels are shown in Fig. \ref{fig:clustering_sn_irac}. The clustering component does not yet appear to decrease as the shot noise is lowered by a factor of $\sim 6$ in analyses using progressively deeper exposures. This has important cosmological implications for proposed models as summarized in the figure caption and discussed further later.
\paragraph{Coherence of new sources between 3.6 and 4.5\mic:}

\begin{figure}[t]
\includegraphics[width=3in]{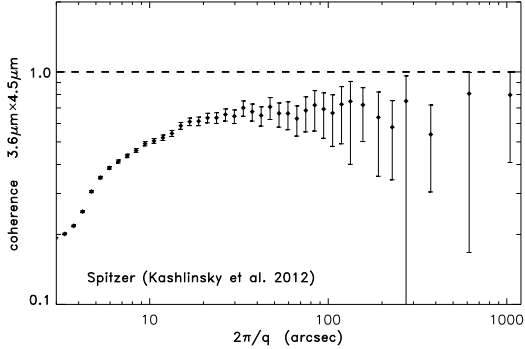}
\caption{\small 
Coherence between in CIB 3.6 and 4.5 $\mu$m. Small angular scales show the incoherent contributions due to differentially removed sources at the two bands. Larger scales are dominated by the coherent CIB from new populations. Adapted from \citet{Kashlinsky:2015a}
}
\label{fig:seds_coherence}
\end{figure}

\begin{figure}[t]
\includegraphics[width=3.5in,trim=1cm 0 0 0]{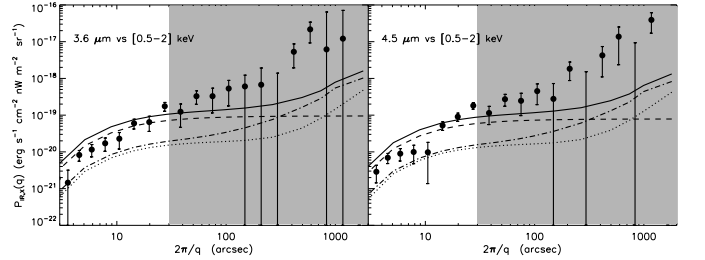}
\caption{\small Cross-power between CIB at 3.6 (left) and 4.5 (right) \mic\ and unresolved soft X-ray CXB at [0.5-2]keV from \citet{Cappelluti:2017a} is shown with filled circles and 1-$\sigma$ errors. Line denote contributions from remaining known sources: shot-noise (dashes) and clustering component (dot-dashed) from galaxies and AGNs, hot gas (dotted) and the total (solid). Shaded regions marks angular scales where clustering component dominates, which is measured to exceed overall for $>20\dasec$ the contributions from known sources at $>5\sigma$ level for both IR bands. See details in \citet{Cappelluti:2017a}.}
\label{fig:cib-cxb}
\end{figure}

Fig.\ \ref{fig:seds_coherence}, derived from the \citet{Kashlinsky:2012} measurements, shows the coherence between the source-subtracted CIB fluctuations from Spitzer at 3.6 and 4.5 \mic. There appears a consistent picture of the CIB measurements obtained with {\it Spitzer} : 1) the coherence is always bounded from above by unity including the errors, which were evaluated using the Fisher transformation, 2) with small scales dominated by the remaining known galaxy populations, which are independently removed at the two bands and so are less coherent than 3) the large scales, where new populations dominate, which cannot be resolved with Spitzer and, hence, were not yet removed.
\paragraph{CIB-CXB cross-power:} appears significant between the source-subtracted CIB in {\it Spitzer} measurements and unresolved soft X-ray CXB as illustrated in 
Fig.\ \ref{fig:cib-cxb} with results from \citet{Cappelluti:2017a}.
If the observed arcminute scale CIB is produced by sources at the epoch of the 
first stars, 
then it arises from 
sources which would have been coeval with or evolve into, or even be, the first 
generation of BHs in the Universe. Since BH accretion 
inevitably produces intense X-ray radiation, one would expect a certain level of coherence between the
fluctuations of the two cosmic backgrounds.
The coherence uncovered in the measurements can also be interpreted as the fraction of the emission due to the common populations so that ${\cal C}_{ij}\approx\zeta_i^2\zeta_j^2$, where $\zeta_i$ and $\zeta_j$ are the fractions of the emissions produced by the common populations between bands $i,j$. \citet{Cappelluti:2013} determined the level of coherence between the source-subtracted CIB and CXB and found it of the order
${\cal C}_{\rm CIB-CXB}\sim 0.05$ at the largest angular scales, so if all the CXB power is produced by sources correlating with the 
CIB then a lower limit on the CIB fluctuations produced 
in association with 
the X-ray sources is 15\%-25\%. We note that the stated coherence represents a {\it lower} limit on the true CIB-CXB coherence of the new sources, since the CXB power they contribute is, while observationally unknown, less than the measured power from the diffuse X-ray maps.
The level of unresolved CXB around 1 keV is $\lsim 1$ keV/cm$^2$/sec/sr \citep[see Table III in][and Fig.\ \ref{fig:fig_guenther1}]{Cappelluti:2017}, corresponding to comoving number density of the X-ray photons at 1 keV of $n_{\rm CXB}\lsim 4\times 10^{-10}$cm$^{-3}$. At the same time the excess CIB of $\sim 1$nW/m$^2$/sr around 3\mic\ requires comoving density of CIB photons at $n_{\rm CIB} \sim 6\times 10^{-4}$cm$^{-3}$. Thus the sources producing the two together should have $n_{\rm CXB}((1+z){\rm keV})/n_{\rm CIB}(3/(1+z)\mic)\lsim 6.6\times 10^{-7}$ 
requiring the X/O ratio (defined as the logarithmic slope from 0.25\mic\ to 2 keV, \citet{Tananbaum:1979}) $\alpha_{\rm X/O}< -2$.

\begin{figure}[t]
\includegraphics[width=2.5in]{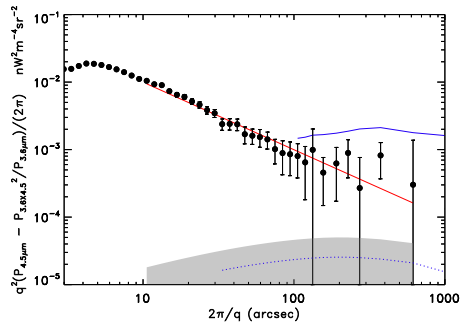}
\caption{\small Adapted from \citet{Kashlinsky:2015a}. The Lyman-break based tomography application to the current {\it Spitzer}/IRAC measurements of \citet{Kashlinsky:2012} at 3.6 and 4.5 \um\ (filled circles). Red solid (straight) line shows the $P\propto q^{-1}$ template that fits the data and is consistent with the non-linear clustering of
known galaxies remaining after differential subtraction at the two bands. Blue line shows the high-$z$ $\Lambda$CDM template that fits the CIB fluctuation data at 4.5 $\mu$m. Filled region is the $1\sigma$ limit on the CIB power remaining for populations at $z>z_{\rm Lyman-break}(4.5\mic) \gtrsim 30$ with blue dotted line shows the central fit. The power left for these populations is $\lsim2\%$ of that measured at 4.5\mic.
}
\label{fig:seds_lybreak_irac}
\end{figure}

\paragraph{Application of Lyman tomography to Spitzer CIB}
was made by \citet{Kashlinsky:2015a} with data analyzed in the IRAC configuration of \citet{Kashlinsky:2012}. As Fig.\ \ref{fig:filters} shows the IRAC filters are adjacent and non-overlapping, presenting a testing ground for the Lyman tomography.
The measured CIB powers at the two IRAC channels, $P_{3.6}, P_{4.5}$, and the cross-power , $P_{3.6\times4.5}$, shown in Fig. \ref{fig:cib_final}, 
were used to construct per eq. \ref{eq:lyman_tomography_p_df} the excess power component that arises where the Lyman-break populations are present at 4.5 \mic, but not at 3.6\um\ ($30\lsim z\lsim 40$ assuming the Ly-break at these pre-reionization epochs due to the Ly$\alpha$ absorption).
The CIB data used consist of two regions of $21'\times21'$ and $8'\times 62'$ of similar integration depth. The regions have full overlap between 3.6 and 4.5\mic.
Fig. \ref{fig:seds_lybreak_irac} shows the resultant $P_{\Delta z}=P_{4.5}-P_{3.6\times4.5}^2/P_{3.6}$ with $1\sigma$ errors. The slope of the fluctuations is close to that of non-linear galaxy clustering produced by differentially removed
sources at the two IRAC bands.
\citet{Kashlinsky:2015a} decompose the data shown in the figure into 1) shot-noise, 2) non-linear clustering from remaining differentially removed galaxies at the two
IRAC bands, assumed to follow $P\propto q^{-1}$, consistent with the 2MASS CIB measurements, and 3) high-$z$ $\Lambda$CDM and evaluate the amplitudes of each component. 
The red solid line in the figure shows the resultant fit from the non-linear clustering component. In the presence of the empirically determined remaining galaxy component, the amplitude of clustering component with the concordance $\Lambda$CDM power template at $z\simeq 30$ is shown at its $1\sigma$ upper limit.
The resultant high-$z$ component is shown in Fig. \ref{fig:seds_lybreak_irac}. Its fitted amplitude implies the contribution to the power measured at 4.5 \mic in the {\it Spitzer} data by \citet{Kashlinsky:2012} to be at most 2\% from $z\gtrsim 30$, setting the best upper limits available to date on emissions
from these epochs. 
\subsection{Measurements at 1--2 \um}
\label{sec:data1-2}

There is significantly less agreement between the various measurements at this wavelength range, and their interpretation is therefore subject to what dataset is assumed to represent reality. The measurements have been done in the following chronological order: 1) CIB analysis by \citet{Kashlinsky:2002,Odenwald:2003} using deep 2MASS data from the ground \cite{Nikolaev:2000} at 1.1, 1.6 and 2.2 \mic, 2) much deeper CIB analysis \cite{Thompson:2007,Thompson:2007a} using space-based NICMOS/{\it HST} data at 1.1 and 1.6 \mic\ over a smaller region \cite{Thompson:2005}, 3) the shallowest of the analysis over a larger area of the sky \cite{Zemcov:2014} using partially overlapping filters center at 1.1 and 1.6 \mic\ with a stratospheric CIBER measurement \cite{Bock:2013}, and 4) a deep analysis using WFC3/{\it HST} data by \citet{Mitchell-Wynne:2015}. 
The lower 4 panels in Fig.\ \ref{fig:filters} show the filters employed in obtaining the results discussed in this section in chronological order.

\subsubsection{Deep 2MASS}
\label{sec:2mass}

\begin{figure}[t]
\includegraphics[width=2.5in, trim= 1cm 0 0 0]{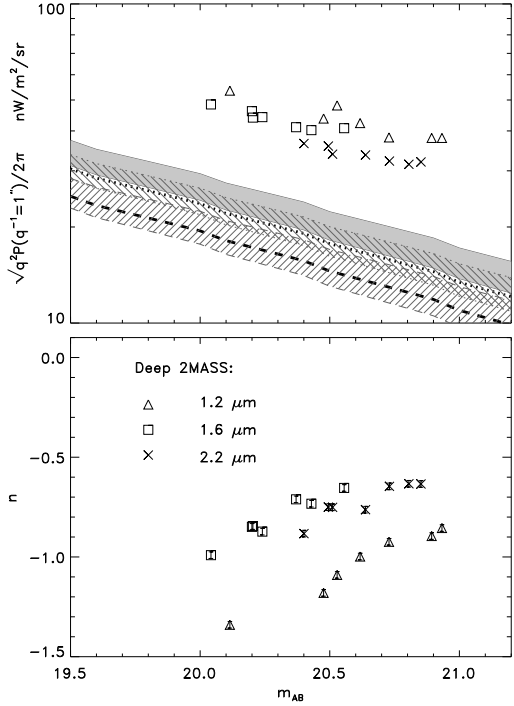}\\
\caption{\small CIB fluctuations from 2MASS \citep[see data in Fig. 2 of ][]{Kashlinsky:2002}. Shaded regions are shot-noise contributions from HRK12 reconstruction from HFE to LFE with thick central line for default model (solid - J, dotted - H, dashed - K).
}
\label{fig:cib_2mass}
\end{figure}

The 2MASS standard star survey \cite{Nikolaev:2000} was used by \citet{Kashlinsky:2002,Odenwald:2003} to develop the required methodology and probe for the first time source-subtracted CIB fluctuations. The analysis was done after the assembled field of $\simeq 8.6'\times 1^\circ$ was divided, in order to eliminate artifacts, into seven square patches of 
$512''\times 512''$ probing CIB in each patch out to angular scales $2\pi/q\sim 200''$. The resolution was limited by atmospheric seeing at about $2''$. Galaxies have
been identified and removed down to Vega magnitude of $\sim 18.7-20$ (AB magnitudes $\sim 20-21$) in the J, H, K$_s$ photometric bands, with each of the patches clipped to its individual depth. As discussed in \citet{Kashlinsky:2002} this leaves CIB from galaxies at $z\gsim 0.6-1$ depending on magnitude and band. The sky fraction 
removed with the resolved sources was less than 10\% allowing robust CIB FT analysis. After analyzing contributions from atmospheric glow and other foregrounds, CIB 
fluctuations were claimed with the clearly non-white-noise spatial spectrum produced by (evolving) non-linear clustering from remaining galaxies with $P\propto q^{-n}$ and
the slope varying between $n$=1.4 for the shallowest removal and $n$=0.6 for the deepest; for reference the present-day non-linear clustering has $n\!\sim$1.3. 
Fig.\ \ref{fig:cib_2mass} shows the amplitude of the resultant source-subtracted CIB fluctuations at the fiducial scale $q^{-1}=1''$ and the effective deduced slope, $n$, 
in the seven 2MASS CIB patches with sources remaining below the flux corresponding to the AB magnitude shown in the horizontal axis.

While it is reported to ``identify the signal as CIB fluctuations from the faint unresolved galaxies", there may be possible systematical biases affecting this analysis which may stem from the required in the data destriping corrections, adopted to cover a narrow width of pixels in the Fourier plane, and which in turn affect the conversion of the remaining $\sigma$'s of the maps to effective magnitudes as discussed in \citet{Odenwald:2003}. In addition, the ground-based observations are significantly affected by the variability of OH-glow. In any event, this study probes the remaining CIB at too shallow a depth (by today's standards) 
to be useful in probing high-$z$ emissions.

\subsubsection{NICMOS/HST}
\label{sec:nicmos}

\begin{figure}[t]
\includegraphics[width=2.5in, trim= 1cm 0 0 0]{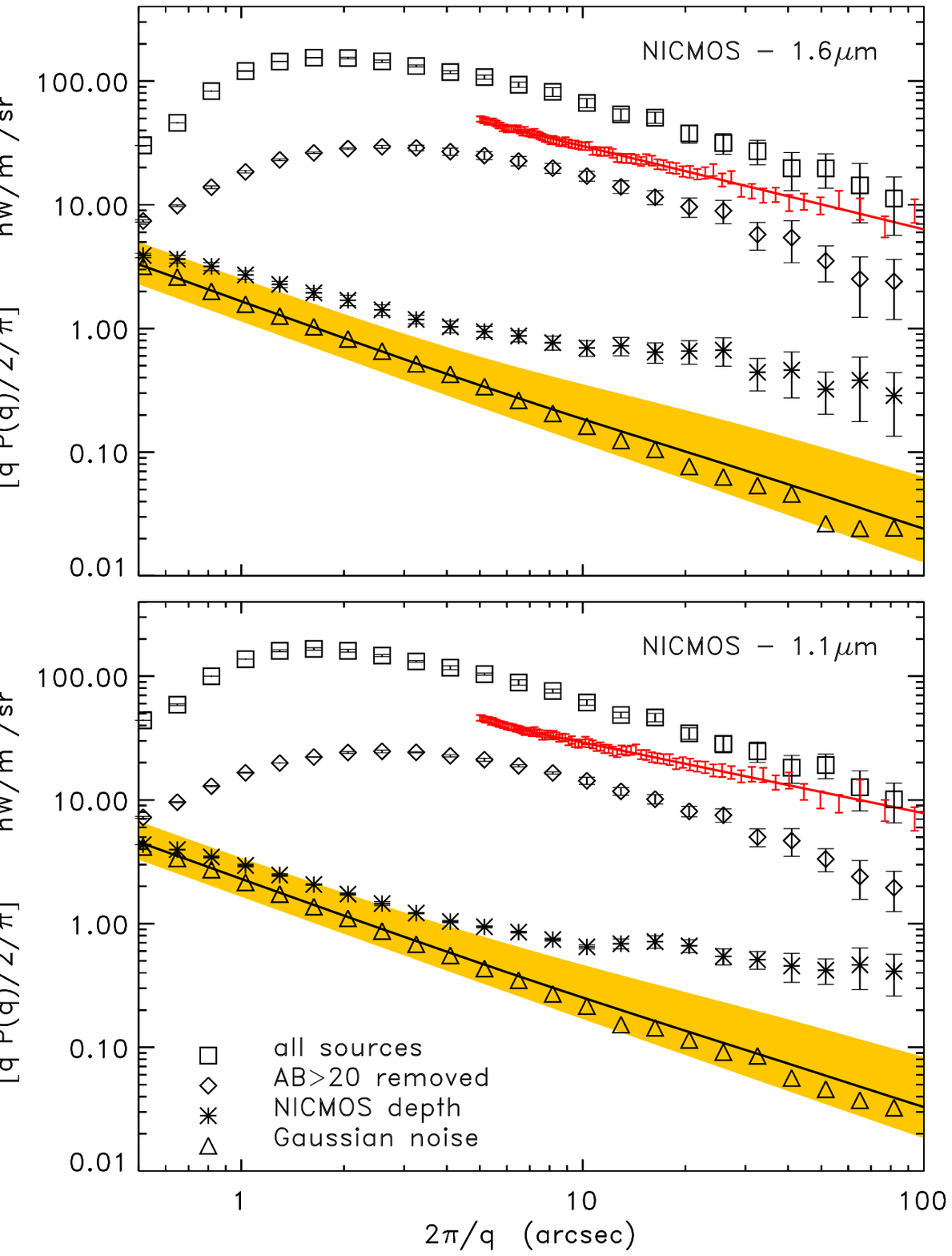}\\
\caption{\small Adapted with modifications from \citet{Thompson:2007,Thompson:2007a}. RMS fluctuations from the auto-power from the NICMOS analysis \citep{Thompson:2007} at different magnitude removal thresholds. Squares mark no removal, diamonds are for removal out to $m_{\rm AB}\simeq 19$ and asterisks are for the maps cleaned of NICMOS sources to the final depth. Triangles mark the NICMOS noise as estimated by \cite{Thompson:2007,Thompson:2007a}. For comparison the 2MASS-based CIB fluctuations  from \citet{Kashlinsky:2002,Odenwald:2003} are shown as red error bars without symbols. The CIB fluctuations from galaxies remaining at the greatest NICMOS depth are shown in yellow shade using the HRK12 reconstruction. 
}
\label{fig:cib_nicmos}
\end{figure}
NICMOS-based source-subtracted CIB fluctuations at 1.1 and 1.6\mic\ were studied by \citet{Thompson:2007,Thompson:2007a} after progressively eliminating galaxies down to much fainter fluxes than in 2MASS using data from the NUDF field of $2'$ on the side \cite{Thompson:2005}. After removing identified sources down to AB magnitude of $\sim 27.7$, 93\% of the map
remained for robustly direct power spectrum evaluation. The sky maps were at the sub-arcsecond resolution of {\it HST}. \citet{Donnerstein:2015} discusses the contributions from remaining outer parts and finds them small. \citet{Thompson:2007} show the Fourier plane of their images to be clean of artifacts from map construction. The resultant CIB fluctuations from that study are plotted with black asterisks in Fig.\ \ref{fig:cib_nicmos} at various depths of removal. At the magnitude
limits corresponding to the depth reached in the 2MASS studies, the NICMOS results do not fully agree with the former study but the difference can be accounted for if one assumes the 2MASS images to be at an effectively brighter removal magnitude due to e.g.\ destriping as discussed above. The asterisks in the figure show the diffuse light fluctuations at the ultimate removal threshold. The remaining diffuse light fluctuations appear significantly in excess of those from remaining known galaxies \cite{Helgason:2012a}.

Based on the color-ratio of the 1.1 to 1.6 \mic\ diffuse fluctuations \citet{Thompson:2007} suggest ``that the 0.8--1.8 \mic\ near-infrared background is due to resolved galaxies in the redshift range $z<8$, with the majority of power in the redshift range of 0.5-1.5''.

\subsubsection{CIBER}
\label{sec:ciber}

\begin{figure}[t]
\includegraphics[width=2.5in, trim = 1cm 0 0 0]{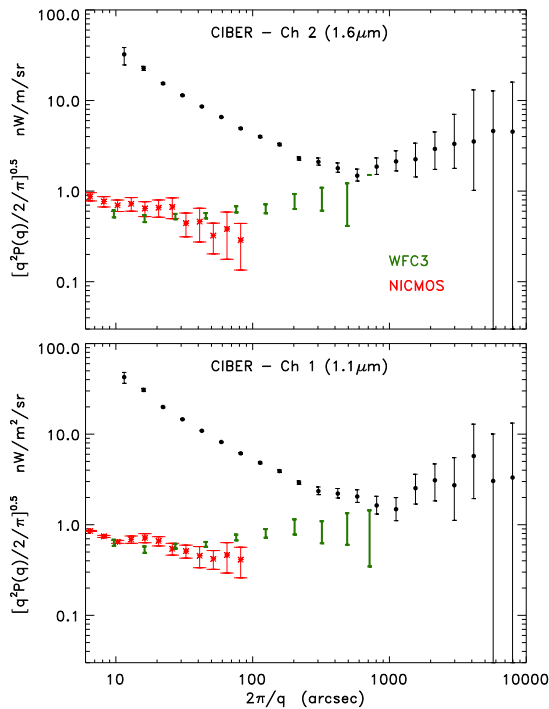}\\
\caption{\small Mean squared fluctuations from the auto-power from the CIBER analysis \cite{Zemcov:2014}. 
For comparison the NICMOS \citep{Thompson:2007} (red stars) and WFC3 \citep{Mitchell-Wynne:2015} (green error bars without symbols) HST-based results are also shown. 
}
\label{fig:cib_ciber}
\end{figure}

The CIBER suborbital rocket-borne experiment \cite{Bock:2013,Zemcov:2014} has recently suggested CIB fluctuations at 1.1 and 1.6 \mic\ shown in Fig.\ \ref{fig:cib_ciber} from \citet{Zemcov:2014}. Its imaging camera probes emissions with $\Delta \lambda/\lambda\simeq0.5$ around the central wavelengths over a square field-of-view of $2^\circ$ on the side 
with $\sim 6''$ pixels. After removing galaxies to Vega magnitude of 17.5 at J band (about 3 magnitudes brighter than deep 2MASS), 
and construction of maps that are the difference of separate fields (to remove common instrumental artifacts), 
only $30-50\%$ of the sky is left for Fourier analysis on the CIBER maps; see e.g.\ Figs. S4--S7 of \citet{Zemcov:2014}. After rejecting some of the data due to the stratospheric air-glow, 
four fields observed over 2 flights formed the basis for the analysis.

Their key assertions are that 1) ``The observed fluctuations exceed the amplitude from known galaxy populations'', 2) since they do not fit the epoch-of-reionization modeling of \citet{Cooray:2012} they ``are inconsistent with EoR galaxies and black holes'', and 3) ``are largely explained by IHL emission'' 
without accounting for the remaining difference between the measurement and the IHL model.

Although the masking approaches 70\% of the pixels in this study, the correlation function has {\it not} been evaluated to substantiate the robustness of the claimed power spectra. Various potential issues with the analysis have been discussed in \citet[][Sec. 2.1.2]{Kashlinsky:2015a}, which are impossible to assess further in the absence of explicit calculation of the correlation function for the heavily masked maps. \citet{Yue:2016a} have questioned the extragalactic origin of the claimed CIBER-{\it Spitzer} cross-power assigning it to the Galactic cirrus instead.
\subsubsection{WFC3/HST}
\label{sec:wfc3}
\citet{Mitchell-Wynne:2015} looked at diffuse background fluctuations in deep {\it HST}/WFC3 (and ACS)
observations of the CDFS.
They applied the self-calibration procedure of \citet{Arendt:2010} to construct 120 arcmin$^2$ maps at
1.25 and 1.6 \mic. 
After masking 47\% of the maps, they use the same methodology for computing the power spectrum from FTs as in \citet{Zemcov:2014}; despite the highly substantial masking their correlation function is not shown. Using the assembled WFC3-based images in conjunction with ACS and IRAC data they fit a multi-component model assuming 1) the existence of IHL with the template from \citet{Cooray:2012}, in addition to 2) remaining known galaxies modeled after \citet{Helgason:2012a}, 3) diffuse Galactic cirrus emission and, 4) a high-$z$ component from \citet{Cooray:2012a}. Assuming these components they conclude that the {\it HST}-based CIB fluctuations at 1.1 and 1.6 \mic\ contain high-$z$ emissions at the luminosity density a factor of $\sim(2-3)$ lower than derived earlier in \citet{Kashlinsky:2007b} at 3.6 and 4.5 \mic\ from {\it Spitzer} CIB measurements. They also obtain
with this fit modeling a cirrus level ``at least a factor of 3 larger than the upper limit" from CIBER.

\begin{figure}[t]
\includegraphics[width=3.in, trim = 1cm 0 0 0]{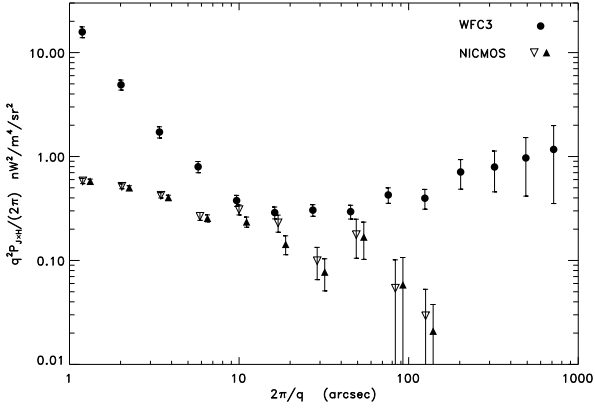}\\
\caption{\small Black circles show the $1.25\times1.6$\mic\ cross-power from WFC3 from \citet{Mitchell-Wynne:2016}. Triangles are the $1.1\times1.6$\mic\ cross-power evaluated for this review from NICMOS maps: uppward triangles are for ``Thompson mask," downward for extended mask; the two are slightly shifted for easy display.
}
\label{fig:cib_wfc3}
\end{figure}
The source-subtracted CIB fluctuations from that study at 1.1 and 1.6 \mic\ are shown in green in Fig.\ \ref{fig:cib_ciber} and can be interpreted as being in reasonable agreement with findings of \citet{Thompson:2007,Thompson:2007a}. However, a significant disagreement emerges - at both small and large scales - when one evaluates the cross-power between the two wavelength maps for the NICMOS images from \citet{Thompson:2007,Thompson:2007a} and that from \citet{Mitchell-Wynne:2015} as shown in Fig.\ \ref{fig:cib_wfc3}. 
\citet{Thompson:2007,Thompson:2007a} did not evaluate this cross power, but the archival images (intensity, 
sigma, and SExtractor detections) are
available\footnote{\tiny\url{https://archive.stsci.edu/pub/hlsp/udf/nicmos-treasury/version2/}}.
We downloaded the 1.1 and 1.6 $\mu$m images, rotated and cropped them, and 
applied appropriate conversions to $\nu I_{\nu}$ in nW/m$^2$/sr.
No model is applied to remove sources. Masking the sources using the regions
indicated by the SExtractor detection image yields a mask that is similar to,
but not exactly the same as, the masking illustrated in \citet{Thompson:2007,Thompson:2007a}.
This mask excludes 10\% of the data, slightly larger than the stated above 7\% exclusion. We also tested a more conservative mask analogous to \citet{Donnerstein:2015}, in which 
we expanded the masked regions by a radius of 7.5 pixels ($0.675''$), which 
leads to excising 31\% of the data. The cross-power does not require noise subtraction if the noise at the two channels in uncorrelated. There appear significant differences
in the cross power of the NICMOS and WFC3-based data sets. Additionally, the auto- and cross-powers for the \citet{Mitchell-Wynne:2015} results appear to lead to coherence exceeding unity at both small and large scales. It is not clear what {\it HST} dataset and diffuse maps better approximate reality.\\[-6mm]
\subsubsection{Current state of CIB fluctuations at 1--2 \um}
\label{sec:state1-2}

Unlike in the 2--5\mic\ range, there appears no mutually agreed upon CIB fluctuations results at 1--2\mic, preventing from any robust cosmological modeling. This will be reflected in our discussions below, although whenever it makes sense we will make brief, if more speculative, excursions into this range of wavelengths. 
\subsection{Integrated CIB excess}
\label{sec:cibexcess}

The measured CIB fluctuation excess at 2--5 \mic\ appears to have only small (within the uncertainties) variations between $\sim1'$ and $\sim 1^\circ$. The power spectrum of such CIB fluctuations from the new populations can be characterized with an amplitude at some fiducial scale and a template. The CIB fluctuations, at say $\sim 5'$ which was used for such normalization in \citet{Kashlinsky:2012}, as measured with {\it Spitzer} and {\it AKARI} can be integrated to give the net integrated CIB flux fluctuations over the wavelengths of the detections leading to:
\begin{eqnarray}
\delta F_{2-5\mic}(5') = \int^{IRAC}_{AKARI}\; 
\left(\frac{q^2P_\lambda}{2\pi}\right)^{1/2}\; \frac{d\lambda}{\lambda} = \nonumber\\
\left[\frac{(4.5/2.4)^\alpha -1}{\alpha}\right] \delta F_{4.5\mic}(5')\simeq 0.09\frac{{\rm nW}}{{\rm m^2 sr}}
\label{eq:cibfluc_bol}
\end{eqnarray}
where $\nu \delta I_\nu \equiv [\frac{q^2P_\lambda}{2\pi}]^{1/2}$ is the CIB flux fluctuation in nW m$^{-2}$ sr$^{-1}$ and we assume per Fig.\ \ref{fig:cib} that it 
scales with wavelength as $\nu \delta I_\nu \propto \lambda^{-\alpha}$ with $\alpha \simeq 3$; for $\alpha=2$ the above expression gives $\delta 
F_{2-5\mic} \simeq 0.065$ nW m$^{-2}$ sr$^{-1}$. In the above expression we have taken the {\it AKARI} and {\it Spitzer}/IRAC filters to have the integrated range of 
2--5 \um\ and the ``nominal'' central values of the filters were plugged into the middle expression above. 

Assigning the relative amplitude of CIB fluctuation for a given template at the fiducial scale of, say $5'$, $\Delta_{5'}\equiv \delta F_{2-5\mic}(5')/F_{2-5\mic}$ would require the new populations to produce a net integrated CIB flux of $F_{2-5\mic}=\delta F_{2-5\mic}(5')/\Delta_{5'} \sim 1$nW/m$^2$/sr for $\Delta_{5'}\sim 10\%$. 
If its $\lambda^{-3}$ SED extends to 1.6 \um, the integrated CIB fluctuation excess from the new populations would be higher 
at $\delta F(5')\sim 0.3$ nW m$^{-2}$ sr$^{-1}$ over the 1.6--5 \um\ range leading to $F_{1.6-5\mic} \lsim 3$ nW m$^{-2}$ sr$^{-1}$ still within the errors of the current 
conservative CIB measurements of \citet{Thompson:2007,Thompson:2007a}. Conversely, if the $\lambda^{-3}$ SED of the CIB excess observed with {\it Spitzer} does not extend to the shortest {\it AKARI} 2.4 \um\ channel the required CIB would be correspondingly smaller.

\section{Implications of CIB fluctuation results}
\label{sec:implications}

\subsection{General implications}
\label{sec:general}
The general implications of the source-subtracted CIB fluctuations stem from 1) the properties of the clustering, in shape and amplitude, that appear 2) at very low shot noise power levels \cite{Kashlinsky:2007b}.

As discussed in Section \ref{sec:basis}, the shot-noise power is $P_{\rm SN}\simeq S_\nu(\bar{m}) F_{\rm tot}(> m_{\rm lim})$,
where $F_{\rm tot}(> m_{\rm lim})$ is the CIB flux from 
remaining sources. 
The measured levels of the shot-noise do not currently reach the regime of 
 attenuation of the large-scale fluctuation from clustering; the point where this happens would then probe the flux of the typical sources responsible for this CIB component. The deepest current limits reached are $P_{\rm SN}=(26,14)$ nJy$\cdot$nW m$^{-2}$ sr$^{-1}$ at (3.6,4.5)\um. Since $P_{\rm SN}\sim S F_{\rm tot}$, these limits coupled with the above, imply the upper limits on the typical fluxes of the sources producing them:
 \begin{equation}
 S_{(3.6,4.5)\mic}\; \lsim \; (26,14)\; \left(\frac{F_{\rm tot}}{\rm nW/m^2/sr}
 \right)^{-1} \;\; {\rm nJy}
 \label{eq:sfromsn}
 \end{equation}
 Such objects would have $m_{\rm AB} \gsim$28--29 and may have fluxes well below what can be probed individually even with the {\it JWST}. 
  
 A {\it lower} limit on the projected surface density, $n_2$, of the new sources can be estimated in a similar manner by writing the shot-noise power from these sources as $P_{\rm SN}\sim F_{\rm CIB}^2/n_2$. The measured shot-noise at $P_{\rm SN} \sim 10^{-11}$nW$^2$/m$^4$/sr \citep{Kashlinsky:2007a} gives an {\it upper} limit on the shot-noise from the new populations, so their number per beam of area $\omega$, ${\cal N}_2$, must {\it exceed}:
 \begin{equation}
{\cal N}_2 \gtrsim 0.1\!\left(\frac{F_{\rm CIB}}{{\rm nW/m^2/sr}}\right)^2\!\!\!\left(\frac{P_{\rm SN}}{10^{-11}{\rm nW^2/m^4/sr}}\right)^{-1} \!\! \frac{\omega}{10^{-12}{\rm sr}}
\label{eq:n2_kamm3}
\end{equation}
Confusion intervenes when there are more than 0.02 sources/beam \citep{Condon:1974}, so this shows that the bulk, perhaps all, of the new populations would be within the confusion noise of the instruments with beams of $\omega\gsim 2\times10^{-13}$sr or effective radii $\gsim 0.05''$. Note that the shot noise in the current measurements is clearly produced by remaining known galaxies and the component contributed by the new sources may be much smaller leading to still stronger constraints from confusion.

\subsection{Known populations}
\label{sec:hrk12}

It is now generally agreed that known populations appear insufficient to explain the source-subtracted CIB signal measured at 2-5 \mic. Its origin is then posited to lie in new sources, either at high $z$ or more recent epochs. It was also shown that even extrapolating from the measured UV LFs \cite{Bouwens:2011} of  the known galaxy and stellar populations to higher $z$ does not explain the CIB \cite{Cooray:2012a,Yue:2013a}, although latter studies of high-$z$ UV LF \cite{Finkelstein:2015} may ease the degree of the disparity somewhat.

\subsection{High-$z$ sources}
\label{sec:hi-z}

The bolometric flux produced by populations containing a fraction $f$ of the baryons in the Universe after they have converted their mass-energy into radiation with efficiency $\epsilon$ at an effective redshift $z_{\rm eff}\equiv 1/\langle(1+z)^{-1}\rangle$ is given by eq.\ \ref{eq:f_cib_theor}.
Populations at high $z
$ are strongly biased, span a short period of cosmic time, and are expected to produce $\Delta_{5'}\sim 10\%$ relative CIB fluctuations around 5$'$ scale. 
Such populations would then require producing about $F_{\rm CIB} \sim 1$ nW m$^{-2}$ sr$^{-1}$ in the integrated flux at near-IR wavelengths 
($2-5\ \mic$)
implying a correspondingly large luminosity density around rest-frame UV at $z\gsim 10$ as argued by
\citet{Kashlinsky:2007b}. \citet{Mitchell-Wynne:2015} derived similar numbers from an assumed multi-component fit, including IHL and high-$z$ sources, to deep {\it Spitzer} and {\it HST} data.
The overall fraction of Universe's baryons needed to explain the CIB is $f_{\rm Halo}f_*$ \citep[see Sec. 2.3.2 in][]{Kashlinsky:2015a}. 
Massive stars can convert matter into radiation with an efficiency of 
$\epsilon \simeq 0.007$, whereas accretion onto BHs can reach 
$\epsilon \lesssim 0.4$. 
If the integrated CIB fluctuation approximates the bolometric flux produced by these sources, the mean fraction of baryons that go into the sources inside each halo, is:
\begin{eqnarray}
f_* =  0.1\left(\frac{f_{\rm Halo}}{0.01}\right)^{-1}\!\left(\frac{\epsilon}{0.01}\right)^{-1}\!\left(\frac{z_{\rm eff}}{10}\right)\! \left(\frac{\Delta_{5'}}{0.1}\right)^{-1} 
\times \nonumber \\
\left(\frac{F_{\rm tot}}{\rm nW/m^2/sr}\right)\left[\frac{F_{\rm CIB}(2\!-\!5\mic)}{F_{\rm tot}}\right]\;\;\;\;\;
\label{eq:fraction_bc}
\end{eqnarray}
Thus in order to produce the measure CIB at $z>10$ with ``reasonable" formation efficiencies ($f_*<10\%$) one requires a large fraction of matter in collapsed halos capable of producing luminous sources.

\subsubsection{First stars}
\label{sec:cibvs1ststars}

Potential CIB contributions from first stars have been discussed by various authors assuming both predominantly massive Pop III stars \cite[e.g.][]{Kashlinsky:2015a,Santos:2002,Salvaterra:2003,Cooray:2004,Kashlinsky:2004} as well as mixed stellar mass functions which include also normal mass stars at high $z$ \cite[e.g.][]{Fernandez:2010,Fernandez:2012,Helgason:2016}.

\begin{figure}[t]
\includegraphics[width=3.in]{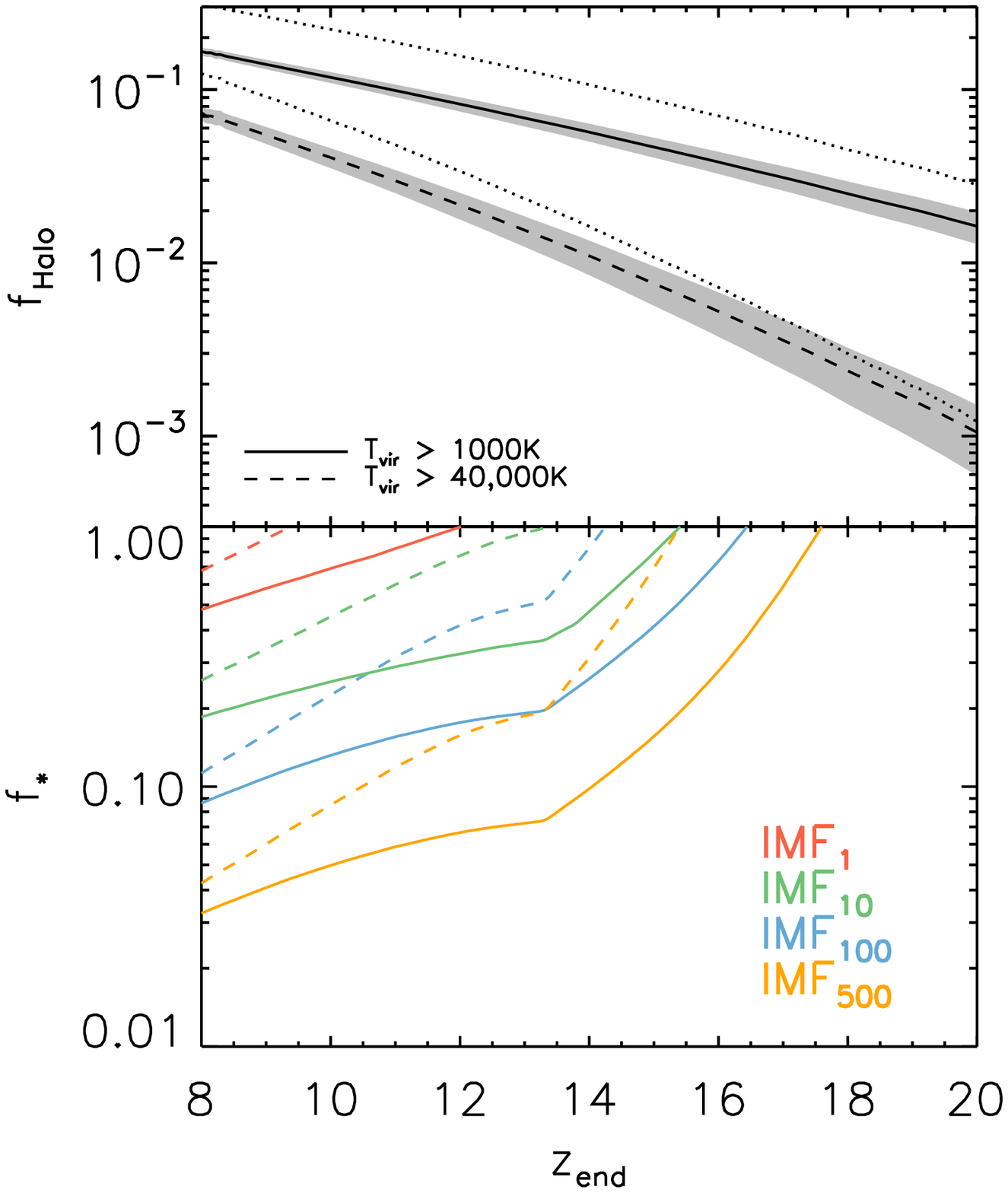}
 \caption[]{\footnotesize{Adapted from \citet{Helgason:2016}. {\bf Top}: the mass fraction in collapsed halos as a function of redshift. Solid and dashed lines correspond to halos with different $T_{\rm vir}$ as marked.  Dotted lines show the classic \citet{Press:1974} prediction compared with solid lines for the variant of \citet{Sheth:2002}. Note that the classic Press-Schechter prescription underestimates the abundance of the most extreme objects. Additional complication may arise through UV radiation from first sources, supernova blast waves, and chemical enrichment of the ambient gas. {\bf Bottom}: the star formation efficiency $f_*$ required to produce the CIB fluctuations by a given redshift, $z_{\rm end}$. The curves assume the entire stellar population forming with IMF$_1$ , IMF$_{10}$ , IMF$_{100}$ , IMF$_{500}$ (red, green, blue, orange; top to bottom) in all halos. The solid lines show the case where minihalos $T_{\rm vir} > 1000$K are included whereas the dashed lines include $T_{\rm vir} > 40,000$ K halos only.}}
\label{fig:fig5_h16}
\end{figure}

\citet{Helgason:2016} conducted an extensive study of the contribution to CIB expected from early stellar populations in standard $\Lambda$CDM cosmology and Fig.\ \ref{fig:fig5_h16} summarizes their results for stellar contributions. The fraction of halos, $f_{\rm Halo}$, collapsing at given $z$ according to several variants of the \citet{Press:1974} prescription, is shown in the upper panel of the figure. Then one can evaluate the net CIB assuming stars of a given stellar mass function form in the collpased halos with mean efficiency $f_*$. Four stellar mass functions were considered:  1) IMF$_1$ with standard Kroupa mass-function in the $(0.1-100) M_\odot$ range, 2) IMF$_{10}$ with lognormal mass function with characteristic mass of $10M_\odot$ and dispersion of $1 M_\odot$ in the $(1-500) M_\odot$ range, 3)
IMF$_{100}$ with Salpeter-type power law $\propto M^{-2.35}$ in the $(50-500) M_\odot$ range, and 4) IMF$_{500}$ with all stars having $500 M_\odot$ emitting in the near-Eddington fashion. All stars were assumed to evolve from single zero-age main sequence objects, using calculations of luminosity and spectra computed from the population synthesis code of \citet{Zackrisson:2011}. The resultant mean efficiency $f_*$ required to explain the observed CIB fluctuations at 2--5\mic\ within the standard $\Lambda$CDM density field (Fig.\ \ref{fig:p_lcdm}) appears high as shown in the lower panel of Fig.\ \ref{fig:fig5_h16}.

\citet{Helgason:2016} discuss further the requirements of high-$z$ sources to produce the observed CIB fluctuations within the conventional, if necessarily simplified, framework of gravitational clustering and spherical collapse of adiabatic $\Lambda$CDM fluctuations. They conclude that 1) first galaxies, if extrapolated to $z>8$ from known UV luminosity functions would produce much less CIB fluctuation power than observed \citep[cf.][]{Cooray:2012a,Yue:2013}, and 2) at still higher $z$ (first) stars would have to i) form inside the collapsed halos at substantial formation efficiencies (converting $f_* \gsim 5\%$ of the available baryons in collapsing halos) and ii) be very massive ($\sim 500M_\odot$) if they are to explain by themselves the observed CIB anisotropies. \citet{Kashlinsky:2015} reproduce the observed {\it Spitzer} signal with massive early stars forming at the mean formation efficiency $f_* \simeq$4\% out to $z=10$. 

The ``high-mean-formation-efficiency" difficulty can ultimately be traced to a relative paucity of high-$z$ collapsed halos - with the parameters considered appropriate for star formation - due to the limited amount of power set by the adiabatic $\Lambda$CDM component of matter fluctuations, which arose from the period of inflation. Later we discuss how the abundance of the halos collapsed at high $z$ is dramatically increased if PBHs constitute the DM, and reduce - by large factors - the efficiencies required to produce the observed CIB anisotropies.

We note that various natural evolutionary modes of first stars, e.g.\ enhanced binary formation in turn leading to high mass X-ray binaries \cite{Mirabel:2011}, would reduce the required efficiency, $f_*$, easing the energetics requirements for producing the observed CIB excess.

\subsubsection{Direct collapse black holes (DCBHs)}
\label{sec:cibvsdcbh}

The motivation to consider first black holes as CIB sources is two-fold: (a) the power from even the faintest reionization sources appears to be insufficient; (b) the CIB-CXB correlation implies the presence of a substantial population of accreting sources. In addition, DCBH ($M_{\rm DCBH} =10^{4-6} M_\odot$) seeds can ease the already mentioned problem of explaining the inferred masses of supermassive BHs. It is then appealing to consider high-$z$ accreting DCBH as additional CIB sources. Such faint  ``AGN" have so far escaped detection from even the deepest X-ray observations \citep{Willott:2011, Cowie:2012} at any stage during their growth. Whereas deep X-ray surveys do not cover enough volume at high redshift, current wide-area studies are simply not deep enough ($L_X > 10^{42.75}$ erg s$^{-1}$, \citet{Fiore:2012}). A possible exception is the discovery of two $z>6$ DCBHs claimed by \citet{Pacucci:2016}, which has raised considerable hope to firmly identify these supermassive BH ancestors.

CIB fluctuations may also arise from DCBHs providing a viable alternative to discover them. The original proposal of this was made by \citet{Yue:2013, Yue:2014} who showed that under some conditions, a high-$z$ DCBH population could explain the observed CIB fluctuations, and most importantly, also the observed CIB-CXB coherence.  The spectrum of accreting black holes formed through the direct collapse of metal free gas in halos with virial temperature $> 10^4$ K are very likely to be Compton-thick. This fact has several important implications:  (a) as most of photons with energy $>13.6$ eV are absorbed by the large column density of surrounding gas, the contribution of these objects to reionization is negligible; (b) for the same reason, the DCBH contribution  to the CXB is reduced significantly; (c) ionizing photons are re-processed into optical-UV bands (free-free, free-bound and 2-photon emission) while Ly$\alpha$ photons are trapped, and finally converted into 2-photon emission. These secondary photons eventually escape the object and considerably boost (by a factor of 10) the contribution of these sources to the CIB fluctuations. 

\begin{figure}[t]
\includegraphics[width=3.3in, trim=5mm 0 0 0]{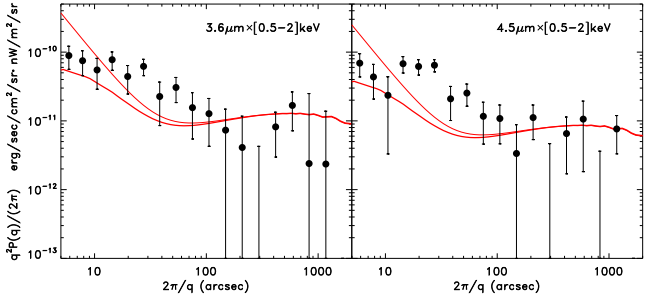}
\caption{Mean squared fluctuation spectrum for the
CIB--CXB$_{0.5-2.0{\rm keV}}$ cross-power at 3.6 \mic\ and 4.5 \mic\ \cite{Cappelluti:2017a}. The predicted signal from DCBHs is shown with red lines for $N_H \simeq 1.5\times 10^{25}$ cm$^{-2}$ as revised from \citet{Yue:2013} with modifications for $N_H$ (B. Yue 2017, private communication). Thick red line shows convolution with the IRAC beam of the underlying DCBH model (thin line).} 
\label{fig:fig-dcbh}
\end{figure}
According to \citet{Yue:2013} predictions, fitting the latest Spitzer observations at 3.6 and 4.5 $\mu$m, the observed CIB fluctuations at $\theta > 100''$ can be explained (Fig.\ \ref{fig:fig-dcbh}) by DCBHs formed in metal-free halos with virial temperature $T_{\rm vir} = (1-5) \times 10^4$ K earlier than $z\simeq 12.5$. These DCBHs are formed with initial masses of the order of $10^{5.8} M_\odot$, and subsequently were able to grow by accreting gas at the Eddington limit for about 30-50 Myr. 

A population of DCBHs with these characteristics would produce a CXB intensity at 1.5 keV that is well below the current observational limits as long as the obscuring gas column density exceeds  $N_H \simeq 10^{25}$ cm$^{-2}$. Analogously, DCBHs contribute only marginally to the CXB angular power spectrum. However, the DCBH signal emerges in the CXB-CIB cross-correlation at scales $> 100''$. For $N_H= 1.5\times 10^{25}$ cm$^{-2}$ the cross-correlation level of the DCBH population is $\simeq 8 \times 10^{-12}$ erg s$^{-1}$ cm$^{-2}$ nW m$^{-2}$ sr$^{-1}$, in tantalizing agreement with recent observations \citep{Cappelluti:2013}, despite the remaining large uncertainties in current data.  

Thus, the near-IR CIB fluctuations and their coherence with the CXB might be the smoking gun of a peculiar population of early intermediate mass BHs; they might also shed light on the challenging questions posed by the rapid formation of SMBHs seen in quasars.
\subsubsection{Primordial black holes (PBHs)}
\label{cibvspbh}
Following the original LIGO discovery of GW150914 from two $\sim$30$M_\odot$ coalescing BHs  \cite{Abbott:2016,Abbott:2016a} and a tentative detection of another similar object, LVT151012 \cite{Abbott:2016c,Abbott:2016d}, two more GW events were announced from a total of $\sim 6-7$ weeks of Advanced LIGO operations: GW151226 \cite{Abbott:2016b} and GW170104  \citep[][]{Abbott:2017}, where there appears a marginal evidence for misaligned spins while no electromagnetic emissions were detected. 
With the current total of 8-10 BHs\footnote{A further GW from two merging BHs of  $\sim$ 20 and 30 $M_\odot$ was announced toward the completion of the aLIGO O2 run \cite{Abbott:2017} after this review has been prepared.}, this indicates the presence of BHs with masses peaking near $M_{\rm BH}\sim$(20--30)$M_\odot$. While the pre-LIGO detection expectations were that the dominant source of detectable GWs would be binary-neutron-star (BNS) mergers \cite{Abadie:2010}, by now a growing population of BHs within the above mass-range, while subject to LIGO-specific selection effects,  appears to dominate the GW emitting sources\footnote{\tiny\url{http://www.virgo-gw.eu/docs/GW170814/BHmassChartGW092017.jpg}}.
If these BHs are primordial making up or dominating DM, the extra Poissonian component of the density fluctuations would lead to much greater rates of collapse at early times, which would naturally produce the observed levels of the CIB fluctuations \citep[][]{Kashlinsky:2016}. 

\begin{figure}[t]
\includegraphics[width=3.5in, trim=0.7cm 0 0 0]{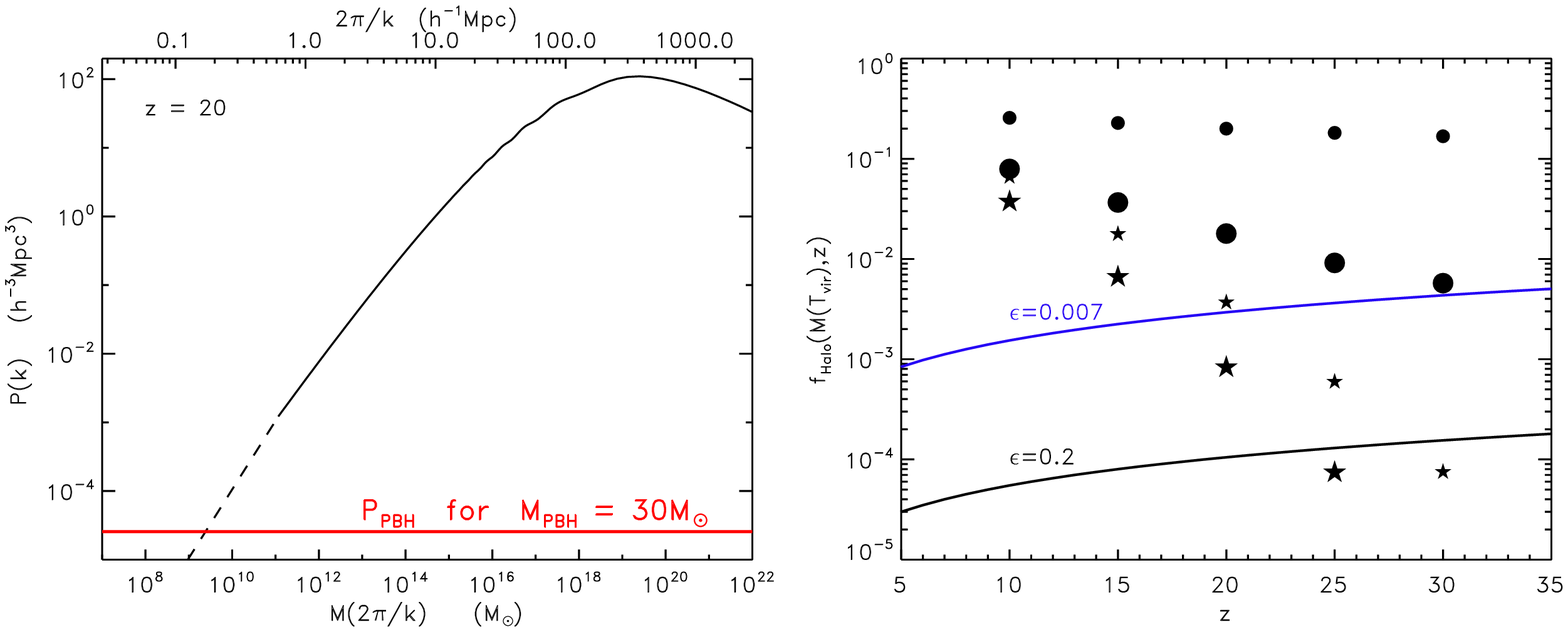}
 \caption[]{\footnotesize{Adapted from \cite{Kashlinsky:2016}. 
 {\bf Left}: Black solid line marks the CMBFAST-computed $\Lambda$CDM power spectrum at $z$=20 vs the mass within the comoving radius $2\pi/k$. Black dashes show the $P_{\Lambda{\rm CDM}}\!\propto \!k^{-3}$ extrapolation to scales not accessible to CMBFAST, but relevant for first halos collapse. Red horizontal solid line is the Poissonian power from DM PBHs of $M_{\rm PBH}$=$30M_\odot$, which clearly dominates the scales relevant for halo collapse at this $z$; $P_{\rm PBH}\!\propto\! M_{\rm PBH}$, with $M_{\rm PBH}$ being a suitably averaged mass in case of PBH mass distribution. {\bf Right}: Fraction of collapsed halos at  $T_{\rm vir}\!>\!10^3$K vs $z$ for standard $\Lambda$CDM power spectrum (small stars) and $M_{\rm PBH}$=$30M_\odot$ (small circles); same for $T_{\rm vir}\!>\!10^4$K shown with large stars/circles. Thick solid lines mark the overall fraction of baryons,   $f_{\rm Halo}f_*$ needed to produce the observed CIB with the H-burning radiative efficiency  $\epsilon$=0.007 (blue) and BH-type efficiency $\epsilon$=0.2 (black). The required mean baryon conversion efficiency into luminous sources inside each halo, $f_*$, would be the ratio of the solid curves to the symbols. {\it $f_*$, while high otherwise, is very modest if the PBHs make up DM}.}}
\label{fig:fig1_k16}
\end{figure}
As discussed earlier, the DM from PBHs will contain an extra (isocurvature) component due to Poissonian fluctuations \cite{Meszaros:1974,Meszaros:1975} with the power component at the time of the PBH formation being $P_{\rm PBH, initial}=n_{PBH}^{-1}$  in comoving units. 
From their formation to today ($z=0$) these isocurvature fluctuations would grow, at wavelengths below the horizon at matter-radiation equality $z_{\rm eq}$, by a scale-independent factor of $\frac{3}{2}(1+z_{\rm eq})$, so the extra power component at redshift $z$  is given by \citep{Afshordi:2003,Kashlinsky:2016}:
\begin{eqnarray}
P_{\rm PBH}(z) = \frac{9}{4}(1+z_{\rm eq})^2n_{\rm PBH}^{-1} [g(z)]^{-2} \nonumber  \simeq \\  2\times10^{-2} \left(\frac{M_{\rm PBH}}{30M_\odot}\right) \left(\frac{\Omega_{\rm CDM}h^2}{0.13}\right) g^{-2}(z)\;{\rm Mpc}^3
\label{eq:p_pbh}
\end{eqnarray}
where $g(z)$ is the linear growth factor of fluctuations from $z$ to today, with $g(0)$=1. We assumed all PBHs to have identical mass of 30$M_\odot$; the discussion can be trivially generalized to any PBH mass distribution with a suitably averaged effective $M_{\rm PBH}$. Fig.\ \ref{fig:fig1_k16}, left shows the extra power component for $M_{\rm PBH}=30M_\odot$ compared to the $\Lambda$CDM power spectrum from the purely adiabatic fluctuation component. The power is plotted vs the mass contained in wavelength $2\pi/k$ which is $M(r) =1.15\times10^{12} (r/1{\rm Mpc})^3 M_\odot$ for the adopted cosmological parameters. This extra power is $\propto M_{\rm PBH}$ and for $M_{\rm PBH}>1M_\odot$ dominates the small scales relevant for collapse of the first halos at $z>10$, but has no impact on the observed CMB anisotropies or baryonic-acoustic-oscillations (BAO) \citep[][]{Eisenstein:1999} which appear in CIB fluctuations on arcminute scales. Moreover, unlike the clustering component, white noise power contributions to the angular CIB power spectrum are not affected by biasing amplification \citep{Kashlinsky:2004}. This shows that there is a dramatic increase in power from the Poissonian PBH component, normalized to the LIGO results, on scales relevant to first halo collapse. 

The higher abundance of collapsed halos in which first sources would form at $z>10$ for the PBH DM case is shown in Fig.\ \ref{fig:fig1_k16} (right) for 1) minihalos where H$_2$ formation is efficient evolve at $T\lsim 10^3$K and 2) where, in the absence of H$_2$, the metal-free gas will be able to cool to $10^4$K and collapse in halos with larger virial temperature will proceed isothermally.  
In this case luminous sources within the much more abundant early collapsed halos would reproduce the observed {\it Spitzer} and {\it AKARI} CIB fluctuations with modest formation efficiency requirements. 
This can be demonstrated by taking population models from \citet{Helgason:2016} and rescaling them by the collapse-efficiency ratio from Fig.\ \ref{fig:fig1_k16}, right. Specifically, {\it Spitzer}-based CIB fluctuations would now be reproduced with only $f_*<0.5\%$ forming out to $z\gsim15$ (instead of 4\% with formation continuing to $z\simeq10$) and the lines in Fig. 5 of \citet{Helgason:2016} need to be rescaled down by the corresponding factors. Additionally the measured CIB-CXB coherence \citet{Cappelluti:2013} would require that at least $\gsim$(10--15)\% of the luminous CIB-producing sources are accreting BHs, broadly consistent with this scenario. The black-body temperature of the emissions arising from the Eddington-accreting BHs is $T_{\rm acc} \propto M_{\rm BH}^{-1/4}$ \citep[e.g.][]{Kazanas:2015}, so PBHs being much less massive than DCBHs  may have CXB component extending to harder X-ray energies.

Gas collapse/evolution in the PBH minihalos may affect the subsequent emitting source formation inside them as outlined in Sec.\ \ref{sec:PBH}. A possibility, discussed in \citet{Yue:2013a} for the DCBH model, whereby the gaseous collapsed halos are Compton thick so the ionizing photons are absorbed and reprocessed into a two-photon continuum, 
may also apply here.

The arguments are valid only if the PBHs make up all, or at least most, of DM, but at the same time the mechanism appears inevitable if DM is made of PBHs. Upcoming extensive aLIGO observing runs, O3 and beyond, planned to start after increasing sensitivity \citep{Abbott:2016e}, and combined with aVIRGO\footnote{\url{http://www.virgo-gw.eu/}}, should be critical in testing this proposition. 

\subsection{New intermediate and low-$z$ sources}
\label{sec:cibvsihl}

\subsubsection{Intrahalo light}
The fits to CIB fluctuations at 3.6 and 4.5 \mic\ according to the original IHL model from \citet{Cooray:2012} are shown with green short dashes in Fig.\ \ref{fig:clustering_sn_irac}. The  revised model fits from \citet{Zemcov:2014} are shown with long green dashes in the figure; in that most recent form, IHL arises mostly at $z<0.5$. While the model can be said to reasonably fit the CIB fluctuations at the highest shot-noise, in its presented forms the IHL fails to account for the data at deeper shot noise levels, available before the introduction of the model, and it remains to be seen whether satisfactory fits to the available data can be constructed by its proponents. In addition, there remain a number of observational and theoretical challenges which make the IHL 
interpretation problematic. All tests that have been conducted so far have failed to reveal any spatial correlation between the fluctuation signal 
and extended emission from detected galaxies. If the IHL were to arise from stars originally formed within galaxies, the unresolved fluctuations should produce a 
measurable spatial correlation with the spatial distribution of resolved galaxies. The apparent absence of such correlations with i) the subtracted outer parts of 
galaxies, ii) artificial halos placed around galaxies, and iii) the insensitivity to the increased area of source masking, all present challenges for the IHL model. These 
observational tests are described in detail in \citet{Arendt:2010} and \citet{Donnerstein:2015}. The IHL also does not account for the measured correlation with the soft X-ray 
background \citep[][]
{Helgason:2014,Mitchell-Wynne:2016, Cappelluti:2017a}.

\subsubsection{Axion decay} 
\citet{Gong:2015} proposed that axions with a mass $\sim 4$ eV decay via two $\gamma$'s with wavelengths in the near-IR band and in the process leave a signature in the EBL power spectrum over the 0.6-1.6 \mic\ range in agreement with data. It is not clear, however, how the measured high coherence levels between the near-IR CIB and unresolved soft X-ray CXB can be explained in this model. 
\subsection{Limitations of current instrumental configurations}
\label{sec:limitations}
\begin{figure}[t]
\includegraphics[width=2.5in]{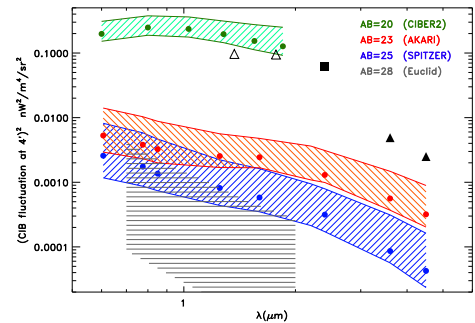}
 \caption[]{\footnotesize{Adapted from \citet{Kashlinsky:2015a}. CIB fluctuations at $2\pi/q=4'$ vs $\lambda$ from known galaxies remaining at marked depths (colored bands top to bottom as per legend) are compared to the measured CIB fluctuations at 3.6 and 4.5\um \citep[][filled triangles]{Kashlinsky:2012,Kashlinsky:2015a} and 2.4\mic\ \citep[][filled square]{Matsumoto:2011}.  Open triangles show an example of $z>10$ contribution at J, H bands, which reproduces {\it Spitzer} data \citep[][see caption to Fig.\ \ref{fig:librae-power} below]{Kashlinsky:2015,Helgason:2016}; it has no emissions below $\simeq 1.1 \mic$. Filled 
circles mark the default HRK12 reconstruction and dashed areas mark the limits due to the HFE and LFE 
extrapolation extremes.  
The fiducial scale of $4'$ is shown since the fluctuations at larger scales are approximately constant until $\sim $20$'$--$30'$ where they approach the HZ regime,  $P\propto q$.
}
\label{fig:fig17_jwst} }
\end{figure}

Current observations of source-subtracted CIB fluctuations, at 2--5 \mic, as discussed above suggest existence of important new cosmological populations, and its coherence with unresolved CXB implies that in part they include BHs. It is important to identify the nature and the epochs of these sources and their influence on the contemporaneous high-$z$
Universe. We identify here these goals, the limitations of the current surveys in their regard, and the observational capabilities required to resolve them:

\paragraph{Probing directly the epochs from Ly-cutoff} is critical to understanding the origin of the new populations responsible for the clustering component of the CIB fluctuations. This can be probed by the implied absence of emissions below the Lyman-cutoff which corresponds to the rest Ly$\alpha$ line in the presence of \HI\ (prior to full reionization).
This cutoff around 0.1$(1+z)$\mic\ provides a critical  marker of the epochs when the CIB originated; at $z\simeq 10$ this corresponds to observer wavelength of $\sim 1\ \mic$. Determining the epochs of the CIB fluctuation sources 
requires availability of both visible and near-IR exposures to sufficiently large depths ($m_{\rm AB}\gsim 24$) ideally on the same instrument. Fig.\ \ref{fig:fig17_jwst} shows with hashed regions the reconstructed CIB fluctuation levels, with their systematic uncertainties, from galaxies remaining in the currently available and shortly upcoming experimental configurations with their depth and wavelengths accessibility related to this:  CIBER2 \citep[][green]{Lanz:2014}, {\it AKARI+HST} (red), {\it Spitzer+HST} (blue) and the {\it Euclid}-based configuration detailed in the following Section \ref{sec:librae}. The CIB fluctuation signal is illustrated with the amplitudes at $4'$ where the source-subtracted CIB fluctuation is theoretically expected to be near its peak. The amplitude of the mean squared CIB fluctuation detected with {\it AKARI} and {\it Spitzer}  is shown with black solid square and triangles respectively. As one can see, CIB/EBL fluctuations from remaining known galaxies are such that, to probe a possible Lyman break of the CIB signal at 3.6/4.5 \mic, one must eliminate sources to fainter magnitudes than feasible in the current experiments \citep[see discussion in][]{Kashlinsky:2015a}.
\\[-10mm]
\paragraph{Probing the CIB cross-power and coherence with the CXB with good energy resolution and statistical precision} is further critically important in that it tells us in what fraction the CIB emissions come from stellar nucleosynthetic processes or are generated via BH accretion as well as proportions of the two kinds of populations.
The use of {\em Chandra} as a probe for the CIB vs. CXB coherence has intrinsic limitations arising from the design of the telescope itself (Sect.\ \ref{CXO}). In fact, its main feature, the high angular resolution, has been obtained at  the price of a relatively low effective area, a limited field of view, and strong vignetting that limit the observatory's survey grasp. The net effects of this design on the {\it Chandra} survey performances and fluctuation studies are: a)   deep coverage is reached only on a small portion of the field of view near the optical axis; b)  images are significantly shallower off-axis; c) modeling sources of noise/foreground is complicated by the broad PSF tails; d) cosmic variance, due to the limited corrected field of view severely affects the large scale measurements of the clustering components; e) Poisson fluctuations due to the low photon surface density number (counts/pix) (i.e. $\delta F_{\rm Poiss, X} \geq \delta F_{\rm CXB}$) affect the maps severely. The net effect is  that $P_{A-B}\gg P_{\rm CXB}$ or, in other words, the X-ray fluctuations signal is background dominated.   
These effects limit the reliability of CXB vs CIB coherence with  {\em Chandra} to $\lesssim 1000''$.  
Suitably designed raster scans, as used for the UDS or the EGS fields (Sect.\ \ref{CXO}), or stacking fields can mitigate these effects \citep{Cappelluti:2013,Cappelluti:2017}.
Another limitation is that P$_{A-B}$ is one order of magnitude larger in the hard X-ray band than in the soft band \citep{Cappelluti:2017}.  
This limits the energy bands where the coherence can be evaluated and hence the precision on the SED of the signal. 
So far significant cross-power has been measured only between CIB and [0.5-2] keV with upper limits derived in the hard X-ray bands,
limiting probing the X-ray spectrum of the sources. 
No measurement has been performed with XMM-Newton yet, despite its much larger collecting area and smaller vignetting compared to {\em Chandra}; XMM-Newton has a rather broad PSF ($15''$ HEW, on axis) that hampers the masking of resolved sources. Moreover XMM-Newton's orbit is such that the instrument suffers from severe background flaring and soft protons, which are difficult to model. For the same reason the larger effective 
area of XMM-Newton in the hard band cannot be fully exploited in fluctuation studies.
Future survey missions, like eROSITA, will address these problems by covering extensive area of sky with a high throughput wide-field telescope with smaller vignetting, which in addition will be cancelled out by the scan geometry. The background in the eROSITA's L2 orbit may be more stable than for XMM-Newton. eROSITA's broad PSF will remain an issue, but below we suggest possible successful strategies in combining eROSITA and Euclid.
\paragraph{To summarize} the following thus appear to be required to resolve these topics adequately based on the discussion above:
{\bf (1)} near-IR sky maps over a large part of the sky integrated deep to $m_{\rm AB}\gsim 24$, 
{\bf (2)} corresponding visible band diffuse maps, integrated 
to depth of $m_{\rm AB}\gsim 25$ or fainter, 
{\bf (3)} corresponding diffuse X-ray maps of large area and good energy resolution between $\lsim 1$ keV and $\gsim 10$ keV in the observer frame, and
{\bf (4)} corresponding microwave diffuse maps covering large sky areas at several frequencies with $\sim1'$ resolution and low instrument noise.
These are required for measurement of the power spectrum and SED of CIB, clustering vs shot-noise, the Lyman break of the sources, cross-powers with other wavelengths, and history of emissions.

\section{LIBRAE: Looking at Infrared Background Radiation Anisotropies with Euclid}
\label{sec:librae}

Looking at Infrared Background Radiation Anisotropies with Euclid (LIBRAE) 
is  a NASA approved project\footnote{\url{http://www.euclid.caltech.edu/page/Kashlinsky\%20Team} and \url{http://librae.ssaihq.com}} to 
probe the CIB using data from
the European Space Agency's 
M-class
mission {\it Euclid}\footnote{\url{http://sci.esa.int/euclid/} and \url{http://www.euclid-ec.org}}. LIBRAE will exploit the {\it Euclid} imaging of the Wide and Deep Surveys at near-IR and visible wavelengths to conduct CIB science with unprecedented precision and scope and will be able to probe both the origin of the CIB and its populations together with the conditions existing at high $z$. 
We discuss the technical prospects and methodology quantifying the science goals of LIBRAE.

\begin{table}
\caption{Euclid survey parameters.}
\begin{tabular} {| p{0.475in} | p{0.35in} | p{0.56in} |  p{0.56in} | p{0.56in} | p{0.56in} |}
 \hline 
{ \footnotesize Survey} & { \footnotesize Area} & {\footnotesize VIS} & { \footnotesize NISP-Y} & {\footnotesize NISP-J} & { \footnotesize NISP-H} \\
{ \footnotesize } & { \footnotesize deg$^2$} & {\footnotesize 0.6--0.9 \mic} & { \footnotesize 0.9--1.2 \mic} & {\footnotesize 1.2--1.5 \mic} & { \footnotesize 1.5--2 \mic} \\
 \hline
{ \footnotesize Wide} & { \footnotesize $2\times 10^4$} & { \footnotesize  $m_{\rm lim}=26$} & { \footnotesize  $m_{\rm lim}=25$} & { \footnotesize $m_{\rm lim}=25$} & { \footnotesize $m_{\rm lim}=25$} \\
{ \footnotesize $F(>\!m_0)$} & & { \footnotesize $1.1_{-0.5}^{+1.4}$} & { \footnotesize $1.1_{-0.5}^{+1.1}$} & {\footnotesize $0.8_{-0.3}^{+0.9}$} &  {\footnotesize $0.6_{-0.3}^{+0.7}$}  \\
$P_{\rm SN}$& & { \footnotesize $46_{-17}^{+31}$} & { \footnotesize $120_{-39}^{+69}$} & {\footnotesize $95_{-34}^{+63}$} &  {\footnotesize $73_{-29}^{+60}$}  \\
\hline
{ \footnotesize Deep} & { \footnotesize 40} & { \footnotesize $m_{\rm lim}=28$} & { \footnotesize $m_{\rm lim}=27$} & { \footnotesize $m_{\rm lim}=27$} & { \footnotesize $m_{\rm lim}=27$} \\
{ \footnotesize $F(>\!m_0)$} & & { \footnotesize $0.5_{-0.3}^{+1.0}$} & { \footnotesize $0.5_{-0.3}^{+0.8}$} & {\footnotesize $0.3_{-0.2}^{+0.6}$} &  {\footnotesize $0.2_{-0.1}^{+0.4}$}  \\
 $P_{\rm SN}$& & { \footnotesize $3.3_{-1.7}^{+4.1}$} & { \footnotesize $8.4_{-4.0}^{+8.8}$} & {\footnotesize $5.9_{-3.0}^{+7.0}$} &  {\footnotesize $4.2_{-2.2}^{+5.7}$}  \\
\hline
\end{tabular}
\begin{flushleft}
Remaining known sources CIB, $F(>m_0)$ in nW/m$^2$/sr, and shot noise, $P_{\rm SN}$ in 
nJy$\cdot$nW/m$^2$/sr. Limiting magnitudes for remaining sources use $\sim$2.5$\sigma$ removal as will be used in CIB studies, which differs from the nominal 5$\sigma$ by $\Delta m=0.75$.
\end{flushleft}
\label{tab:euclid}
\end{table}
\subsection{Euclid configuration and data reduction methodology}

The {\it Euclid} spacecraft will carry a 1.2m telescope to a Sun-Earth L2 
orbit with two instruments: VIS and NISP.
The VIS instrument performs very broadband (0.55-0.90 $\mu$m)
imaging using an array of 36 4k CCD detectors with a pixel scale of $0.1''$.
NISP imaging is done in Y, J, and H bands using an array of 16 2k HgCdTe 
detectors at a pixel scale of $0.3''$. Via a beam splitter, both instruments 
have similar fields of view of $\sim 0.7^\circ \times 0.7^\circ$ 
($\sim0.53$~deg$^2$).

{\it Euclid's} main scientific objectives, studying DE evolution to $z\sim 2$, require the mission to carry out two surveys. The Wide Survey 
aims to cover $\sim20,000$ deg$^2$ at a nominal depth, whereas the Deep Survey 
will total $\sim 40$ deg$^2$
observed to 2 magnitudes deeper than the Wide Survey. Data from 
both surveys should also be useful for studies of the source-subtracted CIB
fluctuations. 
\citet{Laureijs:2011,Laureijs:2014} give a comprehensive overview of the {\it Euclid} primary science goals, telescope, 
instruments, and observing strategy. 

The analysis of source-subtracted CIB fluctuations would involve three
largely separable tasks: 1) construction of source-subtracted 
images of suitable scale and depth, and minimal artifacts, 2) subtraction/masking resolved sources, and
3) evaluating the fluctuations.

The means of producing maps for CIB analysis has varied according to 
the data being used and the researchers performing the study. Default 
processing pipelines are usually more focused on the resolved sources, 
and may not be designed to accurately reconstruct diffuse background 
emission that extends on scales larger than the detector. 
For {\it Spitzer}/IRAC data, self-calibration \citep{Fixsen:2000} has 
proved a useful means of mosaicking individual frames into wider and deeper
mosaic images, while removing fixed-pattern structure that correlates with 
the detector rather than the sky \citep{Arendt:2010}. This technique may 
be applied to {\it Euclid} data, but given the size of 
the {\it Euclid} surveys there are several 
issues of scale which need to be addressed to do this efficiently.

The field size of the data that are self-calibrated and 
analyzed would be limited to sizes up to the maximum scale of interest
for the CIB ($\sim$1$^\circ$). The limited field size retains the ability 
to analyze the 2-D fluctuations without difficulties of the 
mapping projection. For some tests, much larger regions could be mapped
and analyzed in HEALPIX format \citep[][]{Gorski:2005}. For angular scales sampled by the limited fields,
averaging results from many fields should be equivalent to measurement of 
the same contiguous area. Multiple smaller fields also allow a wider capability
to check field to field consistency.

Savings in processing speed and output data volume can be attained by reducing 
the resolution of the data. The smallest angular scale information is 
non-essential, as it primarily reveals the shot-noise level of the CIB, which 
is also revealed at scales of $\sim10''$ and larger (Fig.\ \ref{fig:librae-power}). 
In working with degraded resolution data, it will be useful to remove sources 
from individual exposures {\it before} creating mosaic images rather than after.
Source removal can be based on the size, shape, and brightness of 
identified sources from the standard processing pipelines. Low resolution 
source-subtracted mosaics may also have a decreased fraction of masked pixels,
because the low-resolution pixel need only be masked if {\it all} of the 
underlying full-resolution pixels are masked.

The analysis of the power spectrum using FFTs, or the correlation function 
using slower methods, will be more expedient with smaller, lower resolution 
images and the project will involve both forms of analysis. Processing and 
analysis will be similar for both the Wide and the Deep Surveys, as both 
will use the same observing strategy (exposure times and sequence, 
dithering, etc.). Because the Deep Survey fields are located near the ecliptic 
poles and are revisited regularly, the repeated coverage at constantly 
rotating position angles should lead to better self-calibration result and
improved data quality, beyond the direct increase in sensitivity of the 
observations.
\subsection{Foregrounds: Galactic Stars, ISM  and Zodiacal Light}
\label{sec:librae-foregrounds}

The same foregrounds (stars, DGL, and zodiacal light) 
that can potentially affect the CIB fluctuation measurements
of other spaced-based observations will need to be considered for analysis of the 
{\it Euclid} data as well (Sections V.B.1.c, V.B.2, V.C.3, V.C.4). 

With high angular resolution and good sensitivity (Table \ref{tab:euclid}), 
most Galactic stars can be individually resolved and masked at the 
high latitudes of the {\it Euclid} surveys. The stellar luminosity function
declines at $M_J \gtrsim 8$ \citep[e.g.][]{Bochanski:2010}, but even 
late M stars at 
$M_J ~\sim 11$ can be detected by NISP in the wide survey at distance of 
several kpc. However, cooler and fainter brown dwarfs 
\citep[e.g.][]{Dupuy:2012} will only be individually 
detectable on scales $\lesssim 100$ pc. 

The DGL energy spectrum is 
expected to rise as wavelengths decrease from {\it Spitzer}/IRAC's 
3.6 $\mu$m band to {\it Euclid} NISP's Y band. At the {\it Euclid} 
wavelengths, the DGL should be strongly dominated by scattered light 
with little to no contribution from thermal emission. The wide survey 
will necessarily cover many regions of higher DGL intensity than 
previously studied small deep fields (usually selected in part for 
low ISM column densities). Thus the DGL will usually be stronger than 
in most previous studies, but the wide survey observations will allow 
much more robust correlation of potential DGL against other tracers of ISM, or 
even simply Galactic latitude. Measured relative to the 100 $\mu$m 
thermal dust emission, the mean intensity of high latitude DGL at 
$\sim0.4 - 5$ $\mu$m has been pieced together by a number of studies 
\citep{Brandt:2012, Arai:2015, Tsumura:2013, Sano:2015, Sano:2016}. 
However, these and other studies \citep[e.g.][]{Mitchell-Wynne:2015} also 
report significant variations in the DGL at varied locations. 

To help minimize backgrounds and increase sensitivity, the {\it Euclid} 
surveys avoid low ecliptic latitudes. However the range of allowed
solar elongations is very limited compared to most other facilities, and 
repeat coverage will only exist for the deep survey and calibration fields.
Therefore, detection of zodiacal light influences through temporal tests 
will be limited, and examination of trends vs. ecliptic latitude may be the 
most useful approach for the wide survey. A critical look at the 
colors of the fluctuations can also be useful as CIBER results indicate 
that the mean CIB spectrum is redder than the zodiacal 
light \citep{Matsuura:2017}. At longer wavelengths, 3.6 - 4.5 $\mu$m, 
\citet{Arendt:2016} have shown that the zodiacal light does not affect the 
power from clustering in large scale fluctuations, but it does contribute to the white noise
component of the power spectrum.

\subsection{Probing the power spectrum and its Lyman break}
\label{sec:librae-power}

\begin{figure*}
\includegraphics[width=7in]{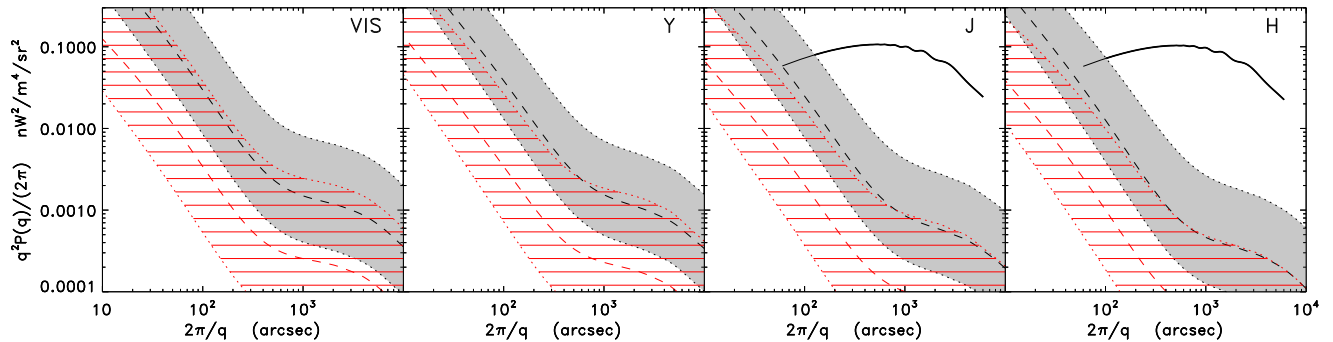}
\caption{\small  
HRK12 reconstruction of fluctuations from known galaxies remaining 
in the {\it Euclid} VIS and NISP bands (Table \ref{tab:euclid}) is shown with grey shaded area for Wide Survey and red lined area for Deep Survey; dashed line denotes the default reconstruction and  dotted lines show the HFE to LFE limits for each band. The thick solid line from \citet{Kashlinsky:2015}  is an example of high-$z$ CIB, which fits {\it Spitzer} 3.6, 4.5 \mic\ CIB fluctuation data, based on the IMF500 model from \citet{Helgason:2016} with $f_*=0.04$ ending at $z_{\rm end}=10$; it has no emissions in Y and VIS bands due to the Lyman cutoff of the source emissions. 
}
\label{fig:librae-power}
\end{figure*}
The configuration of {\it Euclid}'s near-IR and visible bands, and the coverage and depth of the surveys, are all uniquely suitable in probing, highly accurately, source-subtracted CIB fluctuations from new sources at early times. The large area covered by the {\it Euclid}'s Wide Survey enables measuring the fine features of their power spectrum with unprecedentedly high precision and the Deep Survey allows probing the clustering component at unprecedentedly faint depth.

Mask corrections for the evaluated power are not expected to be important with the {\it Euclid} configuration: at 0.3$''$ resolution there would be $\sim 1.44\times 10^{8}$ pix/deg$^2$, whereas at the depth of the Deep survey there would be $\sim 2\times 10^5$ sources/deg$^2$ according to deep counts at the near-IR NISP bands. Even taking conservatively $\sim 30-50$ pixels/source on average, the mask would eliminate only  a few percent of the pixels in the Wide Survey and a bit more in the Deep Survey. However, although the mask fraction is small enough to enable robustly accurate power computation, the images would already be in the confusion limit for their deepest sources, increasing the usefulness of CIB studies.
In this limit the correction for masking, even if necessary in this high-precision measurement, can be done using a methodology of  \citet{Kashlinsky:2012} adopting a high-accuracy template as prior, running it through the mask in simulated maps and comparing the output to measurements. At this resolution removing sources at $2.5\sigma$, or even more aggressively, would be possible; in this limit only $\sim 1\%$ of the noise pixels would be additionally removed at this threshold. This makes the expected total fraction of removed pixels comfortably below 10\%.

Fig.\ \ref{fig:librae-power} shows the advantages provided by the {\it Euclid} configuration for CIB power measurements. The HFE to LFE range of CIB fluctuations from the HRK12 reconstruction of remaining known galaxies is shown with shaded regions for each configuration. Thick line shows a high-$z$ CIB example, which fits {\it Spitzer} 3.6 and 4.5 \mic\ CIB fluctuation data \citep[][]{Helgason:2016,Kashlinsky:2015}. The figure shows that such CIB fluctuation component can be robustly resolved in the presence of the known galaxies remaining here. The bulk of the fluctuation signal is contained between $\sim$1$^\prime$ and a few degrees with the peak near $10'$--$15'$ corresponding to the $\Lambda$CDM power spectrum projected to the distance of the emitting sources. The large total area available for the CIB maps
would enable the CIB power measurement with better than sub-percent statistical accuracy below $1\deg$ assuming that 10,000 deg$^2$ would be useful for CIB analysis. In the 40 deg$^2$ area of the Deep survey, the power spectrum will be measured with better than $\lsim 15\%(\theta/1^\circ)$ statistical accuracy on sub-degree scales. Thus the fine structure of the CIB can be resolved with high statistical accuracy in both {\it Euclid} configurations.

The figure also shows that in this configuration the Lyman break of the high-$z$ CIB component can be probed very robustly. In both the  VIS and Y bands the levels of remaining known galaxies are comfortably below the high-$z$ component normalized to the measured source-subtracted CIB from {\it Spitzer} and which is prominent at the {\it Euclid} J and H bands.

Fig. \ref{fig:clustering_sn_irac} shows that the clustering component of the  CIB fluctuations does not yet appear to decrease, within the measurement errors, with the lower shot noise reached in deeper IRAC integrations. As discussed earlier this sets strong constraints on nature of the individual sources producing these CIB anisotropies and finding the shot noise level which starts affecting (decreasing) the large scale clustering component of the CIB would provide important information about the sources producing it. The Wide and Deep survey of {\it Euclid} appear suitable for probing with good accuracy the clustering component as function of shot noise out to significantly lower depths and larger angular scales than hitherto possible.

To conclude this discussion, the {\it Euclid} parameters, designed for independent dark energy studies, are well positioned to 1) probe the fine structure of the CIB power spectrum highly accurately, 2) determine directly the epochs of the sources producing them from the Lyman cutoff by comparing with the signal at the shorter wavelengths, and 3) probe the behavior of the clustering component as one reaches significantly lower shot-noise levels.

\subsection{Probing BH contribution: CXB-CIB crosspower}
\label{sec:librae-erosita}

In addition to the already operating X-ray satellites, {\it Chandra} and XMM-{\it Newton}, 
the expected 2018 launch of eROSITA
will be of significant importance for the LIBRAE measurements of the CXB-CIB cross-power. 
eROSITA\footnote{\url{http://www.mpe.mpg.de/eROSITA}} (extended Roentgen Survey with an Imaging Telescope Array) is an 
instrument developed by the Max-Planck-Institute for Extraterrestrial Physics together
with the German Space Agency DLR, to fly on Russia's space mission 
Spektrum-RG (SRG). 
It consists of an array of 7 Wolter-Type I nested mirror systems with 7 X-ray CCD detectors in the focal planes. 
It will perform an all sky X-ray survey in the [0.1-12] keV range. In the [0.1-2] keV band, the survey 
will be $\sim$30 times deeper than the ROSAT All-Sky-Survey, while in the [2-12] keV band
eROSITA will perform the first ever all-sky survey with a focusing X-ray telescope.
The eROSITA active field of view will be $\sim$1 deg$^2$, which together with the 
large collecting area gives eROSITA a grasp of $\sim$ 1000 cm$^2$ deg$^2$ at [0.5-2] keV, 
about 3$\times$ larger than the combination of the 3 
XMM-Newton telescopes \citep{Merloni:2012}. The eROSITA PSF will have a half power
 diameter (HPD) of $\sim15''$ on axis and 28$''$ in survey mode. 

\begin{figure}[t]
\includegraphics[width=3.2in,height=1.6in]{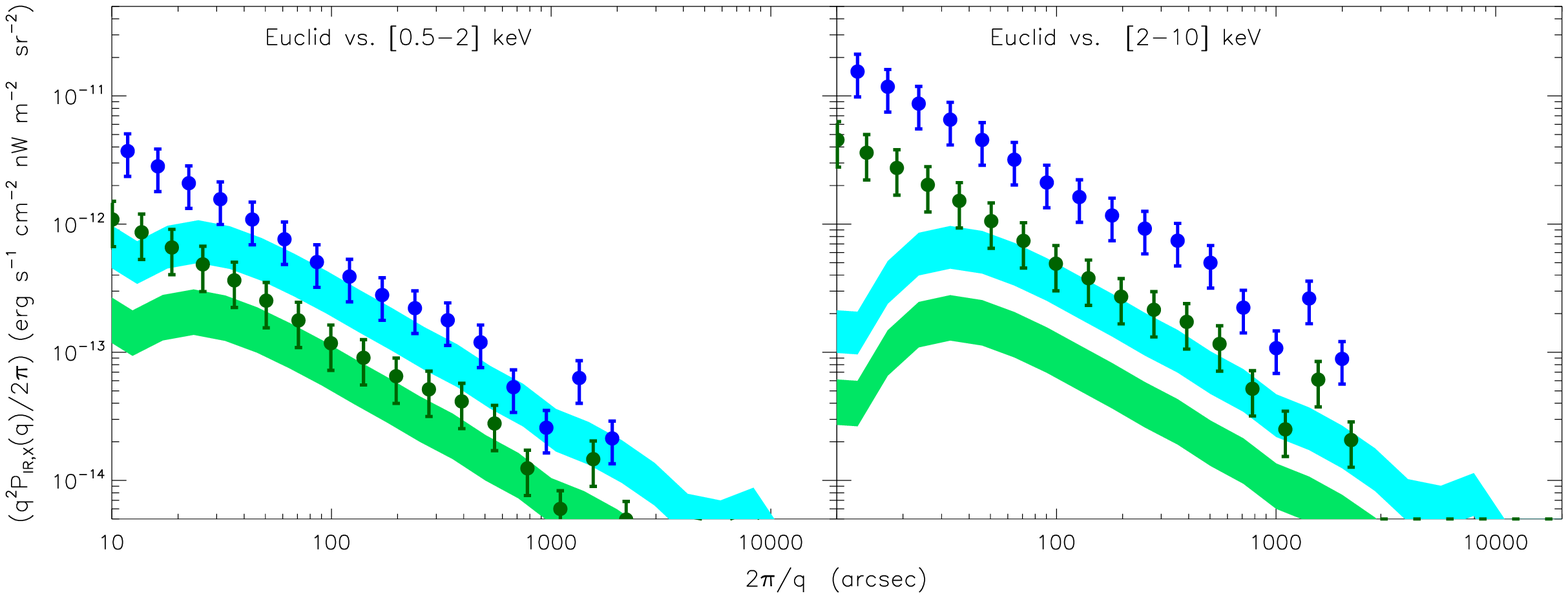}
\caption{\small {\bf Left}: estimates of the Euclid vs. [0.5-2] keV cross-power noise floor.
 Solid bands represent the 
Euclid vs. eROSITA configuration for 40 deg$^2$ survey described here. 
The symbols represent the 3$\sigma$ detection floor expected
 from the HRK12 reconstruction of contributions from
known galaxies remaining (Table V) in the shown binning.  Cyan (upper) shading shows the floor for the {\it eROSITA} 
vs  {\it Euclid}-VIS cross-power; green  {\it eROSITA} 
vs  {\it Euclid}-H. 
Blue and dark green (upper and lower)  dots (mean value from Table IV) with intervals (HFE and LFE),
 represent configurations 
 for  {\em Chandra} vs {\it Euclid}/VIS  and  {\em Chandra} vs. {\it Euclid}/H respectively. 
{\bf Right}: Same as left for [2-10] keV band.}
\label{fig:fig_erosita}
\end{figure}

The eROSITA All-Sky Survey (eRASS) will map the entire sky with a cadence of 6 months and an average exposure of $\sim$2.5 ks, plus two $\sim$100 deg$^2$ regions of deep survey, with an exposure of the order of 100 ks at the North and South Ecliptic poles (NEP, SEP), corresponding to limiting sensitivity of 1.1$\times$10$^{-14}$ erg cm$^{-2}$ s$^{-1}$ and 2$\times$10$^{-15}$ erg cm$^{-2}$ s$^{-1}$, respectively.
In the deep survey the source density will be of the order $\sim$400 deg$^{-2}$. By masking the X-ray sources with a radius of the HPD, about 90\% of the image pixels will be available for the analysis, but still a significant fraction of the source flux will leak outside the mask, which will contaminate the diffuse light estimate. 

Since the uncertainty on the cross-power spectrum is proportional to 
the square root of 
the survey area, eROSITA and Euclid will provide results to better than $\lsim10\%$ statistical uncertainties out to 1\deg--2\deg scales. 
LIBRAE will  cross-correlate the Euclid 
bands with eROSITA X-ray bands and, on smaller areas with {\em Chandra} or XMM-{\em Newton}. In theory, if the observed CXB-CIB cross-power is due to accreting BHs at high-$z$, we do not expect to measure similarly strong coherence between the CXB and the (VIS, Y) as with J, and H bands if the typical sources lie at $z>$(4, 6.5).

LIBRAE will measure the average X-ray spectrum of the sources contributing to the excess fluctuations by cross-correlating multiple X-ray bands with one or more near-IR bands. The eROSITA range could be divided into energy bands compatible with the X-ray CCD energy resolution, that reaches 60 eV at low energies and  $>$100-150 eV at high energies ($>$2 keV). However, in a tradeoff between energy resolution and signal-to-noise ratio a reasonable  energy resolution in the X-ray band will be of the order of 0.5 to 1.5 keV. The X-ray spectrum carries information about the amount of intrinsic absorption in the host galaxy (measured with the depression of the low energy signal) and the accretion rate (measured with the spectral index). This information, combined with a precise measurement of the clustering properties, 
may constrain the number density of the sources. These are very important quantities since, if these are high-$z$ BHs, the number density, accretion rate and columns density are key descriptors of these BH populations. 
If the sources are instead at low-$z$, these are important to characterize the nature of these  populations. 

How well can {\it Euclid} and eROSITA (and {\it Chandra}) measure any CXB-CIB cross-powers? To provide an estimate of the noise floor for these forthcoming measurements by using realistic assumptions of the instrument configurations, we have simulated Euclid deep survey fluctuation maps of 40 deg$^2$ where the signal is only produced by the shot-noise from unresolved discrete sources. The shot-noise levels used are from Table \ref{tab:euclid}  and we  varied them according to assumptions on  the faint end of the luminosity function adopted in the extrapolation to faint {\it Euclid} limiting magnitudes \citep{Helgason:2012a}. 
For eROSITA we simulated two deep survey maps of 40 deg$^2$, each, with 100 ks exposure in the [0.5-2] keV and [2-10] keV bands, respectively. The maps have an average count-rate of 2.14 and 0.92 ct/s/arcmin$^2$ \citep{Merloni:2012} in the two bands, respectively. The count rates have been converted into count maps with an exposure map and Poisson noise was added. The maps have been then transformed into surface brightness maps with an energy conversion factor given by the instrumental response and finally into fluctuation maps in units of erg cm$^{-2}$ sr$^{-1}$. The same procedure has been adopted to create simulated noise maps with the typical configuration of a Chandra medium-deep field of $\sim0.2$ deg$^2$, and an exposure of 400 ks, by using the background values in the EGS field analysis by \citet{Cappelluti:2013,Cappelluti:2017}. 
 
From those maps we estimated the 3$\sigma$ noise floor of the cross-power spectrum for every combination of band and instrument configuration. In Fig.\ \ref{fig:fig_erosita} we show the computed 3$\sigma$ upper limit on the noise power, i.e.\ $3 \sqrt{P_{IR}(q)P_X(q)/N(q)}$, where $P_{IR}(q)$ and $P_X(q)$ are the IR and X-ray power spectra, respectively. The eROSITA vs.\ {\it Euclid} noise floors will be systematically lower (up to one dex) than those of {\em Chandra} and {\it Euclid}. Such a  difference is driven mostly by the larger area sampled by eROSITA, despite the shallower depth compared to {\em Chandra}. Noteworthy in this context is the much higher hard X-ray band sensitivity of eROSITA than {\em Chandra}.
 
The expected mean squared amplitude $q^2P_{IR,X}/2\pi$ for the noise floor is  $\lsim$0.5-5$\times$10$^{-13}$ nW m$^{-2}$ erg cm$^{-2}$ sr$^{-2}$ on scales $\sim100''-500''$, assuming binning of $\Delta \log(q)=0.15$. The observed 3.6/4.5$\mic$ vs [0.5-2] keV cross power in the same angular range is $\sim (3-5)\times10^{-11}$ nW m$^{-2}$ erg cm$^{-2}$ sr$^{-2}$. This means that we can measure drop-outs in the cross-power between X-ray versus 3.6 $\mu$m and X-ray versus {\it Euclid} H -- VIS bands of the order $\sim10^2$. This corresponds to [3.6]--[H] through [VIS] colors of up to $\sim 5$ magnitudes, which are sufficient to obtain significant measurements of the Lyman break. However, this means that, regardless of the nature of the sources, it will be possible for the first time to infer the properties of source populations with a detailed measurement of the broad band SED of the EBL fluctuations. 

\subsection{Probing IGM at pre-reionization: CMB-CIB crosspower}
\label{sec:librae-cmb}

At high-$z$, the early sources, responsible for CIB fluctuations, would have ionized and heated up the 
surrounding gas which, in principle, would generate secondary anisotropies in
the CMB via the TSZ
effect. Given that {\it Euclid} will cover $\sim 20,000$ deg$^2$ with sub-arcsecond resolution at three near-IR 
channels, this weak signal may be teased out of the noise, after 
suitable construction of a comparably large-area, low noise, multifrequency 
CMB maps at roughly arcminute resolution which are expected to be available 
in the near future. \citet{Atrio-Barandela:2014} show how such measurements can lead to a highly 
statistically significant result. At the same time, the CIB signal from high-$z$ should have no correlation with the 
diffuse emission maps obtained from the {\it Euclid} VIS channel if the sources' epochs are such that the Ly$\alpha$ line is redshifted beyond 0.9 \mic; this would facilitate isolating the CMB-CIB cross-power from high $z$. 

For example, massive Population III stars have approximately constant 
surface temperature 
$T_*\sim 10^5$K producing a large number of ionizing photons with
energy $\geq 13.6$eV, and resulting in a constant ratio of the ionizing photons 
per H-burning baryon in these objects. There would be $\sim 10^{62} M_*/M_\odot$  ionizing photons produced 
over the lifetime of these stars ($\sim3\times10^6$yr) \citep[][]{Bromm:2001,Schaerer:2002} by a halo containing $M_*$ 
in such sources. If $\kappa$ ionizing photons are required to ionize a H atom, 
around each halo containing $M_*$ in stars there will be a bubble of 
$M_{\rm ion} \sim 10^5 \kappa^{-1}M_*$ ionized gas, heated to a temperature of
$T_e\equiv T_{e,4} 10^4$K. If the electron temperature $T_e$ and density $n_e$ 
are constant, the Comptonization parameter averaged over the solid angle 
$\omega_B$ subtended by the bubble would be 
$Y_{C,B}=(4/3)\sigma_Tn_eR_{\rm ion}(kT_e/m_ec^2)$, 
where $R_{\rm ion}$ is the radius of the ionized cloud. Each ionized bubble would generate a CMB mean distortion 
over an area of solid angle 
$\omega$ given by $t_{\rm TSZ,B}=G_\nu Y_{C,B} \frac{\omega_B}{\omega}T_{\rm CMB}$
where $G_\nu$ is the frequency dependence of the SZ effect.
The net distortion will be the added contributions of all bubbles
in the CMB pixel,
$T_{\rm TSZ}= n_2\omega t_{\rm TSZ,B}$, where $ n_2\omega$ is the total number
of bubbles along the line of sight on a pixel of solid angle $\omega$. 

Since the shot noise power
$P_{\rm SN} \sim F_{\rm CIB}^2/n_2$, the sky density of 
these sources is given by eq.\ \ref{eq:n2_kamm3}, leading to
\begin{eqnarray}
T_{\rm TSZ} & \simeq
& \frac{4}{\pi}G_\nu T_{\rm CMB} \frac{k_BT_e}{m_ec^2}
\frac{\sigma_T}{d_A^2}\frac{M_{\rm ion}}{\mu m_H}\frac{F^2_{\rm CIB}}{P_{\rm SN}} \simeq
\nonumber
 \\
 & & 200G_\nu\left(\frac{0.5{\rm Gpc}}{d_A}\right)^{2}
\frac{M_*}{10^4\kappa\mu M_\odot}T_{e,4}
\label{eq:tsz_bubble}
\end{eqnarray}
Here, $F_{\rm CIB}$ is the net CIB flux from these sources in nW/m$^2$/sr, $\mu$
is the mean gas molecular weight and $k_B$ the Boltzmann constant. $M_*$ 
corresponds to a conservative choice for the mass of the ionizing sources 
in each early halo and the proper angular diameter distance $d_A$=0.5--0.9 Gpc at $z$=20--10. 
For the above, the effective Thomson optical 
depth due to the reionized medium 
$\Delta \tau \equiv200 {\rm nK}/[T_{\rm CMB}(k_BT_e/m_ec^2)]$=0.044 is below
the measurement values in Table \ref{table:tau}.

\begin{figure}[t]
\centering
\includegraphics[width=3in]{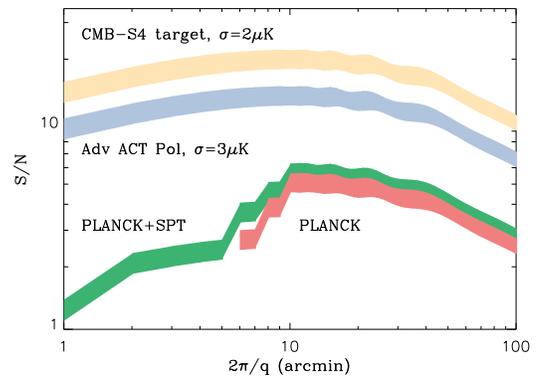}
\caption[fig:fig1]{\small Adapted with modifications from \citet{Atrio-Barandela:2014}. 
Filled regions show the range of the S/N of the CIB-TSZ cross power 
over the {\it Euclid} Wide Survey region covered by the model parameters for 
different experimental configurations for $T_{e,4}$=1. {\it Planck} parameters are for 2 yrs of integration.
At 5$'$ only 353--217 GHz difference maps would be useful, at 7$'$ we 
also add 143--217 GHz, and at $>9'$ we can add the data from 100--217 GHz.
SPT has lower S/N, but can probe angular scales as low as $\sim 1'$. In 
its current configuration the ACT does not add appreciably to the 
measurement, but that can be improved with Advanced ACT and CMB Stage 4 
experiments as shown with the blue band.
}
\label{fig:fig1_abk}
\end{figure}
Due to variation in the number density of bubbles with a relative number
fluctuation of $\Delta\simeq 0.1$, the CMB distortion $T_{\rm TSZ}$ would 
generate CMB temperature fluctuations. The 
TSZ temperature anisotropies would have amplitude $\sim T_{\rm TSZ}\Delta$
that is potentially detectable {\it by cross-correlating the produced CMB 
anisotropies with CIB fluctuations}. 
For bubbles coherent with CIB sources,
the cross-power between CIB and TSZ is $P_{\rm CIB \times TSZ} 
\simeq \sqrt{P_{\rm CIB}}\sqrt{P_{\rm TSZ}}$. To compute this
cross-correlation, the sub-arcsecond {\it Euclid} CIB 
and arcminute resolution CMB maps will be brought to a common 
resolution. When measuring the cross-power from 
IR and microwave maps (mw) of $N_{\rm pix}$ CMB pixels, the error is 
$\sigma_{P_{\rm CIB \times TSZ}}\simeq \sqrt{P_{\rm IR}}\sqrt{P_{\rm mw}}/\sqrt{N_{\rm pix}}$, since at $\gsim 1^\prime$ the 
{\it Euclid} CIB maps will have negligible noise with $P_{\rm IR}$=$P_{\rm CIB}$. From the {\it Euclid} Wide Survey
the CIB power on arcminute scales will
be measurable by LIBRAE to sub-percent statistical accuracy. 
If primary CMB is removed, the foreground-reduced
microwave maps would be dominated by instrument noise $\sigma_{\rm n}$, foreground
residuals $\sigma_{\rm f,res}$ and, more importantly, the TSZ of the unresolved 
cluster population $\sigma_{\rm cl,unr}$. With $N_\nu$ microwave frequency channels the 
variance of the microwave map would be 
$\sigma_{\rm mw}^2=\sigma_n^2/N_\nu+\sigma_{\rm f,res}^2+\sigma_{\rm cl,unr}^2$.
The signal-to-noise would be
S/N$\simeq T_{\rm TSZ}\Delta\sqrt{N_{\rm pix}}/\sigma_{\rm mw}$, reaching 
S/N$\gg 1$ for certain experimental configurations discussed below. Specifically
\begin{equation}
{\rm S/N} = 7 \;\;\frac{T_{\rm TSZ}}{200nK}\;\frac{\Delta}{0.1}\;
\left(\frac{\sigma_{\rm mw}}{5\mu\rm K}\right)^{-1}\left(\frac{N_{\rm pix}}{3\times10^6}
\right)^{\frac{1}{2}}\;
\label{eq:s2n_cmb_cib}
\end{equation}
where $N_{\rm pix}=3\times10^6$ is the expected sky coverage
of the {\it Euclid} Wide Survey at the native {\it Planck} resolution of $5'$. At the same time, emissions from early times, $z\gsim 10$, should exhibit no correlations  at VIS and, likely, Y bands with CMB enabling the measured cross-power from the {\it Euclid}'s longest bands to be uniquely interpreted.

The CMB-CIB cross-power peaks around $\sim 10'$, scales which can be probed
with the forthcoming CMB instruments that plan to cover large areas of the sky with noise of $\sigma_n\lsim$ a few $\mu$K. Also important in eliminating the contribution of primary CMB and the KSZ terms to the measured signal is availability of multiple frequencies covering both sides of the TSZ zero frequency at $\sim$ 217 GHz; any components having black-body energy spectrum can then be eliminated in taking $T$-differences at different frequency pairs as proposed in \citet{Atrio-Barandela:2014}.  
The CMB data from the forthcoming experiments planned to complete by the time of the {\it Euclid} surveys would reach higher S/N of the CMB-CIB cross-power than the combined 
Planck+SPT data shown in Fig. \ref{fig:fig1_abk}.
In its first two years of observation, the ACTPol camera observed 
$\sim 600$ deg$^2$ at 149 GHz with a noise level of 17 $\mu$K-arcmin and
a resolution of $1.3'-2'$ for the different arrays \cite{Sherwin:2016}. 
The NSF-supported new Adv ACTPol camera 
will observe in five
bands spanning the 25-280 GHz range with a resolution of $\sim 1.5'$ \cite{Ward:2016},
similar to the currently operating ACTPol. As of this writing the AdvACTPol configuration is planned to map 
$\sim 10^4$ deg$^2$ with sub-arcminute resolution of $1.3'$ FWHM 
at frequencies of $\sim 97$ and $\sim 148$ GHz \cite{Thornton:2016}. 
At the smallest angular scales, the Silk-Michie damping suppresses the primary 
CMB temperature anisotropies to an amplitude $\delta T\lsim 1\mu$K 
leaving the variance of the microwave map dominated by the instrument
noise and possible foreground residuals. 
 The larger frequency coverage 
will allow to efficiently remove foregrounds. Although the final sensitivity 
and total observing area are yet to be determined, the noise is a factor of $\sim$5--6 lower than in the current data,
and the new camera can observe $\gsim 2\times 10^3$ deg$^2$, covering $\sim 3\times10^6$ pixels in a reasonable
amount of time. Then the cross-correlation of a single 
map with the source subtracted CIB data
will reach the S/N shown in Fig. \ref{fig:fig1_abk}, if foreground residuals 
are negligible. These results are easily scalable to other noise levels and different configurations, since 
$S/N\propto(N_{\rm pix}/\sigma_{\rm n}^2)^{1/2}$ where $N_{\rm pix}$ is the 
number of pixels in the survey area, and $\sigma_{\rm n}^2$ is the noise 
variance of the observations.
The CMB-CIB cross-power can be determined with a statistical S/N$\sim 25$ if the CMB-S4 generation of
experiments currently being designed reach their noise target of $\le 2\mu$K-arcmin with an angular
resolution of  $\lsim2^\prime$ \cite{Abitbol:2017}.

\begin{figure*}[t!]
\includegraphics[width=6.5in]{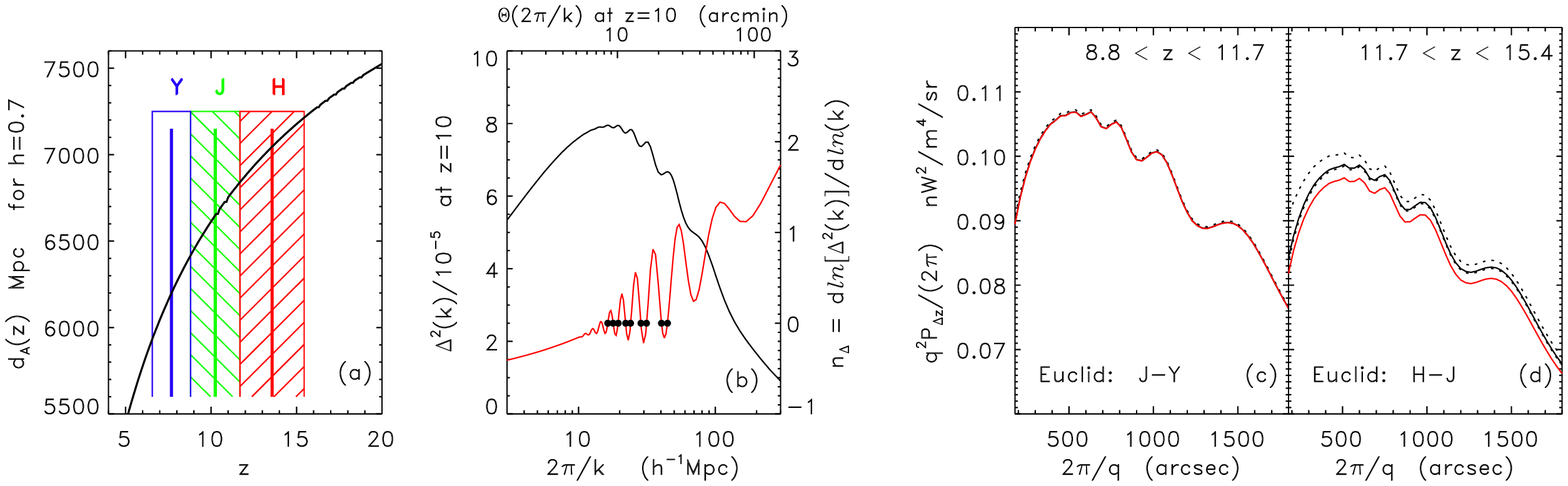}
\caption[]{ \footnotesize 
Adapted from \citet{Kashlinsky:2015}. {\bf (a)} Solid line shows the $d_A$ span vs. $z$ over the Y, J, and H Euclid filters; vertical lines mark the central wavelength of each filer.  At these $z$ the Lyman-break corresponds to Ly$\alpha$ at 0.1216 $\mu$m. {\bf (b)}  Solid line shows $\Delta^2(k)$ at $z$=10. Red line and right axis show the spatial spectral index, $n_\Delta$, of $\Delta^2(k)$ with solid dots marking its extrema. For the HZ regime,  $n_\Delta$=3, which is reached at larger scales.{\bf (c),(d)} The Lyman-tomography reconstruction of the history of emissions and BAOs for {\it Euclid}'s (Y,J,H) filters and Wide Survey depth at each redshift range displayed in red.  Red line (starting near bottom on left) shows the underlying CIB fluctuations by sources in the marked $z$-range from high-$z$ stellar populations reproducing {\it Spitzer} measurements. Black lines show the reconstructed one with contributions by known remaining galaxies from HRK12 with dotted showing the HFE to LFE limits.}
\label{fig:fig_librae_lyman}
\end{figure*}
\subsection{History of emissions from Lyman tomography}
\label{sec:librae-lytomography}

 The NISP filters, depth and sky coverage available for LIBRAE with the {\it Euclid} mission appear particularly useful for the application of the Lyman tomography method, Sec.\ \ref{sec:ly-tomography}, \cite{Kashlinsky:2015}. 
 The large areas covered by the {\it Euclid} surveys enable high precision measurement of the source-subtracted CIB power spectra in each of the bands which then allow accurate construction of the quantity $P_{\Delta z}\equiv P_2-P_{12}^2/P_1$ in each of the two NISP band pairs [J--Y,H--J], with subsequent wavelengths ordered $\lambda_2>\lambda_1$. In terms of $z$-range, the Y--J configuration then covers CIB emissions over $8.8<z<11.7$ and the J--H isolates CIB from $11.7<z<15.4$; the upper/lower redshifts of these ranges will be denoted $z_1/z_2$. This assumes the Lyman-cutoff at rest Ly$\alpha$ of $\lambda_{\rm Ly}=0.12\ \mic$, appropriate for the pre-full ionization conditions at these epochs.
Fig.\ \ref{fig:fig_librae_lyman}a shows that these {\it Euclid}  filters isolate emissions over a narrow range of distances about $\simeq 5-7\%$ in comoving $d_A$, centered at $d_0$. For reference, $2\pi/q=10'$ corresponds to $\ell$=2,160 and subtends comoving scale of $\simeq 20h^{-1}$Mpc at $z$=10, scales that are in highly linear regime at those epochs.

The power spectrum of the emitting sources would be proportionally related to the  underlying $\Lambda$CDM 
one, since the relevant angular scales subtend tens of comoving Mpc where the density field was highly linear. Because the 
procedure isolates a narrow shell in $d_A(z)$ around $d_0$, the comoving angular distance to the central filter wavelength, one can further expand $\Delta^2_{\Lambda CDM}(q/d_A) \simeq \Delta^2_{\Lambda CDM}(qd_0^{-1})\{1-n_\Delta(qd_0^{-1})[\frac{\delta d_A}{d_0}]\}
$, where $n_\Delta(k)\equiv d\ln\Delta^2_{\Lambda CDM}(k)/d\ln k$ is the spatial spectral index of the $\Lambda$CDM template and $\delta d_A\equiv d_A(z)-d_0\lsim (5-7)\%d_0$. Fig.\ \ref{fig:fig_librae_lyman}b shows the template expected from the concordance $\Lambda$CDM model and the spectral index $n_\Delta$ then. Given the narrow range of $d_A$ spanned by each {\it Euclid} filter for the Lyman tomography and the values of $n_\Delta$ the power
from sources over the narrow range of epochs can be approximated as:
\begin{equation}
\frac{q^2P_{\Delta z}}{2\pi}\simeq \Delta^2_{\Lambda CDM}(qd_0^{-1}) 
\int_{z_{\rm Ly}(\lambda_1)}^{z_{\rm Ly}(\lambda_2)} \left(\frac{dF_{\lambda_2^\prime}}{dz}\right)^2
 dz
\label{eq:p_df_1}
\end{equation}
Eq.\ \ref{eq:p_df_1}  shows that 1) history of emissions over $z_{\rm Ly}(\lambda_1)<z<z_{\rm Ly}(\lambda_2)$ is recoverable in the {\it Euclid} adjacent filter configurations and 2) the resultant $P_{\Delta z}$ preserves information about underlying parameters over these $z$. Both can be recovered in the LIBRAE CIB measurements.

Figs.\ \ref{fig:fig_librae_lyman}c and d illustrate the potential accuracy of this procedure in recovering the history of CIB emissions with an example normalized to reproduce {\it Spitzer} fluctuations at 3.6 and 4.5 \mic.  It is taken from \citet{Kashlinsky:2015} using 1) an IMF500 modeling \cite{Helgason:2016} and 2) a {\it Euclid}-specific reconstruction of the contribution from remaining known galaxies from HRK12. Red lines show the true history of the emissions inside halos collapsing according to the standard $\Lambda$CDM model. Incoherence due to remaining known galaxies is explicitly incorporated in this example and the history recovered with this method is shown with black lines covering the span of systematic uncertainties of the reconstruction. Except a slight bias upward of a few percent, the history of emissions appears recovered accurately with the {\it Euclid} configuration, even preserving the BAO features in the underlying power spectrum in this example. \citet{Kashlinsky:2015} show that good accuracy is achieved even when only a few percent of the {\it Spitzer}-based CIB fluctuations originate at high $z$. In practice, with the {\it Euclid} CIB measurements the true $z$ will be verified by measuring distance from fitting the angular template which appears accurately recoverable.

\subsection{Probing BAOs  and dark energy at $10< z < 16$}
 \label{sec:librae-bao}

{\it Euclid}'s goal is to explore the Universe's expansion history 
to understand the origin of the current accelerated period and the nature
of DE by measuring the clustering of galaxies out to $z\simeq2$ and the
weak lensing distortion out to $z\le 3$. \citet{Amendola:2016} summarize the main observables to be extracted from the
data to forecast future performance of the satellite
in testing the various models.  The Ly tomography described
above 
will, in principle, be possible using all 4 Euclid filters, VIS, Y, J, H. The results will contribute to these goals by exploring the BAOs and cosmological parameters at redshifts $6\lsim z\lsim16$, much
higher than those available with the standard techniques \cite{Kashlinsky:2015}. 

\begin{figure}[t]
\centering
\includegraphics[width=3in]{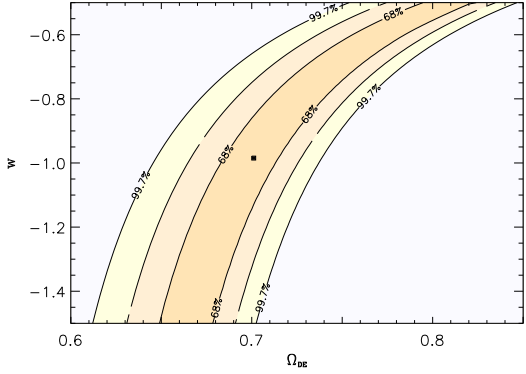}
\caption{
Constraints on the DE equation of state
parameter $w$ and DE density $\Omega_{DE}$
from the Ly$\alpha$ tomography with the Euclid diffuse maps using differencing configurations of [Y--VIS], [J--Y] and [H--J]. Contours correspond
to the 68, 95.5, 99.7\% ($1,2,3\sigma$) confidence levels as marked. 
}
\label{fig:bao}
\end{figure}
From the derived CIB maps used in the tomographic reconstruction, [Y--VIS],  
[J--Y] and  [H--J], isolating populations over $\Delta d_A\ll d_A$, 
LIBRAE will test the expected  
$\Lambda$CDM template at this new range of $z$. The location and amplitude of the
maximum and BAOs imprinted in the matter
power spectrum can be used then to determine cosmological parameters.
To that purpose, the power spectrum needs to be sampled with
sufficient angular resolution. In the frequency domain the resolution 
$\Delta q$, is set by maximum size, $\Theta_0$, of the region being analyzed:
$\Delta q=2\pi/\Theta_0$. To achieve a resolution of $\Delta\theta=0.5^\prime$
requires $\Theta_0\sim 20^\deg$ \cite{Kashlinsky:2015}. In Fig.~\ref{fig:bao} 
we plot the power spectra of $\Lambda$CDM models with various 
$\Omega_\Lambda$=[0.635,0.68,0.725]. The data are centered around the 
$\Lambda$CDM matter power spectrum with $\Omega_\Lambda$=0.68. These cosmic variance errors do not include instrumental noise and systematic 
effects; the data are taken from a rectangular patch of area $20^\circ\times10'$. 
Since the power is measured from CIB fluctuations that are biased
with respect to the underlying matter power spectrum, the data
constrains the overall shape but not its amplitude. 
Consequently, all cosmological parameters that modify the shape and location
of the acoustic peaks, such as $\Omega_{\rm bar}$, $\Omega_K$, massive neutrino energy
density, etc, can be constrained by the tomographic reconstruction of the power spectrum.

BAOs encode information about the sound horizon at recombination,
whose value is $r_s=144.81\pm 0.24$~Mpc \cite{Planck_Collaboration_XVI:2014}. 
The angular scale subtended by the sound horizon can be measured from
the correlation function of galaxies to 
derive angular diameter distances at epochs probed by
galaxy catalogs \cite{Eisenstein:1998}. The technique can also be applied
to the frequency domain \cite{Percival:2010} to constrain 
$\Omega_{\rm DE}$, the DE equation of state, $p=w\rho c^2$, and/or the interactions 
within the dark sector \cite{Wang:2016}. The dynamical evolution of 
DE affects reionization of the Universe \cite{Xu:2017} and Ly
tomography will provide angular diameter distances to that $z$, allowing
us to test the effect of models on an epoch that
can not be probed with current techniques \cite{Aubourg:2015}.
Fig~\ref{fig:bao} shows constraints on the equation of state
$w$ and energy density $\Omega_{\rm DE}$ 
at the $68, 95$ and $99.7\%$ confidence levels
derived applying the Lyman tomography technique to {\it Euclid} data.
While the method may not constrain the parameters as well as other techniques, 
it extends the BAO regime to epochs not yet tested and could be complemented
with measurements at lower $z$ such as those of e.g. \citet[][]{Hemantha:2014,Wang:2014}
to put stronger constraints on DE properties \citep[][]{Wang:2006}.
In addition, it gives an important self-consistency test and could supply 
valuable information to resolve the current discrepancy in BAO measurements at
$z\!<$1 and $z\!\simeq$2--3 \cite{Aubourg:2015}.
\subsection{LIBRAE summary}
\label{sec:librae-summary}
\begin{figure}[t]
\hspace{-1.cm}
\includegraphics[width=2.75in,angle=-90]{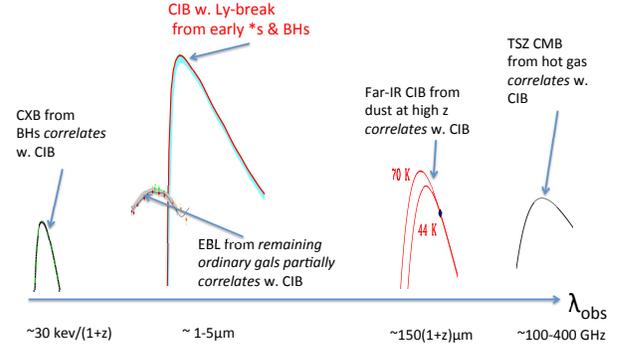}
\caption{\small  Panoramic diagram of the LIBRAE goals.
}
\label{fig:lbiraegoals}
\end{figure}
LIBRAE's goals are summed panoramically in Fig.\ \ref{fig:lbiraegoals}:
\begin{itemize}
\item Measuring power spectrum of source-subtracted CIB fluctuations at near-IR to sub-percent statistical accuracy with the Wide Survey's NISP data.
\item Probing epochs of sources producing the CIB fluctuations by cross-correlating with diffuse light from the Wide Survey's VIS data.
\item Probing the CIB properties as function of depth from the Deep Survey.
\item Determining the nature of the sources (BHs vs. stars) by cross-correlating with X-ray data assembled for this project.
\item Probing the condition of the IGM at pre-reionization by cross-correlating source-subtracted CIB from the Wide Survey with multi-frequency all-sky CMB data.
\item Probing the history of emissions at 1$0<z<20$ using Lyman tomography
\item Probing BAOs and DE evolutions at $10<z<20$ using the Lyman tomography.
\end{itemize}
LIBRAE will thus identify the net emissions from the first stars era, lead to a better understanding of the IGM at that epoch, isolate the contributions from the first BHs and probe the history of emissions at $9\lsim z \lsim 16$ and the cosmological parameters at those times.

\section{Other forthcoming experimental configurations}
\label{sec:otherconfigs}

$\bullet$ {\bf JWST} with its near-IR camera, NIRCam, will identify individual sources to much fainter fluxes than either {\it Spitzer} or {\it AKARI} creating an opportunity to measure cumulative CIB emissions produced at still earlier epochs or from fainter sources. The NIRCam wavelength coverage spans seven wide overlapping filters from 0.7 to 5\mic\ and so will have a built-in capability to directly probe the Lyman-break of the unresolved populations, provided the instrument noise, astronomical foregrounds and foreground galaxy populations can be isolated.  \citet{Kashlinsky:2015a} identify an experimental configuration of the {\it JWST} which, together with the strategies developed there, can provide critical insight into the origin of the source-subtracted CIB fluctuations detected in {\it Spitzer} and {\it AKARI} measurements, identify the epochs where the fluctuations arise, probe the fluxes of the sources producing them and reconstruct/constrain the history of the emissions via the proposed adjacent two-band Lyman tomography. They show that the CIB science dictates a configuration with  400 hrs of NIRCam mapping for all 7 wide NIRCam filters of 1 deg$^2$ contiguous area to $m_{\rm AB}\simeq 28$ in a low cirrus region, e.g. the Lockman Hole or CDFS. With that setup one would  be able to address important questions pertaining to the details and nature of populations that led the universe out of the ÒDark Ages.Ó They also discuss the effects of the open configuration of the {\it JWST} on the CIB study arguing that the potential stray light effects may be mitigated to yield a fundamental constraint on the otherwise inaccessible range of epochs (and fluxes) of the CIB sources that are expected to lie in the confusion noise of the {\it JWST} beam.
This measurement will supply additional important data for cross-correlating with the CIB to be measured by LIBRAE, and expand the {\it Euclid}'s reach to the greater depth and wavelength coverage available with NIRCam.

$\bullet$ {\bf WFIRST} is a flagship NASA mission\footnote{\url{https://wfirst.gsfc.nasa.gov/}} that will provide further venues for accurate measurements of source-subtracted CIB fluctuations from its deep coverage of 2,000 deg$^2$ of the sky in the planned wide survey mode. The survey will employ 4 (out of six) near-IR bands at 1, 1.3, 1.6 and 1.8 \mic\ and will have deeper, than {\it Euclid}, integrations  to $m_{\rm AB}\sim 27.5$ (2.5$\sigma)$. In addition, as of this writing, the mission is planned to have two extra channels centered at 0.6 and 0.87 \mic.
The visible channel, available around 0.6 \mic\ (J. Kruk, private communication) in the planned Guest Observer  program would allow probing the Lyman break of source-subtracted CIB fluctuations with  the {\it WFIRST} data alone, although probably from mapping a smaller area than the wide survey. 
The net sky area covered by {\it WFIRST} is an order of magnitude smaller than {\it Euclid}'s Wide Survey, but would still allow probing the CIB power spectrum with sub-percent statistical accuracy at arcminute scales. Its deeper exposures will enable probing the evolution of the CIB clustering component at still lower shot-noise levels than with {\it Euclid}. The large area of the survey, mapped at 4 near-IR channels to greater depths than {\it Euclid}, will provide an opportunity for further application of the Lyman tomography and BAO study at the $6.5\lsim z\lsim 15.5$ epochs.

$\bullet$ {\bf CIBER-2} is planned to probe EBL in six bands at $[0.6,0.8,1,1.3,1.6,1.9]\mic$ with a 28.5 cm Cassegrain telescope after removing sources to $m_{\rm AB}\sim 19$ (1.3\mic, J-band) over a $\sim\!1^\circ\times2^\circ$ FoV \cite{Lanz:2014}. The instrument will be flown to suborbital altitudes on board a series of sounding rockets. CIBER-2\footnote{\url{https://cosmology.caltech.edu/projects/CIBER2}} is tasked to ``explore cross correlations on both sides of the Lyman break to distinguish between low and high redshift components of the EBL" fluctuations. However, Fig.\ \ref{fig:fig17_jwst} shows the challenges and problems when trying to achieve this goal with CIB/EBL anisotropies in such a shallow configuration where high-$z$ component appears subdominant compared to, and is much smaller than the uncertainties in, the CIB fluctuations from remaining known galaxies. Other configurations designed to similarly shallow exposures would be subject to similar limitations.

$\bullet$ {\bf SPHEREx} is a proposed NASA MIDEX mission selected for Phase A study\footnote{\url{http://spherex.caltech.edu/}}. If selected further, it will employ Linear Variable Filters to carry out an all sky spectral survey with spectral resolution $R\simeq 41$ at [0.75--4.18]\mic\ and $R\simeq 135$ at [4.18-5]\mic. The survey will have angular resolution of $6.2''$ and depth $m_{\rm AB}\sim$ 18--19 (5$\sigma$). Although, one of the planned goals of the survey is to probe the origin of CIB fluctuations
, the shallow depth, and hence the uncertainties of the substantial component of the remaining known galaxies, would preclude reliably isolating high-$z$ CIB fluctuations as discussed in  Sec. \ref{sec:limitations} and illustrated with Fig. \ref{fig:fig17_jwst}. Additionally, the low angular resolution of the instrument, would remove large fraction of the sky at $\sim 50\%$, on a par with CIBER; this will require development and application of the correlation function tools to verify any FT-based CIB fluctuation analysis.

$\bullet$ {\bf 21 cm SKA}. Cross-correlating the CIB with the \HI\ 21cm line signal from EoR can provide additional information. To first order, if galaxies were the CIB and reionization sources, one would expect a CIB-21cm anti-correlation produced by ionized bubbles around the sources \citep{Fernandez:2014, Mao:2014}. If instead CIB sources are obscured black holes from which only X-rays can escape (i.e.\ no UV emission) the situation can be very different. With their long mean free paths and efficient IGM heating, X-rays could dramatically boost the 21 cm signal during the early EoR stages \citep{Haiman:2011,Mesinger:2013}. The anticorrelation is the strongest when the ionization fraction is about 50\%. Although there are free parameters in these models, the cross-correlation signal is rather insensitive to their variation, as many of the same parameters (as, e.g.\ the star formation efficiency, stellar mass, metallicity) affect both the infrared and the 21 cm line emission.  Cross-correlations can also reduce some of the limitations of both types of experiments, like the lack of redshift information for CIB sources. If detected, the CIB-21cm correlation will inform us precisely on the redshift distribution of the sources \citep{McQuinn:2013}. This will be made possible by the forthcoming Square Kilometer Array\footnote{\url{https://www.skatelescope.org/}} (SKA) data.

\section{Outlook for the future}
\label{sec:future}
This review summarized current observational status of the near-IR CIB anisotropy measurements and their cosmological implications. Following many new measurements and observations this novel field has recently come from a relative obscurity to significant, rapid development, to become a subject of lively scientific debate. The coming years will bring more accurate CIB fluctuation measurements with new upcoming missions. We discussed these here with a particular emphasis on the LIBRAE project which will utilize data from the {\it Euclid} Dark Energy mission, currently planned for launch in late 2020/early 2021,  for source-subtracted CIB measurements. To achieve decisive interpretation, one needs diffuse light measurements with experimental configurations that 1) reach deep exposures to be able to identify the potential Lyman cutoff of the high-$z$ CIB sources, 2) combined with availability of the space-borne data in visible bands, 3) measured over wide area to reach high accuracy determination of the source-subtracted CIB power spectrum, and to simultaneously be able to correlate the measured CIB with 4) suitable X-ray background data (from eROSITA and {\it Chandra}) to probe the contributions from accreting BHs from nucleosynthetic sources at high $z$, and 5)  multi-frequency CMB data over large areas of the sky with low noise and high angular resolution, such as planned from the currently planned surveys (AdvACTPol and CMB-S4), to identify condition of high-$z$ IGM. 
Newly developed methodologies will enable precision science with the future CIB data.

\clearpage
\section{Appendix: acronyms and abbreviations}
\label{sec:appendix}

Table \ref{tab:abbreviations} lists the common acronyms and abbreviations used throughout the review. 
\begin{table}
\caption{Common acronyms  and abbreviations.}
\begin{tabular}{|p{0.65in} | p{2.85in} |}
\hline
Acronym & Full description\\
 \hline 
ACIS-I & {\footnotesize{Advanced CCD Imaging Spectrometer - imaging arrays}}\\
AEGIS~XD & {\footnotesize{All-wavelength Extended Groth strip International Survey - X-ray, deep}}\\
AGB & {\footnotesize{Asymptotic giant branch}}\\
AKARI & {\footnotesize{A Japanese infrared satellite}}\\
ALMA & {\footnotesize{Atacama Large Millimeter/submillimeter Array}}\\
AOR & {\footnotesize{Astronomical observing request}}\\
BAO & {\footnotesize{Baryonic acoustic oscillation}}\\
BH & {\footnotesize{Black hole}}\\
BNS & {\footnotesize{Binary Neutron Star}}\\
CCD & {\footnotesize{Charge-coupled device}}\\
CIB & {\footnotesize{Cosmic infrared background}}\\
CLEAN & {\footnotesize{A deconvolution algorithm}}\\
CMB & {\footnotesize{Cosmic microwave background}}\\
COB & {\footnotesize{Cosmic optical background}}\\
COBE & {\footnotesize{Cosmic Background Explorer}}\\
CXB & {\footnotesize{Cosmic X-ray background}}\\
DAOPHOT & {\footnotesize{Dominion Astrophysical Observatory stellar photometry package}}\\
DCBH & {\footnotesize{Direct collapse black hole}}\\
DE & {\footnotesize{Dark energy}}\\
DGL & {\footnotesize{Diffuse Galactic light}}\\
DIRBE & {\footnotesize{Diffuse InfraRed Background Experiment}}\\
DM & {\footnotesize{Dark matter}}\\
EBL & {\footnotesize{Extragalactic background light}}\\
EoR & {\footnotesize{Epoch of Reionization}}\\
eRASS & {\footnotesize{eRosita all-sky survey}}\\
eROSITA & {\footnotesize{extended Roentgen Survey with an Imaging Telescope Array}}\\
FFT & {\footnotesize{Fast Fourier transform}}\\
FIRAS & {\footnotesize{Far Infrared Absolute Spectrophotometer}}\\
FoV & {\footnotesize{Field of view}}\\ 
FT & {\footnotesize{Fourier transform}}\\
FWHM & {\footnotesize{Full width half maximum}}\\
GOODS & {\footnotesize{Great Observatories Origins Deep Survey}}\\
GP & {\footnotesize{Gunn-Peterson (effect) \cite{Gunn:1965}}}\\
GW & {\footnotesize{Gravitational wave}}\\
\HI\ & {\footnotesize{Neutral hydrogen}}\\
\HII\ & {\footnotesize{Ionized hydrogen}}\\
HFE & {\footnotesize{High faint end (of HRK12 reconstruction)}}\\
HPD & {\footnotesize{Half power diameter}}\\
HRK12 & {\footnotesize{Helgason, Ricotti \& Kashlinsky \cite{Helgason:2012a}}}\\
HST & {\footnotesize{Hubble Space Telescope}}\\
HZ & {\footnotesize{Harrison-Zeldovich}}\\
 \hline
\end{tabular}
\label{tab:abbreviations}
\end{table}

\addtocounter{table}{-1}
\begin{table}
\caption{- continued}
\begin{tabular}{|p{0.65in} | p{2.85in} |}
\hline
Acronym & Full description\\
 \hline 
ICL & {\footnotesize{Intra-cluster light}}\\
IGM & {\footnotesize{Inter galactic medium}}\\
IHL & {\footnotesize{Intra-Halo light}}\\
IMF & {\footnotesize{Initial mass function}}\\
IRAC & {\footnotesize{InfraRed Array Camera}}\\
IRAF & {\footnotesize{Image Reduction and Analysis Facility}}\\
IRC & {\footnotesize{InfraRed Camera (on {\it AKARI})}}\\
IRTS & {\footnotesize{infraRed Telescope in Space}}\\
IRX & {\footnotesize{InfraRed excess}}\\
ISM & {\footnotesize{Interstellar medium}}\\
JWST & {\footnotesize{James Webb Space Telescope}}\\
KSZ & {\footnotesize{Kinematic SZ (effect)}}\\
LBG & {\footnotesize{Lyman Break Galaxy}}\\
$\Lambda$CDM & {\footnotesize{Lambda cold dark matter}}\\
LDDE & {\footnotesize{Luminosity-dependent density evolution}}\\
LF & {\footnotesize{Luminosity function}}\\
LFE & {\footnotesize{Low faint end (of HRK12 reconstruction)}}\\
LIBRAE & {\footnotesize{Looking at Infrared Background Radiation Anisotropies with {\it Euclid}}}\\
LIGO & {\footnotesize{Laser Interferometer GW Observatory}}\\
LW & {\footnotesize{Lyman-Werner}}\\
NEP & {\footnotesize{North ecliptic pole}}\\
NFW & {\footnotesize{Navarro-Frenk-White \cite{Navarro:1997}}}\\
NICMOS & {\footnotesize{Near Infrared Camera and Multi-Object Spectrometer (on {\it HST})}}\\
NIRCam & {\footnotesize{Near Infrared Camera (on {\it JWST})}}\\
NISP & {\footnotesize{Near Infrared Spectrometer and Photometer (on {\it Euclid})}}\\
NUDF & {\footnotesize{NICMOS ultra deep field}}\\
OH-glow & {\footnotesize{OH (molecular) emission}}\\
PBH & {\footnotesize{Primordial black hole}}\\
PDF & {\footnotesize{Probability distribution function}}\\
PIB & {\footnotesize{particle internal background}}\\
PSF & {\footnotesize{Point-spread function}}\\
ROSAT & {\footnotesize{R\"ontgensatellit}}\\
SED & {\footnotesize{Spectral energy distribution}}\\
SEDS & {\footnotesize{Spitzer Extended Deep Survey}}\\
SEP & {\footnotesize{South ecliptic pole}}\\
SRG & {\footnotesize{Spectrum R\"ontgen Gamma}}\\
SMBH & {\footnotesize{Supermassive black hole}}\\
SZ & {\footnotesize{Sunyaev-Zeldovich (effect) \cite{Sunyaev:1972}}}\\
TSZ & {\footnotesize{Thermal SZ (effect)}}\\
VIS & {\footnotesize{Visible instrument (on {\it Euclid})}}\\
WMAP & {\footnotesize{Wilkinson Microwave Anisotropy Probe}}\\
 \hline
\end{tabular}
\label{tab:abbreviations}
\end{table}

\newpage
\section{Acknowledgments}
\label{sec:acknowldegments}
We thank the other LIBRAE team members - Matt Ashby, Volker Bromm, Kari Helgason, Harvey Moseley - for many past and future contributions, collaborations and discussions related to this ongoing project. AK warmly thanks John Mather, Harvey Moseley and Sten Odenwald for their contributions and collaborations over the years to the CIB-related science results from DIRBE, 2MASS and {\it Spitzer}. The following colleagues are thanked alphabetically for useful discussions and information during preparation of this review: Marco Ajello, Jordan Camp, Giovanni Fazio, Alexis Finoguenov, Kari Helgason, Bob Hill, Demos Kazanas, Jeff Kruk, John Mather, Toshio Matsumoto, Harvey Moseley, Jeremy Perkins, Massimo Ricotti, Johannes Staguhn, Kohji Tsumura, Rogier Windhorst, Ed Wollack, Bin Yue. We acknowledge support from NASA/12- EUCLID11-0003 ``LIBRAE: Looking at Infrared Background Radiation Anisotropies with Euclid". We acknowledge critical support in our results reported here from past awards: NSF AST 04-06587, NASA Spitzer NM0710076 and Cycle 8 1464716, NASA ADAP NNH10ZDA001N, NASA ADAP NNX16AF29G, Chandra AR2-13014B and AR6-17017C. FA-B acknowledges financial support from grant FIS2015-65140-P (MINICO/FEDER). Literature search for this review ended on 6/6/2017.




\bibliography{references_review}

\end{document}